\documentclass{jpp}

\usepackage{graphicx}
\usepackage{natbib}

\ifCUPmtlplainloaded \else
  \checkfont{eurm10}
  \iffontfound
    \IfFileExists{upmath.sty}
      {\typeout{^^JFound AMS Euler Roman fonts on the system,
                   using the 'upmath' package.^^J}%
       \usepackage{upmath}}
      {\typeout{^^JFound AMS Euler Roman fonts on the system, but you
                   dont seem to have the}%
       \typeout{'upmath' package installed. JPP.cls can take advantage
                 of these fonts, if you use 'upmath' package.^^J}%
       \providecommand\upi{\pi}%
      }
  \else
    \providecommand\upi{\pi}%
  \fi
\fi

\ifCUPmtlplainloaded \else
  \checkfont{msam10}
  \iffontfound
    \IfFileExists{amssymb.sty}
      {\typeout{^^JFound AMS Symbol fonts on the system, using the
                'amssymb' package.^^J}%
       \usepackage{amssymb}%
         
         \let\geq=\geqslant
      }{}
  \fi
\fi

\ifCUPmtlplainloaded \else
  \IfFileExists{amsbsy.sty}
    {\typeout{^^JFound the 'amsbsy' package on the system, using it.^^J}%
     \usepackage{amsbsy}}
    {\providecommand\boldsymbol[1]{\mbox{\boldmath $##1$}}}
\fi

\providecommand\bcdot{\boldsymbol{\cdot}}

\newsavebox{\astrutbox}
\sbox{\astrutbox}{\rule[-5pt]{0pt}{20pt}}

\def\Alfven{Alfv$\acute {\mathrm{e}}$n}
\def\Alfvenic{Alfv$\acute {\mathrm{e}}$nic}
\newcommand{\diff}{\mathrm{d}}
\def\const{\mathrm{const.}}
\def\nablau{\boldsymbol{\nabla }_{\boldsymbol{u}}}
\newcommand{\jump}[1]{\ensuremath{[\![#1]\!]} }
\newcommand{\Ljump}[1]{\ensuremath{\left[\!\!\left[#1\right]\!\!\right]} }
\newcommand{\average}[1]{\ensuremath{\langle#1\rangle} }

\def\vh{\hat{v}}
\def\ph{\hat{p}}
\def\vecBth{\hat{\boldsymbol{B}_t}}
\def\Bth{\hat{B}_t}
\def\gM{\gamma M_0^2}
\newcommand{\vsp}{\vspace{.3em}}

\newcommand\bm[1]{\boldsymbol{#1}}

\title[Regular \& non-regular solutions of the MHD Riemann problem]{Regular and non-regular solutions of the 
Riemann problem in ideal magnetohydrodynamics}

\author[K. Takahashi and S. Yamada]%
{K.\ns T\ls A\ls K\ls A\ls H\ls A\ls S\ls H\ls I$^1$%
  \thanks{Email address for correspondence: ktakahashi@heap.phys.waseda.ac.jp}\ns
\and S.\ns Y\ls A\ls M\ls A\ls D\ls A$^{2,3}$}

\affiliation{$^1$Department of Physics, Waseda University,
3-4-1 Okubo, Shinjuku, 169-8555, Japan\\[\affilskip]
$^2$Science \& Engineering, Waseda University, 3-4-1 Okubo, Shinjuku, 169-8555, Japan\\[\affilskip]
$^3$Advanced Research Institute for Science and Engineering, Waseda University, 3-4-1 Okubo, Shinjuku, 169-8555, Japan}

\pubyear{?}
\volume{?}
\pagerange{?--?}
\date{?; revised ?; accepted ?. - To be entered by editorial office}
\begin{document}

\maketitle

\begin{abstract}
We have built a code to numerically solve the Riemann problem in ideal magnetohydrodynamics (MHD) for an arbitrary initial condition 
to investigate a variety of solutions more thoroughly. The code can handle not only 
regular solutions, in which no intermediate shocks are involved, but also all types of non-regular solutions if any. As a first 
application, we explored the neighborhood of the initial condition that was first picked up by \citet{BW88} and has been frequently employed 
in the literature as a standard problem to validate numerical codes. Contrary to the conventional wisdom that there will always be a 
regular solution, we found an initial condition, for which there is no regular solution but a non-regular one. The latter solution has 
only regular solutions in its neighborhood and actually sits on the boundary of regular solutions. This implies that the regular solutions 
are not sufficient to solve the ideal MHD Riemann problem and suggests that
at least some types of non-regular solutions are physical. We also demonstrate 
that the non-regular solutions are not unique. In fact, we found for the Brio \& Wu initial condition that there are uncountably many 
non-regular solutions. This poses an intriguing question: why a particular non-regular solution is always obtained in numerical simulations? 
This has important ramifications to the discussion of which intermediate shocks are really admissible.

\end{abstract}

\begin{PACS}
\end{PACS}

\section{Introduction}
The Riemann problem for a system of 1st-order, hyperbolic partial differential equations is an initial value problem, in which two distinct
constant states are separated by a discontinuity initially. By hyperbolic we mean that all eigenvalues of the Jacobian matrix that 
characterizes small perturbations are real. It is well known that the Riemann problem in ideal MHD has in general non-unique solutions for 
a given initial condition even if one discards manifestly unphysical ones such as those including rarefaction shock waves, in which entropy 
is decreased. This is in sharp contrast to the hydrodynamical counterparts and is due to the facts that the system of equations that 
describes ideal MHD is not strictly hyperbolic, i.e. some of the eigenvalues of the Jacobian matrix are coincident with each other, and that
the characteristic fields are neither linear nor genuinely non-linear~\citep[see \S~\ref{sec.Riemann problems} and e.g.][]{Lax1957, JT64, PD90}. 
In fact, the MHD Riemann problem has a far greater variety of solutions including so-called intermediate shock waves. Although these 
intermediate shocks are deemed to be unphysical in some text books, some interplanetary experiments have reported the detections of 
intermediate shocks~\citep{Chao95, FW08, FWC09} and they are actually realized in almost all numerical solutions. Their reality is hence 
still an issue of controversy as will be described below more in detail. 

Solutions of the Riemann problem consist of discontinuities and centered self-similar simple waves in general (see 
\S~\ref{sec.Riemann problems} for more details). Various shock waves appearing in the solutions of Riemann problem
are usually characterized by the flow speeds upstream and downstream in the shock-rest frame: in ideal MHD, for example, the fast 
shock has a flow velocity that exceeds the fast-wave velocity (termed super-fast hereafter) upstream whereas the downstream flow velocity 
is sub-fast, i.e. lower than the fast-wave velocity, but super-\Alfvenic, i.e. higher than the \Alfven-wave velocity 
(see \S~\ref{sec.MHD} for the definitions of these wave velocities); assigning the numbers $1$, $2$, $3$ and $4$ to the super-fast, 
sub-fast and super-\Alfvenic, sub-\Alfvenic \ and super-slow, and sub-slow, respectively, we designate the fast shock as $1 \rightarrow 2$ 
shock; other shocks are referred to just in the same way. 

The intermediate shock waves in ideal MHD have coplanar configurations, i.e. the upstream transverse magnetic field is 
anti-parallel to the downstream one and they are characterized by the flow velocities that are super-\Alfvenic \ upstream and sub-\Alfvenic \ 
downstream. They are normally assigned to one of the following four types: $1 \rightarrow 3$, $1 \rightarrow 4$, $2 \rightarrow 3$ or 
$2 \rightarrow 4$ shock. There are some boundary types as well: for example, the $2\rightarrow 3,4$ shock is a boundary type that has 
a downstream flow speed that is equal to the slow speed as indicated by its name. As mentioned above, these intermediate shock waves
are considered to be unphysical in some textbooks \citep[see e.g.][]{Lax1957, JT64, PD90} because they do not satisfy the so-called 
evolutionary conditions, which require the existence of perturbed states of the same type;
 for ordinary shock waves (such as the fast 
shock in ideal MHD), one of the characteristics runs into the discontinuity from both sides; this is not the case for some of the 
intermediate shocks; for instance, the $1 \rightarrow 3$ intermediate shock is a shock wave whose flow speeds in the shock-rest frame 
are super-fast upstream and sub-\Alfvenic \ and super-slow downstream; hence the fast and \Alfven \ characteristics go into 
the shock both upstream and downstream; in the case of the $2\rightarrow 3$ shock, only the \Alfven \ characteristic converges to
the shock wave; however, another evolutionary condition on the linear independence of eigenfunctions is violated and it is classified as unphysical shock.

In spite of the evolutionary conditions, the intermediate shocks are commonly observed as a stable constituent wave in numerical solutions 
of the MHD Riemann problems \citep{Wu87, BW88, Wu88a, Wu88b, Wu90, Wu&Kennel92}. One of the best known cases is the one discussed by
\citet{BW88}, in which the numerical solution contains a so-called compound wave that consists of a $2\rightarrow 3,4$ shock followed by 
a slow rarefaction wave. \citet{Wu87} also showed numerically that the $2\rightarrow 4$ shock-like structure could be obtained by  
steepening of a continuous wave, an indication that the resultant structure is physical. In addition he demonstrated that 
a perturbed states obtained for a slightly different initial condition retains a similar structure. Moreover,
\citet{Wu88a} computed interactions between the $2\rightarrow 4$ shock-like structure and an \Alfven \ wave and found that 
the structure remains for a while before it breaks to other waves. These results indicate that the intermediate shock is stable
to the perturbation in a sense. \citet{Wu88a} found that the shock-like structures observed in the computations with perturbations
are time-dependent and do not satisfy the Rankine-Hugoniot condition in fact. He claimed that these new types of shocks are the neighboring
states for the intermediate shocks. In yet another simulation \citet{Wu88b} showed that rotational discontinuities,
which are not intermediate, break into some other waves that include an intermediate shock. This is a suggestion that it is not intermediate shocks 
but particular types of evolutionary discontinuities that are unstable and unphysical. On the other hand, \citet{Wu90} studied 
the structures inside shock waves by introducing dissipations and found that they are not uniquely determined by the asymptotic 
states and there remain extra degrees of freedom. He concluded that the stability of the intermediate shocks to the interactions 
with \Alfven \ waves is originated from these degrees of freedom. Incidentally, it was also demonstrated that all types of 
intermediate shocks can be produced by steepening processes in dissipative MHD. 

Completely opposite and equally convincing arguments were made by \citet{FK97,FK01}. They pointed out that if perfect planarity or 
coplanarity is imposed initially, the solution retains the symmetry and magnetic fields cannot rotate in the evolutions; then the
problem is equivalent to solving the reduced system of MHD equations, in which it is assumed that magnetic fields are confined in a
plane and, as a consequence, the \Alfven \ waves do not exist; this changes the number of characteristics of the system and 
modifies the evolutionary conditions; in fact, the $180^{\circ}$-rotational discontinuity becomes non-evolutionary whereas 
$1\rightarrow 3$ and $2\rightarrow 4$ shocks become evolutionary for this reduced system. They also demonstrated numerically 
that the intermediate shock does not emerge if coplanarity of the solution is broken by inserting a thin layer within which 
the magnetic field rotates continuously. \citet{BKP96}, on the other hand, found numerically that the compound wave breaks into 
a rotational discontinuity and a slow shock if the exact coplanarity is perturbed. Although \citet{FK01} agreed with  \citet{Wu90}
that the temporary survival of $1\rightarrow 3$, $2\rightarrow 4$ and $1\rightarrow4$ shocks in their interactions with \Alfven \ waves
is due to the non-unique shock structures, it was claimed that they should be regarded as transients.

The evolutionary conditions themselves have been reconsidered in the mean time~\citep{Hada94, Markovskii98, I&I07}. These authors
took dissipations into account, considering that the waves that emerge in dissipative MHD should be physical. 
With dissiptations shocks are no longer discontinuities and the analysis is facilitated.
They found new modes that do not exist in ideal MHD and argued that on account of these modes the intermediate shocks are indeed evolutionary.

Putting aside the reality of the intermediate shocks, \citet{T02, T03} studied when the intermediate shocks appear in solutions of 
the MHD Riemann problem more in detail. In his paper~\citep{T03}, he defined at first the regular and non-regular shocks: the former
has $N+1$ incoming characteristics, where $N$ is the number of equations, whereas the latter includes all the rest.
The definition is a bit vague for those shocks that have a characteristic speed equal to the shock velocity, since it is not obvious 
in this case that the characteristic is incoming or outgoing and, in fact, there has been a controversy in the literature~\citep{JT64}.
Although there is no explicit mention on this issue, the author of \citep{T02, T03} regarded them as non-regular. We will follow this
practice hereafter in this paper\footnote{More precisely, we define the regular shocks as the shocks that satisfy the evolutionary 
condition (see \S\ref{evo}). We treat those characteristics that have the same velocity as the shock velocity as incoming. All the
other shocks are defined to be non-regular. The solutions are referred to as regular if they include no non-regular shocks and called
non-regular otherwise.}. \citet{T03} referred to the solutions without non-regular shocks as regular solutions and all
the others are called non-regular solutions. Believing that there is always a regular solution, he concluded that if a certain condition 
is satisfied, there is a unique regular solution to the ideal MHD Riemann problem. Note, however, his analysis was not complete, since  
he assumed that only a left-going or right-going $1 \rightarrow 3$ or $2 \rightarrow 4$ intermediate shock exists and he ignored the cases in which the 
transverse magnetic field vanishes. Furthermore, he did not prove that there is 
always a unique regular solution. 

We address these issues in this paper, extending the approach in \citet{T02, T03}. We believe that the detailed study of particular
solutions of the MHD Riemann problem together with their neighboring ones are also helpful to get some more insights into the reality
of intermediate shocks. For this purpose we coded a program to produce both regular and non-regular solutions for arbitrary initial 
conditions. We have extended the algorithm proposed by \citet{T02, T03} so that we could handle all combinations of constituent waves,
including all types of intermediate shocks. Note that \citet{ATJweb} are providing a code to solve the MHD Riemann problem that is 
available online. However, it can handle only the regular solutions. It should be emphasized that numerical codes to compute temporal 
evolutions are useless for the current purpose, since they produce a single solution for a given initial condition and we do not know
how numerical dissipations that are inevitably inherent to numerical simulations affect the outcomes. Hence we need to construct 
solutions in all possible ways to satisfy the given jump condition.

The parameter space of the MHD Riemann problem is vast (15 dimensions indeed) and exploring the entire space is almost impossible. We hence 
need to pay attention to those initial conditions and their neighborhoods, which appear of particular interest to us in the context 
explained above. In this paper we investigated two cases: the first one is the initial condition first picked up by \citet{BW88} that 
includes the compound wave in one of the solutions as mentioned earlier; we found that there is a regular solution in this case and that
there are in fact uncountably many non-regular solutions as well; this poses an interesting question why the non-regular solution with
the compound wave is always realized in numerical simulations; in the second case, we found that there is no regular solution; this provides
a counter-example to the conventional wisdom that there is always a regular solution. 
Note that the regular solutions normally consist of fast waves, slow waves, rotational discontinuities and a contact surface with
an exception that there exit so-called switch-off rarefaction waves that nullifies the transverse magnetic fields. 
Then rotational discontinuity does not appear in the solutions.

The paper is organized as follows.
In \S~\ref{sec.Riemann problems}, we briefly review the general theory of Riemann problem.
In \S~\ref{sec.MHD}, the shock waves and simple waves in ideal MHD, which are the constituent waves in the solutions of 
MHD Riemann problem, are presented. In \S~\ref{sec.method}, we describe how to construct various solutions that satisfy the
jump conditions. In \S~\ref{sec.results}, we apply the code to two particular initial conditions mentioned above and show both the 
regular and non-regular solutions. We discuss possible implications of the results in \S~\ref{sec.conclusions}.

\section{Theory of Riemann problem} \label{sec.Riemann problems}
\subsection{Hyperbolic equations and weak solutions}
In this section we consider general conservation equations in one spatial dimension.
Since the Riemann problem assumes plane-symmetry, this is sufficient. Then the coupled equations are expressed as 
\begin{equation}
\frac{\partial \bm{u}}{\partial t} +\frac{\partial \bm{f}}{\partial x} = \bm{0},
\label{cons}
\end{equation}
where $\bm{u}$ is a vector of conserved variables  and $\bm{f}(\bm{u})$ is a flux vector.
The system (\ref{cons}) is called hyperbolic if the Jacobian matrix, which is defined as 
\begin{equation}
\label{Jacobian}
A_{ij} := \frac{\partial f_i}{\partial u_j},
\end{equation}
has $N$ real eigenvalues $\lambda _k(\bm{u})$ $(k = 1, 2, \dots , N)$  and corresponding $N$ linealy independent right eigenvectors, 
$\bm{r}_k(\bm{u})$ \citep[e.g.][]{FK01}. In particular, if all eigenvalues are different from each other for any $\bm{u}$, the system is called 
strictly hyperbolic.

The $k$-th characteristic field is called genuinely non-linear if the following condition is satisfied for any $\bm{u}$:
\begin{equation}
\bm{r}_k \bcdot \nablau \lambda _k \neq 0.
\end{equation}
Here $\bm{\nabla }_{\bm{u}}$ denotes the operator, ${}^t(\partial/\partial u_1, \dots, \partial/\partial u_{N})$, where ${}^t(\dots)$ stands for 
transposition. On the other hand, The $k$-th characteristic field is called linear if
\begin{equation}
\bm{r}_k \bcdot \nablau \lambda _k = 0
\end{equation}
for all $\bm{u}$. Note that the characteristic field may be neither linear nor genuinely non-linear.

In the system described by Eq.~(\ref{cons}), discontinuities are ubiquitous. Indeed they may be spontaneously produced 
from continuous initial conditions by wave-steepening. Since Eq.~(\ref{cons}) is differential equations, these discontinuities cannot 
be treated as they are and some extensions are needed. The weak solutions of Eq.~(\ref{cons}) are defined as the functions
that satisfy the following equations instead of Eq.~(\ref{cons}) in a finite domain $D(x,t)$ \citep[e.g.][]{JT64}:
\begin{equation}
\label{weak}
\int ^{\infty}_{-\infty} \bm{w}(x,0)\bcdot \bm{\phi}(x) \diff x + \int ^{\infty}_{0}\diff t \int^{\infty}_{-\infty}\diff x \left( \frac{\partial \bm{w}}{\partial t} \bcdot \bm{u} +\frac{\partial \bm{w}}{\partial x} \bcdot \bm{f}\right) = 0,
\end{equation}
where $\bm{\phi }(x)$ is an initial condition and $\bm{w}(x,t)$ is a vector of test functions, which are differentiable with respect to $x$ 
and $t$ as many times as is required and are identically zero outside $D(x,t)$. This equation is derived by multiplying the original 
equations by $\bm{w}$ and integrating the product by parts. Note that differentiability of weak solutions is not required. If the solution 
is differentiable and satisfies Eq.~(\ref{cons}), then Eq.~(\ref{weak}) is also satisfied. Hence ordinary solutions are indeed weak solutions.
It should be remarked that uniqueness of weak solutions for a given initial condition is no longer guaranteed.

\subsection{Solutions of Riemann problem: simple waves and discontinuities}
The Riemann problem is initial value problems for a system of conservation equations, in which two constant states are separated initially by
a discontinuity, which is located, e.g., at $x=0$:
\begin{equation}
\bm{u}(x,0) = \left\{
\begin{array}{ll}
\bm{u}_\mathrm{R} & (x>0) \\
\bm{u}_\mathrm{L} & (x<0) .
\end{array}\right.
\end{equation}
We seek weak solutions to this problem. As remarked at the end of the previous section, the weak solutions are not unique in general and it is
one of the important issues in the Riemann problem to determine which solution is physically meaningful. Since there is no typical time and 
length scales in the problem, the solution should be self-similar. It is well known that solutions of the Riemann problem consist of a couple of 
centered simple waves and discontinuities. The latter satisfies the Rankine-Hugoniot relations \citep[e.g.][]{JT64}.

The simple waves are defined as waves, in which there is only one independent component in $\bm{u}$ and all the other components are its functions. 
They are related with the eigenvalues of the Jacobian matrix (\ref{Jacobian}) in such a way that all quantities in the simple wave are constant 
along each characteristic given by 
\begin{equation}
x = \lambda _k t +\mathrm{constant}.
\end{equation}
The simple wave that corresponds to eigenvalue $\lambda _k$ is referred to as the $k$-th simple wave.

\begin{figure}
 \begin{center}
 \includegraphics[scale=0.5]{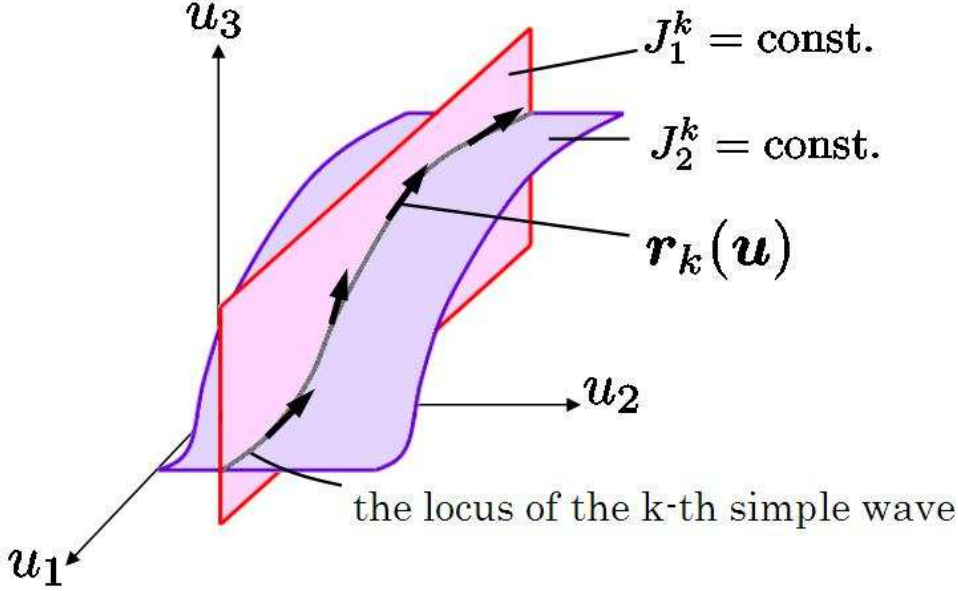}
 \caption{A schematic picture of the $k$-th Riemann invariants and corresponding locus of the $k$-th simple wave
 in the case of $N=3$. There are two $k$-th Riemann invariants: $J_1^k$ and $J_2^k$. 
 The red surface is defined by $J_1^k = \const $ and the blue one is given by $J_2^k = \const $
 The intersection is a line called the locus of the $k$-th simple wave, to which the $k$-th right eigenvector, $r_k(\bm{u})$, is tangential.}
 \label{Riemann invariants}
 \end{center}
\end{figure}

Associated with the $k$-th simple wave are so-called the $k$-th Riemann invariants denoted by $J_l^{k} (l = 1, 2, \dots, N-1)$, which are constant 
across the $k$-th simple wave. They are defined to be the quantities that satisfy the following relation with the $k$-th right eigenvector, $\bm{r} _k$:
\begin{equation}
\bm{r} _k \bcdot \bm{\nabla }_{\bm{u}} J_l^{k}  = 0.
\end{equation}
The $k$-th Riemann invariants define $N-1$ hypersurfaces in phase space and their intersection is a line, to which the $k$-th right 
eigenvector, $\bm{r} _k$, is tangential, and is called the locus of the $k$-th simple wave. See a schematic picture drawn in Fig. \ref{Riemann invariants}.

The Rankine-Hugoniot relations, which are satisfied by the quantities on both sides of a discontinuity, are expressed as 
\begin{equation}
\label{R-H}
s\jump{\bm{u}}  -\jump{\bm{f}} = \bm{0},
\end{equation}
where $s$ is the velocity of the discontinuity and $\jump{X} := X_1 -X_0$ denotes a jump in quantity $X$ across the discontinuity and 
$X_0$ and $X_1$ are the values ahead of and behind the discontinuity, respectively. Since this is a system of $N$ equations, we obtain a 
one-parameter family of solutions for a state given on one side of the discontinuity. Not all solutions are physical as mentioned repeatedly. 
Rarefaction shock waves for hydrodynamical equations are well known examples. Since entropy decreases across such shock waves, they
are certainly unphysical and rejected. This entropy condition is not sufficient to eliminate all unphysical solutions in general and 
the so-called evolutionary conditions are introduced.

\subsection{Evolutionary conditions\label{evo}}
Intuitively put, the evolutionary conditions require that physical solutions should be stable against splitting into other waves 
\citep[see e.g.][]{FK01, PD90, Landau, JT64, Lax1957}. To be more precise, when a small-amplitude wave is injected to the discontinuity, 
the resultant state, which is determined uniquely, is required to still consist of the same type of discontinuity and outgoing 
small-amplitude waves. This is satisfied if the number of characteristics emanating from the discontinuity is $N+1$ and 
if the initial jump, $\jump{\bm{u}}$, is linearly independent of the eigenvectors that correspond to the outgoing characteristics.

\section{Ideal MHD} \label{sec.MHD}
In plane symmetry, the ideal MHD equations are given by
\begin{eqnarray}
\label{mass}
\frac{\partial \rho}{\partial t} +\frac{\partial }{\partial x}(\rho v_n) = 0, \\
\frac{\partial }{\partial t}(\rho v_n) +\frac{\partial }{\partial x} \left( \rho v_n^2 +p +\frac{\bm{B}_t^2}{2}\right) =0, \\
\frac{\partial }{\partial t}(\rho \bm{v}_t) +\frac{\partial }{\partial x} ( \rho v_n\bm{v}_t -B_n\bm{B}_t) =\bm{0}, \\
\frac{\partial \bm{B}_t}{\partial t} +\frac{\partial }{\partial x}(v_n \bm{B}_t -B_n \bm{v}_t) = \bm{0}, \\
\label{energy}
\frac{\partial e}{\partial t} +\frac{\partial }{\partial x}\left[ \left( e +p +\frac{\bm{B}^2}{2}\right) v_n -B_n\bm{B}\bcdot \bm{v} \right] = 0,
\end{eqnarray}
where $\rho$, $p$, $\bm{v}$ and $\bm{B}$ are density, pressure, flow velocity and  magnetic field, respectively \citep{Landau}. 
The subscripts $n$ and $t$ indicate the normal component, i.e. $x$-component, and transverse component, i.e. $y$ or $z$-component, respectively.
The total energy density is denoted by $e = p/(\gamma -1) +\rho \bm{v}^2/2 +\bm{B} ^2/2 $, where the equation of state for ideal gas is
assumed and $\gamma$ is the ratio of specific heats. The normal component of magnetic field, $B_n$, is constant owing to the divergence-free 
condition.

\subsection{Simple waves \label{smp}}
The eigenvalues of the Jacobian matrix for the system of Eqs.~(\ref{mass})-(\ref{energy}) are
\begin{eqnarray}
\label{eigenvalues}
v_n \mp c_f, \quad
v_n \mp c_A, \quad
v_n \mp c_s, \quad
v_n,
\end{eqnarray}
where $c_f$, $c_A$ and $c_s$ are called the fast, \Alfven \ and slow speeds, respectively, and are expressed as 
\begin{eqnarray}
\label{fast}
c_{f,s} := \left[ \frac{1}{2}\left(\frac{\bm{B}^2}{\rho} +a^2\right) \pm \sqrt{\frac{1}{4}\left( \frac{\bm{B}^2}{\rho} +a^2\right)^2 -a^2\frac{B_n^2}{\rho} } \right]^{1/2}, \\
\label{Alfven}
c_A := \sqrt{\frac{B_n^2}{\rho}}.
\end{eqnarray}
In the above expressions, $a = \sqrt{\gamma p/\rho}$ is the acoustic speed. In Eq.~(\ref{eigenvalues}), the minus (plus) sign is applied to 
the left-going (right-going) waves. The simple waves corresponding to these eigenvalues are referred to as the fast, \Alfven , slow and entropy waves, 
respectively. The right eigenvectors for fast and slow waves are given as 
\begin{eqnarray}
\label{fseigenvec}
\bm{r}_{f,s}^\mp =  \xi _{f,s} \left[
\begin{array}{ccccc}
-\rho \\
-\gamma p \\
\pm c_{f,s}   \vsp \\
\displaystyle
\pm \frac{c_{f,s}}{1-(c_{f,s}/c_A)^2}\frac{\bm{B}_t}{B_n} \vsp \\
\displaystyle
\frac{\bm{B}_t}{(c_A/c_{f,s})^2 -1} \\
\end{array} \right], 
\end{eqnarray}
in which the new variables, $\xi _f$ and $\xi _s$, are introduced as follows:
\begin{eqnarray}
\xi _f := \sqrt{\frac{a^2 -c_s^2}{c_f^2 -c_s^2}}, \quad
\xi _s := \sqrt{\frac{c_f^2 -a^2}{c_f^2 -c_s^2}}.
\end{eqnarray}
These factors are necessary to ensure that the eigenvectors do not vanish for any $\bm{u}$. In deriving the above expressions of right 
eigenvectors we assume that $\bm{u} = (\rho , p, v_n, \bm{v}_t, \bm{B}_t)$. Since we do not use the right eigenvectors for the \Alfven \ 
and entropy waves in solving the MHD Riemann problem, we do not give their explicit forms here. Note that the \Alfven \ and entropy waves 
are linear whereas the fast and slow waves are neither linear nor genuinely non-linear.

The eigenvalues are degenerate in the following two cases:
\begin{eqnarray}
B_n  = 0 & : & \quad c_f = \sqrt{a^2+\frac{\bm{B}_t^2}{\rho}}, \quad c_s = c_A = 0 , \\
B_n \ne 0,\ \bm{B}_t = \bm{0} & : & \quad c_f = \mathrm{max}(a, c_A), \quad c_s = \mathrm{min}(a, c_A).
\end{eqnarray}
In the latter case the limits of the right eigenvectors for fast and slow waves as $\bm{B}_t \rightarrow \bm{0}$ depend on 
the magnitudes of the acoustic and \Alfven \ speeds. For $a > c_A$ we obtain  
\begin{eqnarray}
\bm{r}_f^\mp \rightarrow \left[
\begin{array}{ccccc}
-\rho \\
-\gamma p \\
\pm a \\
\bm{0} \\
\bm{0}
\end{array} \right]
, \quad
\bm{r}_s^\mp \rightarrow a \left[
\begin{array}{ccccc}
0 \\
0 \\
0 \\
\pm \mathrm{sgn}(B_n) \, \bm{e}_t \\
\sqrt{\rho} \, \bm{e}_t
\end{array} \right],
\end{eqnarray}
where $\bm{e}_t$ is a unit vector that has the same direction as the transverse magnetic field.
Note that $\bm{r}_f^\mp$ is reduced to the eigenvectors for the rarefaction waves in the ordinary hydrodynamics.
In the opposite case, i.e. $a < c_A$, we get 
\begin{eqnarray}
\label{sw-off eigenvector}
\bm{r}_f^\mp \rightarrow a \left[
\begin{array}{ccccc}
0 \\
0 \\
0 \\
\mp \mathrm{sgn}(B_n) \bm{e}_t \\
-\sqrt{\rho} \bm{e}_t
\end{array}\right]
,\quad
\bm{r}_s^\mp \rightarrow \left[
\begin{array}{ccccc}
-\rho \\
-\gamma p \\
\pm a \\
\bm{0} \\
\bm{0}
\end{array}\right],
\end{eqnarray}
in which $\bm{r}_s^\mp$ is reduced to the eigenvectors for the ordinary rarefaction waves in hydrodynamics.
Finally in the case of $a = c_A$, we find 
\begin{eqnarray}
\bm{r}_f^\mp \rightarrow \frac{1}{\sqrt{2}} \left[
\begin{array}{ccccc}
-\rho \\
-\gamma p \\
\pm a \\
\mp a\bm{e}_t \\
-a\sqrt{\rho} \bm{e}_t 
\end{array}\right]
,\quad
\bm{r}_s^\mp \rightarrow \frac{1}{\sqrt{2}} \left[
\begin{array}{ccccc}
-\rho \\
-\gamma p \\
\pm a \\
\pm a\bm{e}_t \\
a\sqrt{\rho} \bm{e}_t 
\end{array}\right].
\end{eqnarray}
As mentioned earlier, the right eigenvectors are chosen in our code so that these degenerate cases could be 
properly handled as the limits of non-degenerate cases. 

In the fast rarefaction wave the magnitude of transverse magnetic field is decreased and, as a limiting case, 
it vanishes behind the so-called switch-off rarefaction wave. Since the fast rarefaction wave cannot reverse the direction 
of the transverse magnetic field, the switch-off rarefaction is the end point of the fast rarefaction locus.

\subsection{Discontinuities}
As mentioned earlier, discontinuities are another important element in the solutions of Riemann problem. The quantities on
both sides of a discontinuity satisfy the Rankine-Hugoniot relations, which in ideal MHD are expressed as
\begin{eqnarray}
\label{mass flux}
m = \const , \\
m^2\jump{v} +\Ljump{ p+\frac{\bm{B}_t^2}{2} } = 0, \\
m\jump{ \bm{v}_t } -B_n\jump{ \bm{B}_t } = \bm{0}, \\
m\jump{ v\bm{B}_t } -B_n\jump{ \bm{v}_t } = \bm{0}, \\
\label{energy jump}
m\left( \Ljump{\frac{pv}{\gamma -1} } +\average{p}\jump{v} +\frac{1}{4}\jump{ v }\jump{ \bm{B}_t }^2 \right) = 0,
\end{eqnarray}
in the rest frame of the discontinuity. In the above expressions, $m := \rho _0 {v_n}_0 = \rho _1 {v_n}_1$ is the mass flux, 
$v := 1/\rho$ is the specific volume and $\average{X} := (X_0 +X_1)/2$ stands for the arithmetic mean of upstream
and downstream quantities. The solutions can be classified into the fast, slow or intermediate shock or into the rotational, contact 
or tangential discontinuity. Details of the classification are found in the literature \citep[e.g.][]{JT64,PD90,T02,T03}.
In what follows, we summarize only those features that are needed for later discussions.

Following \citet{T02,T03}, we normalize all quantities with those upstream as
\begin{eqnarray}
\label{vh}
\hat{v} := \frac{v_1}{v_0}, \quad
\hat{p} := \frac{p_1}{p_0}, \quad
\hat{\bm{B}}_t := \frac{ {\bm{B}_t}_1}{\sqrt{p_0}}, \\
\bm{A} := \frac{ {\bm{B}_t}_0}{\sqrt{p_0}}, \quad
B := \frac{B_n}{\sqrt{p_0}}, \quad
\label{M_0}
M_0 := \frac{{v_n}_0}{a_0},
\end{eqnarray}
and employ in the following the dimensionless MHD Rankine-Hugoniot relations, which are obtained by substituting 
Eqs.~(\ref{vh})-(\ref{M_0}) and eliminating $\jump{\bm{v}_t}$ in Eqs.~(\ref{mass flux})-(\ref{energy jump}): 
\begin{eqnarray}
\label{Euler_normal_hat}
\ph -1 +\gM (\vh -1) +\frac{1}{2}(\vecBth ^2-\bm{A}^2) = 0, \\
\label{Euler_trans_hat}
\gM (\vh \vecBth -\bm{A}) -B^2(\vecBth -\bm{A}) = 0, \\
\label{energy_hat}
M_0\left[ \frac{1}{\gamma -1}(\ph \vh -1) +\frac{1}{2}(\vh -1)(\ph +1) +\frac{1}{4}(\vh -1)(\vecBth -\bm{A})^2 \right]= 0.
\end{eqnarray}
Fixing the upstream quantities, $A$, $B$ and $M_0$, we solve Eqs.~(\ref{Euler_normal_hat})-(\ref{energy_hat}) and use 
Eqs.~(\ref{vh})-(\ref{M_0}) to obtain $\vh$, $\ph$ and $\vecBth$. The other downstream quantities  
 can be calculated as
\begin{eqnarray}
\label{vnormal}
{v_n}_1 = \vh {v_n}_0, \\
\label{vtrans}
{\bm{v}_t}_1 = {\bm{v}_t}_0 \pm \frac{a_0 B}{\gamma M_0}\jump{\vecBth}.
\end{eqnarray}
In Eq.~(\ref{vtrans}), the plus and minus signs correspond to the left- and right-going discontinuities, respectively.

\subsubsection{contact, tangential and rotational discontinuities}
The solutions of Eqs.~(\ref{Euler_normal_hat})-(\ref{energy_hat}) that have a vanishing mass flux, i.e. $M_0=0$ but a non-vanishing 
normal component of magnetic field, i.e. $B \ne 0$, are called the contact discontinuity and satisfy the following relations:
\begin{eqnarray}
\vh = \mathrm{arbitrary}, \quad \ph = 1, \quad \vecBth = \bm{A}, \\
\jump{\bm{v}} = \bm{0}.
\end{eqnarray}
The solutions with $M_0 = 0$ and $B = 0$, on the other hand, are named the tangential discontinuity, for which the following 
relations hold:
\begin{eqnarray}
\label{tolpre}
\vh = \mathrm{arbitrary}, \quad  \ph -1 +\frac{1}{2}(\vecBth^2 -\bm{A}^2) = 0, \\
\jump{v_n} = 0,\quad \jump{\bm{v}_t} = \mathrm{arbitrary}.
\end{eqnarray}

The solutions with $M_0 \ne 0$ and $B \ne 0$ are either a linear wave ($\vh = 1$) or a shock wave ($\vh > 1$).
The former is referred to as the rotational discontinuity, since the transverse component of magnetic field rotates, not 
varying its magnitude during its passage. The rotational discontinuities meet the following conditions:
\begin{eqnarray}
\label{rot}
\vh = 1, \quad \ph = 1, \quad \vecBth ^2= \bm{A}^2, \quad M_0^2 = \frac{B^2}{\gamma}, \\
\jump{v_n} = 0, \quad \jump{\bm{v}_t} = \pm \frac{1}{\sqrt{\rho}}\jump{\bm{B}_t},
\end{eqnarray}
where the plus and minus signs correspond to the left- and right-going waves, respectively. 
The last relation in Eq.~(\ref{rot}) implies the upstream Mach number is equal to the ratio of the \Alfven \ velocity to the 
acoustic speed there.

All the above discontinuities satisfy the evolutionary conditions except for the rotational discontinuity in which the transverse 
magnetic field rotates by $180^{\circ}$. The latter is sometimes called weakly evolutionary in the literature \citep{JT64}, since 
the neighboring rotational discontinuities are all evolutionary. Following this point of view, we treat the solutions with weakly 
evolutionary discontinuities as regular solutions in this paper.\footnote{This may need reconsideration, however, and will be 
addressed elsewhere.}

\subsubsection{shock waves}
The solutions of Eqs.~(\ref{Euler_normal_hat})-(\ref{energy_hat}), for which $M_0 \ne 0$ and $\vh > 1$, i.e. 
matter is compressed as it passes through the discontinuities, are called shock waves. Their notable feature is that magnetic 
fields are either planar or coplanar. This is apparent from Eq.~(\ref{Euler_trans_hat}). Indeed, recalling $\vh > 1$, we obtain
\begin{equation}
\label{eq_bth}
\vecBth = \frac{\gamma M_0^2 -B^2}{\gamma M_0 ^2\vh -B^2}\bm{A}.
\end{equation}
Substituting ${v_n}_1 = \vh{v_n}_0$ to eliminate $\vh$, we also obtain the following relation:
\begin{equation}
\vecBth = \frac{{v_n}_0^2 -{c_A}_0^2}{{v_n}_1^2 -{c_A}_1^2} \bm{A},
\end{equation}
which shows immediately that transverse magnetic fields are coplanar if and only if the upstream flow velocity is super-\Alfvenic \ whereas
the downstream speed is sub-\Alfvenic.

The shocks with planar transverse magnetic fields are either fast or slow shocks, the former of which amplifies the magnitude of transverse
magnetic fields whereas the latter reduces it. The shocks that change the direction of transverse magnetic fields are referred to as  
intermediate shocks. The terminology employed in the literature is rather confusing. It is remarked that the intermediate shock in this
paper is different from the intermediate discontinuity defined in \citet{JT64}. In fact, the latter is used for the discontinuities 
that satisfy the evolutionary conditions whereas the intermediate shocks in this paper are non-evolutionary as will be evident shortly.

Recalling $c_f \geq c_A \geq c_s$, we assign $1$ to the states with super-fast velocities, $2$ to those with sub-fast and 
super-\Alfvenic \ velocities, $3$ to those with sub-\Alfvenic \ and super-slow velocities and $4$ to those with sub-slow velocities
in the shock-rest frame. With this allocation, the fast shock is denoted by $1 \rightarrow 2$ shock, since the upstream velocity is
super-fast (state $1$) whereas the downstream speed is sub-fast and super-\Alfvenic (state $2$). Similarly the slow shock is designated 
as $3 \rightarrow 4$ shocks. The intermediate shocks normally belong to one of the following four types: $1 \rightarrow 3$, 
$1 \rightarrow 4$, $2 \rightarrow 3$ and $2 \rightarrow 4$ shocks. The $1 \rightarrow 3$ and $2 \rightarrow 4$ intermediate shocks are 
called over-compressive shocks and $9$ out of $14(=7 \times 2)$ characteristics run into these shock waves. To the $1 \rightarrow 3$ 
and $2 \rightarrow 4$ shocks converge the fast and \Alfven \ characteristics and the \Alfven \ and slow characteristics, respectively.
The $1 \rightarrow 4$ intermediate shock is doubly over-compressive and $10$ characteristics  of all types go into the shock.
In the case of $2\rightarrow 3$ shock, only the \Alfvenic \ characteristic converges to the shock wave and the number of in- and out-going 
waves are right. However, the other evolutionary condition on the linear independence of eigenfunctions is violated and the shock is 
hence classified as non-evolutionary.
In some cases, the flow velocity coincides with one of characteristic velocities. We employ a pair of numbers to specify those states:
"$1,2$ ", "$2,3$" and "$3,4$" represent those states whose flow speed are equal to the fast, \Alfven \ and slow speeds, respectively.
The shock wave with the upstream velocity being super-\Alfvenic \ and the downstream speed being equal to the slow velocity, for example, 
is designated as the $2 \rightarrow 3,4$ shock. Of our special concern among these intermediate shocks are the so-called 
switch-on ($1 \rightarrow 2,3$) and switch-off ($2,3 \rightarrow 4$) shocks, the details of which will be given shortly.

\subsubsection{fast and slow loci}
We regard the shock solutions of Eqs.~(\ref{Euler_normal_hat})-(\ref{energy_hat}) as functions of the upstream Mach number ($M_0$), 
normal ($B_n$ or $B$) and transverse ($\vecBth$ or $\bm{A}$) components of magnetic field, divide them into two families and look into 
their loci in some detail. Since magnetic fields in shock waves are 
either planar or coplanar as pointed out earlier, in the following we assume without loss of generality that magnetic fields are confined in 
the $(x,y)$-plane and treat $\Bth$ and $A \ (\geq 0)$ as scalar variables.

\begin{figure}
\begin{tabular}{cc}
\begin{minipage}{0.45\hsize}
\begin{center}
\includegraphics[scale=0.26]{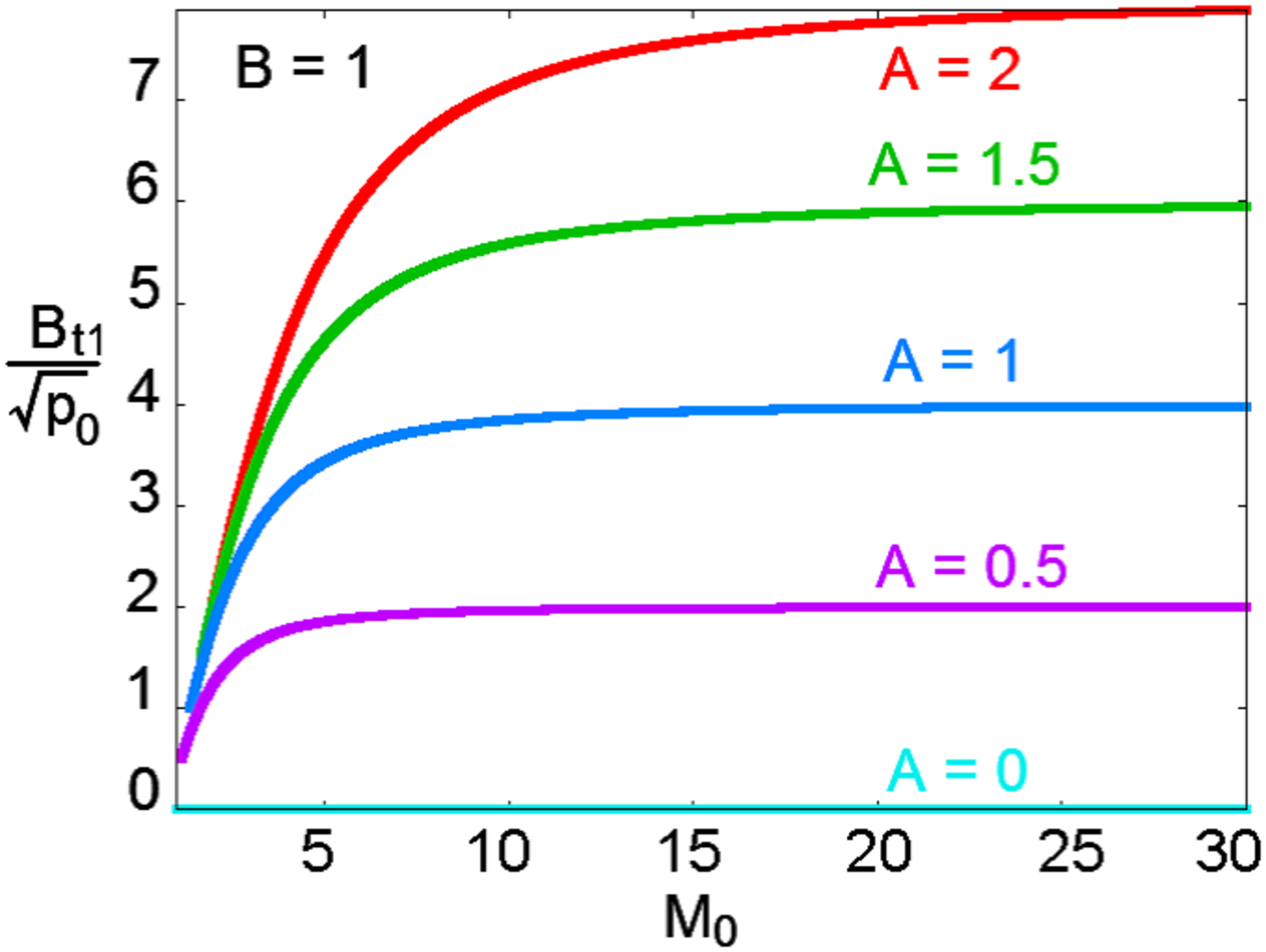}
\end{center}
\end{minipage} &
\begin{minipage}{0.45\hsize}
\begin{center}
\includegraphics[scale=0.26]{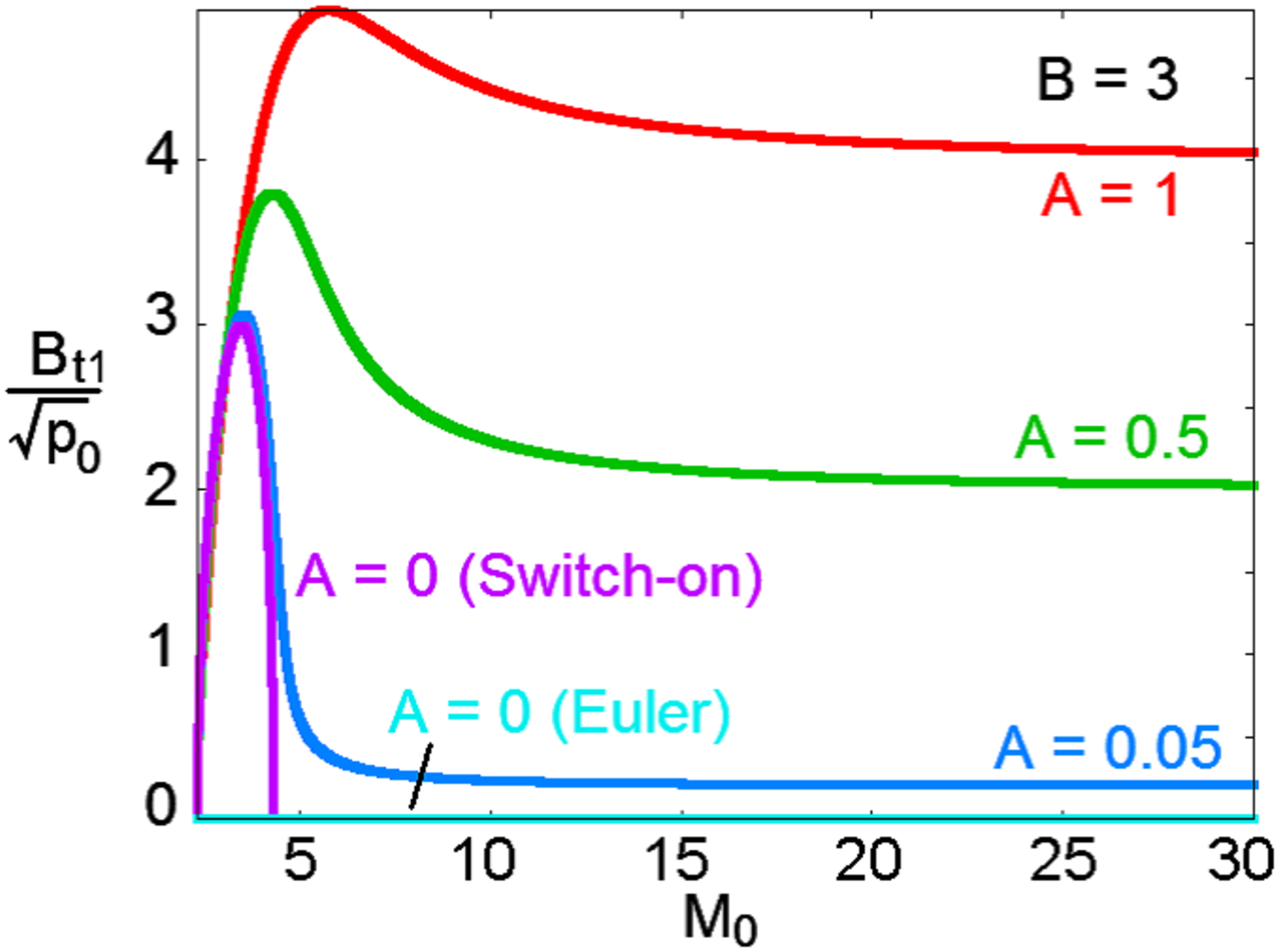}
\end{center}
\end{minipage} \\
\begin{minipage}{0.45\hsize}
\begin{center}
\includegraphics[scale=0.26]{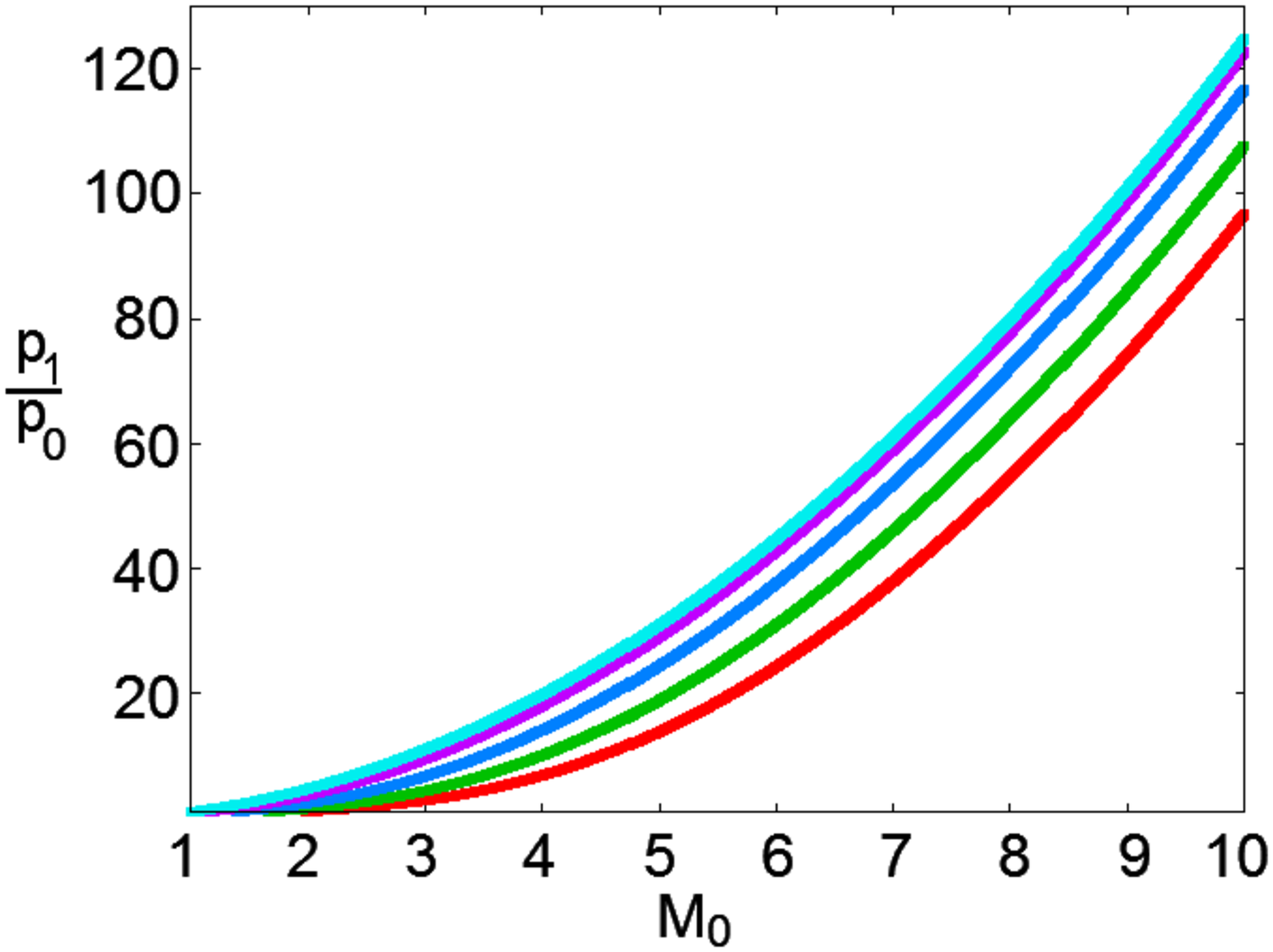}
\end{center}
\end{minipage} &
\begin{minipage}{0.45\hsize}
\begin{center}
\includegraphics[scale=0.26]{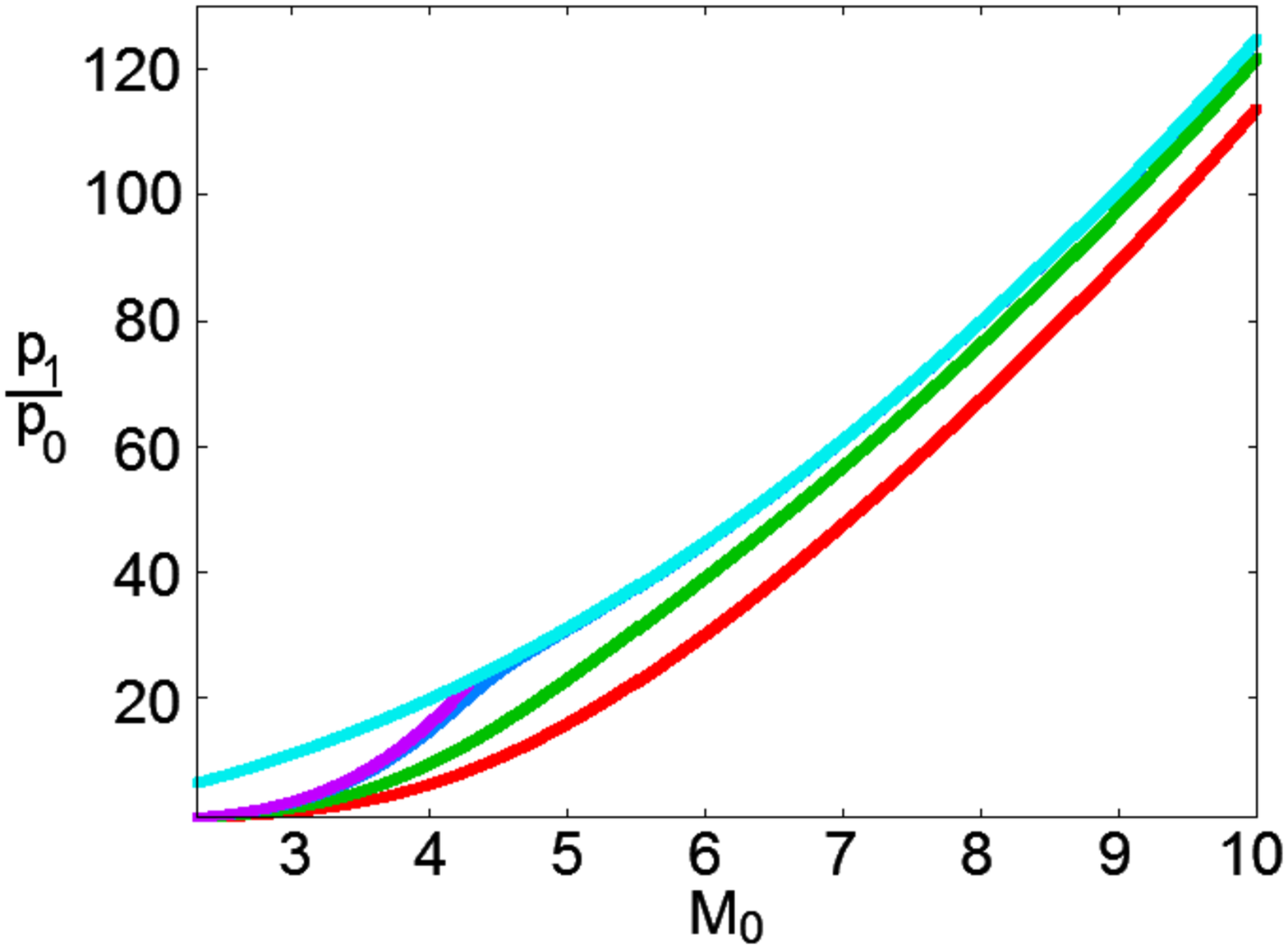}
\end{center}
\end{minipage} \\
\begin{minipage}{0.45\hsize}
\begin{center}
\includegraphics[scale=0.26]{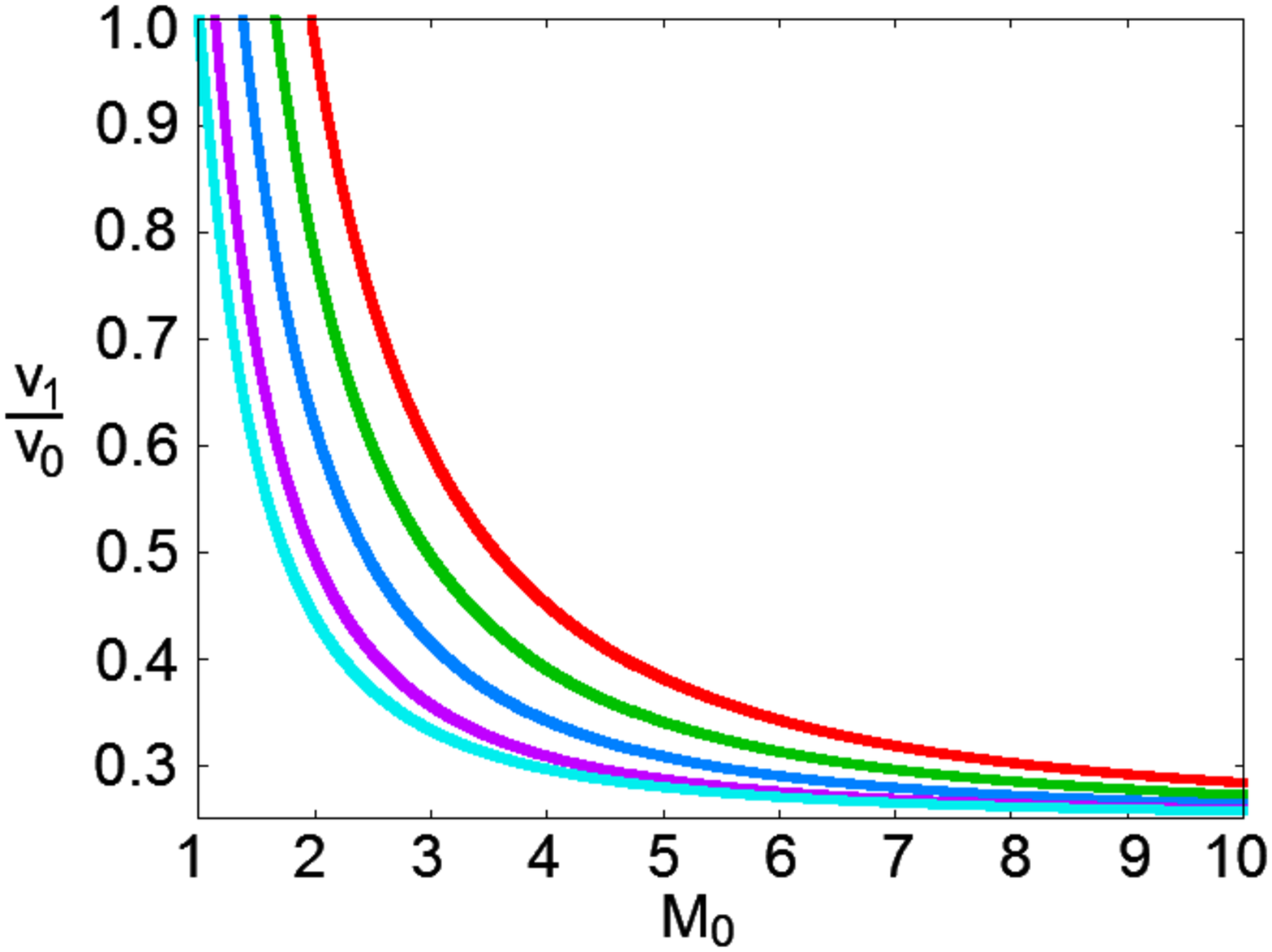}
\end{center}
\end{minipage} &
\begin{minipage}{0.45\hsize}
\begin{center}
\includegraphics[scale=0.26]{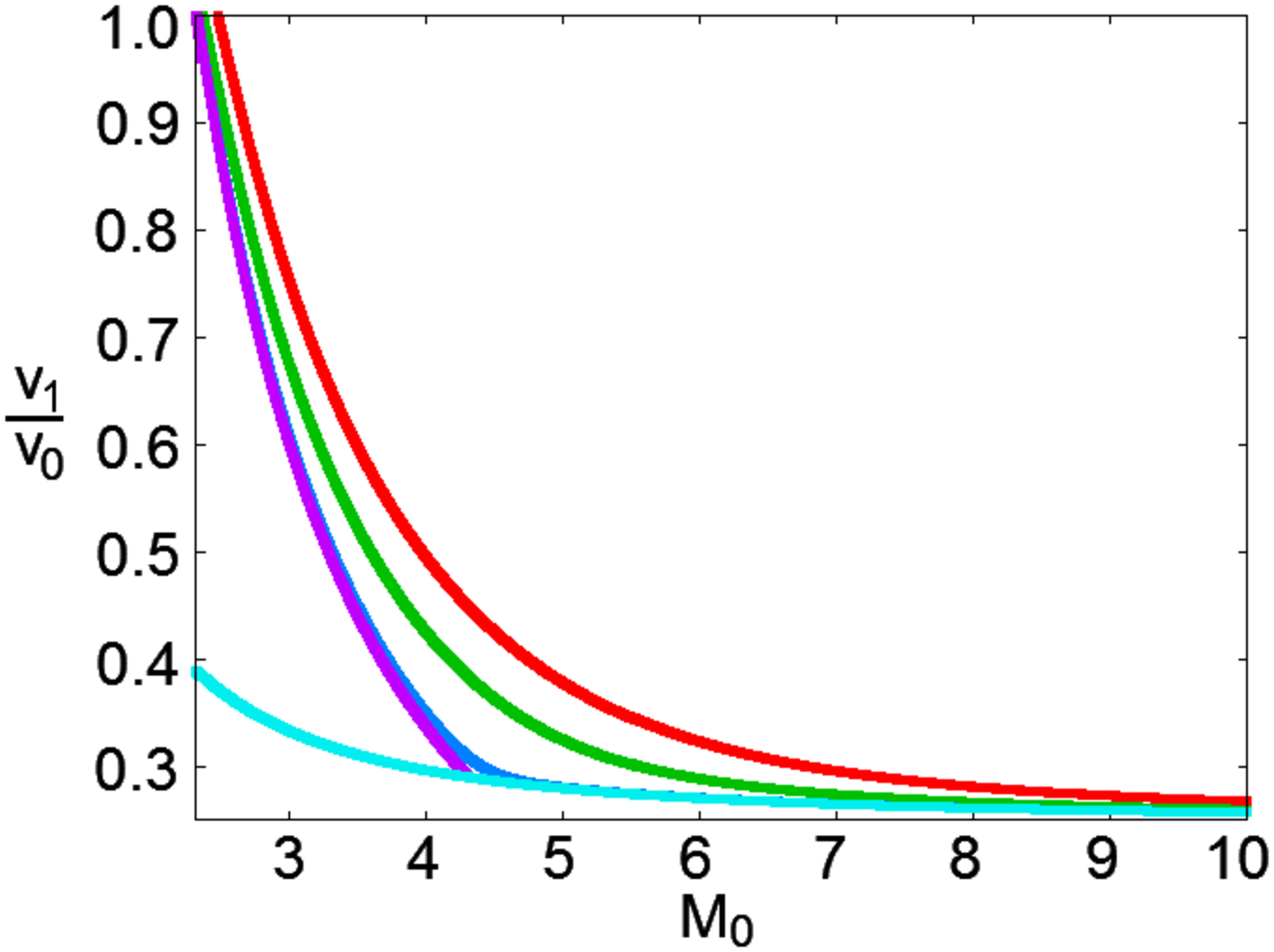}
\end{center}
\end{minipage} 
\end{tabular}
\caption{The fast loci for different combinations of the normal ($B$) and transverse ($A$) component of magnetic field. 
The left panels: $B=1$ and $A=2$ (red), $1.5$ (green), $1$ (blue), $0.5$ (purple) and $0$ (light blue). 
The right panels: $B=3$ and $A = 1$ (red), $0.5$ (green), $0.05$ (blue), $0$ (purple, switch-on shock) and $0$ 
(light blue, Euler shock). The switch-on shock does not exist for $B=1$. See the text for details.}
\label{fig.locus1}
\end{figure}

Eliminating $\ph$ and $\Bth$ from Eqs.~(\ref{Euler_normal_hat})-(\ref{energy_hat}) we obtain the following cubic equation for 
the specific volume, $\vh$:
\begin{eqnarray}
&&(\gM \vh) ^3 -\left[ \frac{2}{\gamma +1} +\frac{\gamma -1}{\gamma +1} \gM +2B ^2+\frac{\gamma A^2}{\gamma +1}\right] ( \gM \vh )^2 \nonumber \\
\label{fast branch}
&&+\left[2B^2 \left( \frac{2\gamma}{\gamma +1} +\frac{\gamma -1}{\gamma +1}\gM \right) +B^4 -\frac{2 -\gamma}{\gamma +1}\gM A^2 +A^2B^2\right] \gM \vh \\ 
&&-\left[ B^4 \left( \frac{2\gamma}{\gamma +1}  +\frac{\gamma -1}{\gamma +1}\gM \right) +\frac{\gamma -1}{\gamma  +1}\gM A^2B^2 \right] = 0. \nonumber
\end{eqnarray}
In the deriving this equation, we used the assumption of $\vh \ne 1$. The family of the fast shock is the solutions characterized 
by the feature that matter is compressed and the transverse magnetic field is amplified by the passage of shock wave. It is then found
that this branch of solutions satisfies the inequality $\vh _\mathrm{min} < \vh < 1$, where the minimum is given by 
\begin{equation}
\vh _\mathrm{min} = \mathrm{max} \left( \frac{B^2}{\gM}, \frac{\gamma -1}{\gamma +1} \right).
\end{equation}
The loci of the solutions are shown in Fig.~\ref{fig.locus1} as a function of the upstream Mach number ($M_0$) for some combinations of 
the upstream normal ($B$) and transverse ($A$) components of magnetic field.

As seen in the right panel of  Fig.~\ref{fig.locus1}, the fast locus is divided into two branches for the special case of $A = 0$, i.e. 
the vanishing upstream transverse magnetic field, with Mach numbers satisfying the following inequalities:
\begin{equation}
\label{sw-on_ineq}
\hat{c}_{f0} < M_0 < \sqrt{\frac{\gamma +1}{\gamma -1}\frac{B^2}{\gamma} -\frac{2}{\gamma -1}},
\end{equation}
where $\hat{c}_{f0} := c_{f0}/a_0$ is the normalized fast velocity. One of the branches that generates non-vanishing transverse magnetic 
fields by the shock passage is called the switch-on shock branch and the other is referred to as the Euler shock branch, in which
the transverse component of magnetic field remains zero. The post-shock specific volume and transverse magnetic field are given by
\begin{eqnarray}
\vh & = & \frac{2+(\gamma -1)M_0^2}{(\gamma +1)M_0^2}, \quad\frac{B^2}{\gM }, \\
\Bth & = & 0, \quad \sqrt{\frac{\gM - B^2}{B^2}\left[ (\gamma -1) \left( \frac{\gamma +1}{\gamma -1}B^2 -\gM \right) -2\gamma \right]},
\end{eqnarray}
respectively. In the above expressions, the first options correspond to the Euler shock and the second ones to the switch-on shock.
The requirement that the quantity in the square root be non-negative gives the inequality (\ref{sw-on_ineq}).
It is noted that the flow speed behind the switch-on shock is equal to the \Alfven \ speed. The switch-on shock is hence designated as
the $1 \rightarrow 2,3$ shock and is non-regular. It is also noteworthy that the Euler shock is essentially a hydrodynamical shock
wave and its locus is extended to $M_0 < \hat{c}_{f0}$, where it is smoothly connected to the slow-shock counterpart.
The Euler shock is evolutionary except the range within which the switch-on shock branch appears, i.e. the range satisfying the inequality (\ref{sw-on_ineq}).

\begin{figure}
\begin{tabular}{cc}
\begin{minipage}{0.45\hsize}
\begin{center}
\includegraphics[scale=0.26]{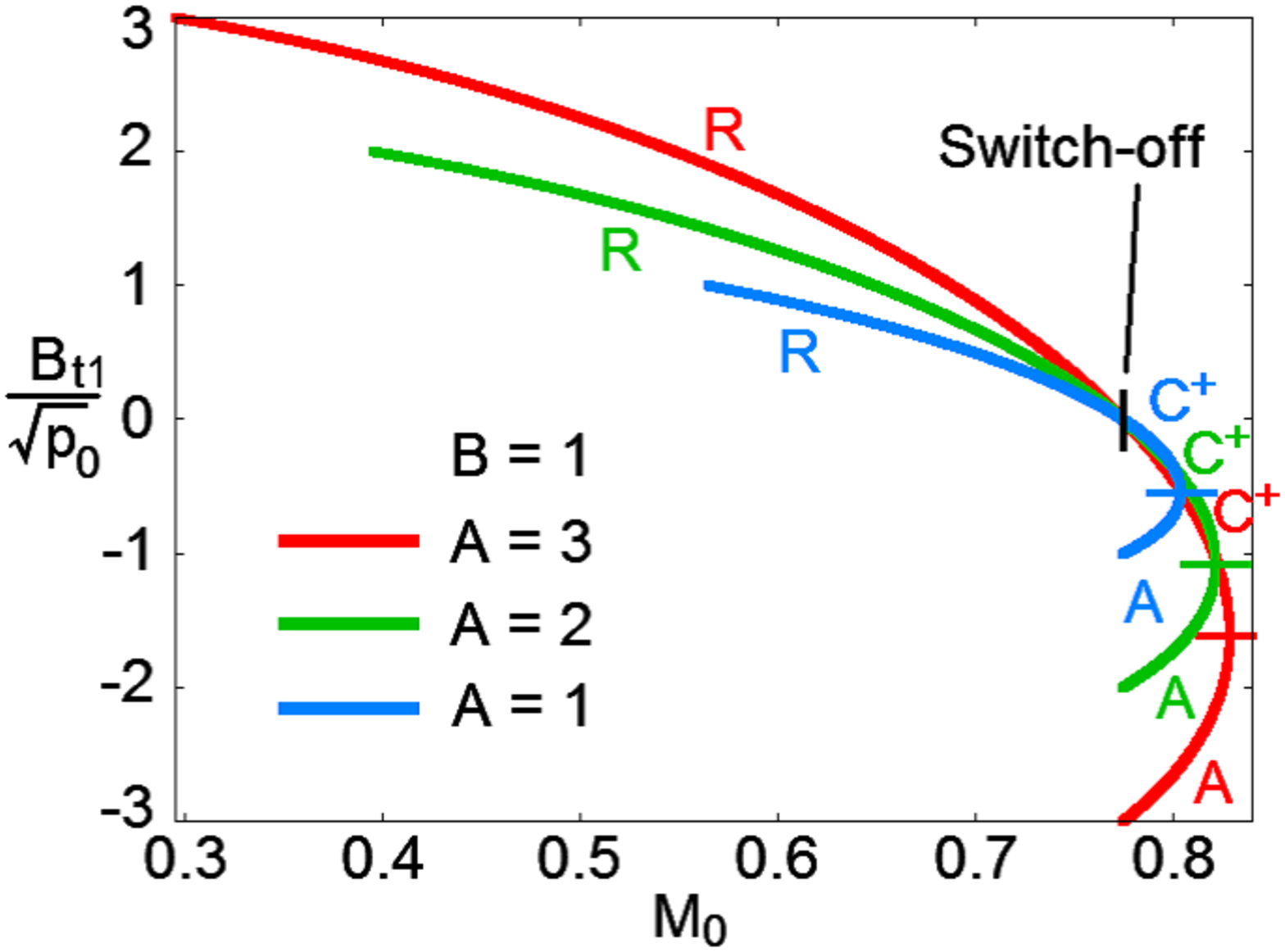}
\end{center}
\end{minipage} &
\begin{minipage}{0.45\hsize}
\begin{center}
\includegraphics[scale=0.26]{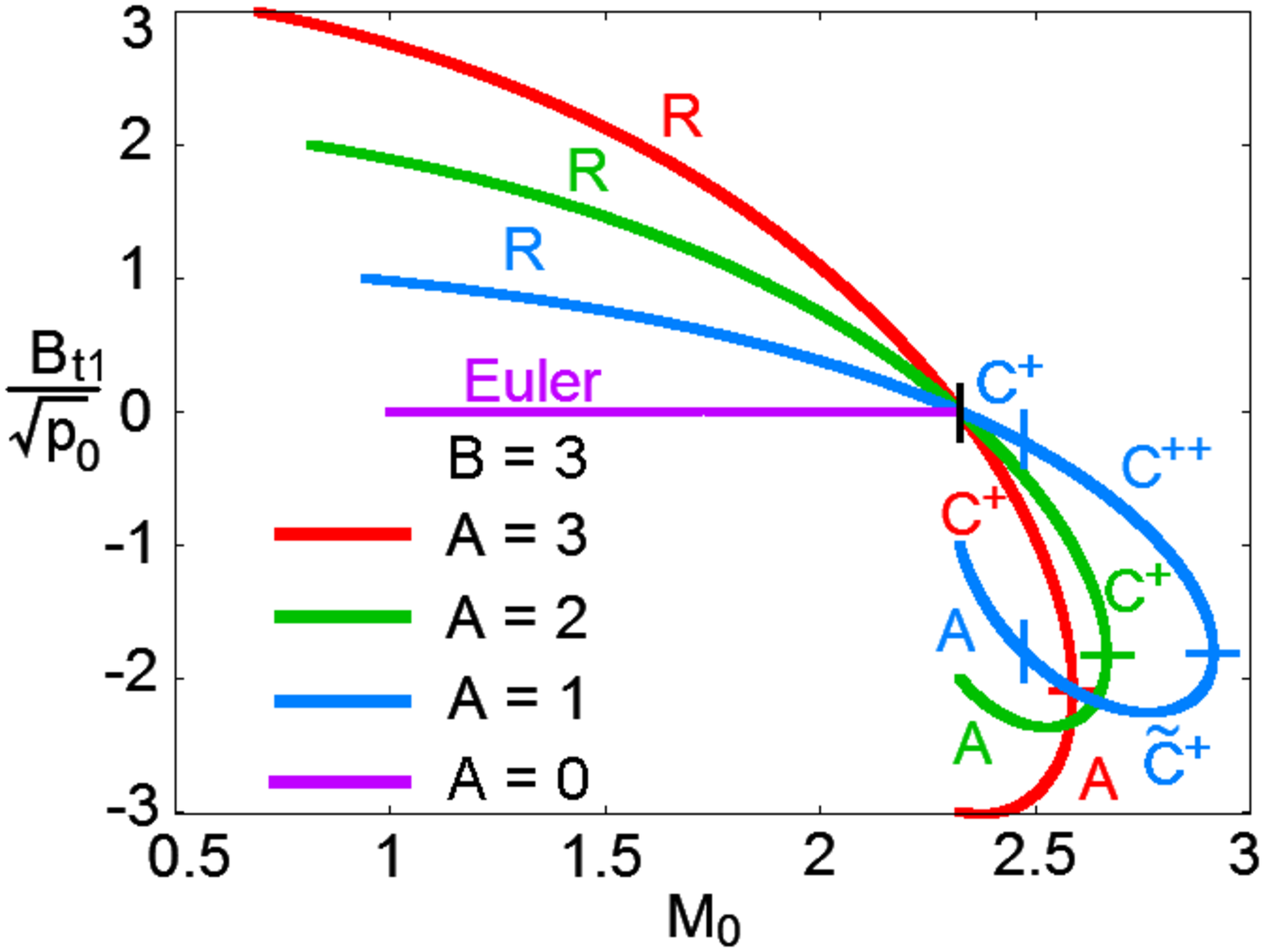}
\end{center}
\end{minipage} \\
\begin{minipage}{0.45\hsize}
\begin{center}
\includegraphics[scale=0.26]{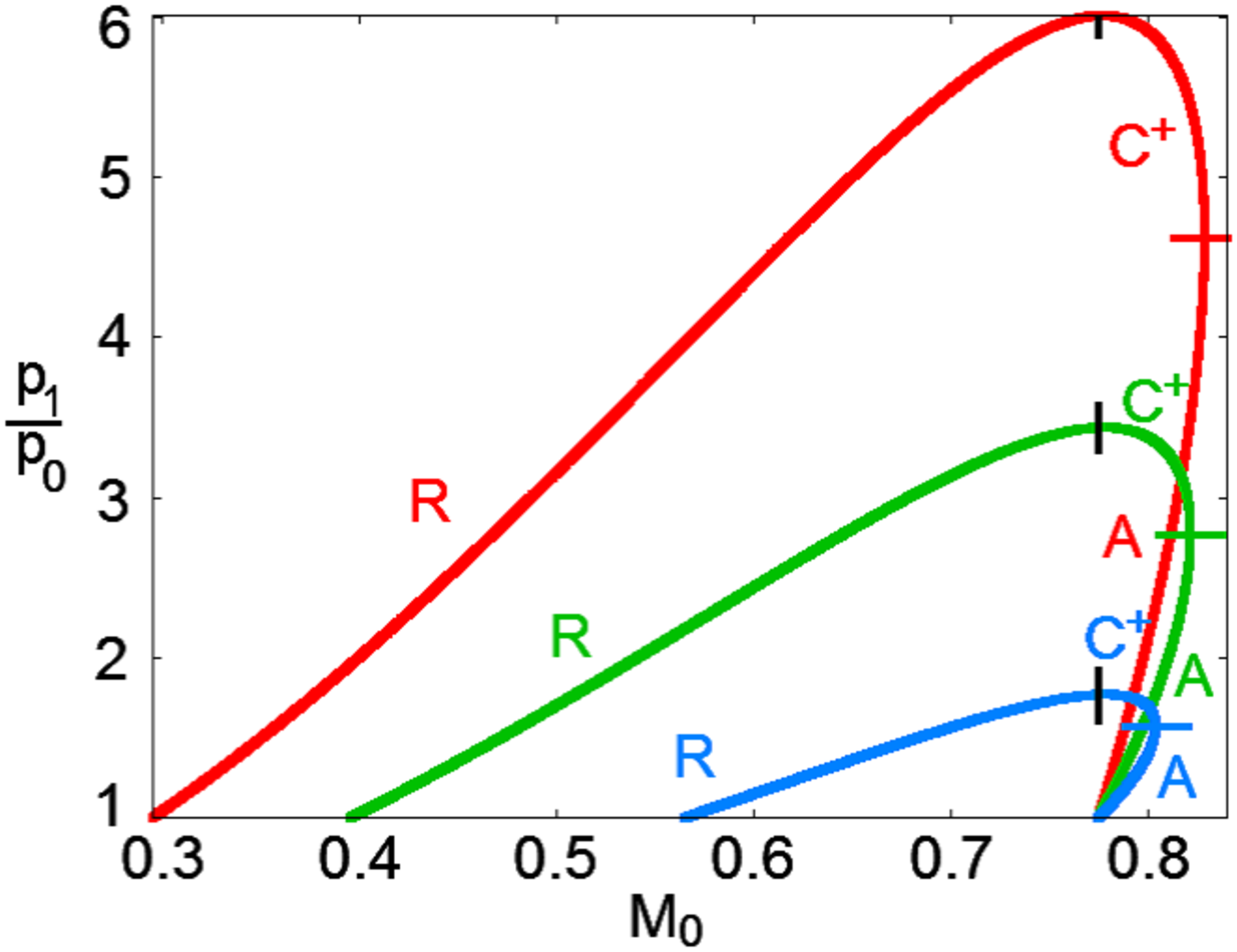}
\end{center}
\end{minipage} &
\begin{minipage}{0.45\hsize}
\begin{center}
\includegraphics[scale=0.26]{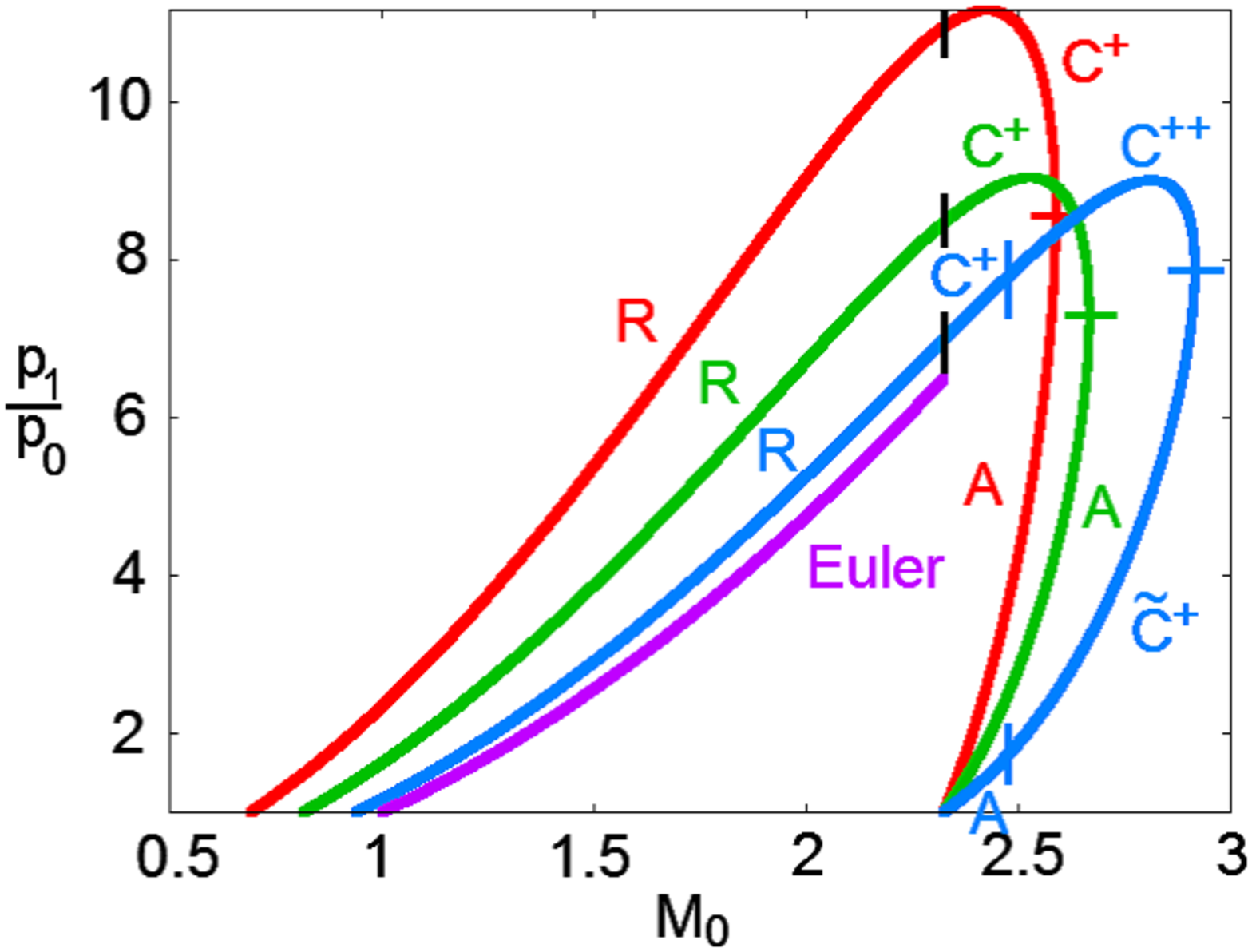}
\end{center}
\end{minipage} \\
\begin{minipage}{0.45\hsize}
\begin{center}
\includegraphics[scale=0.26]{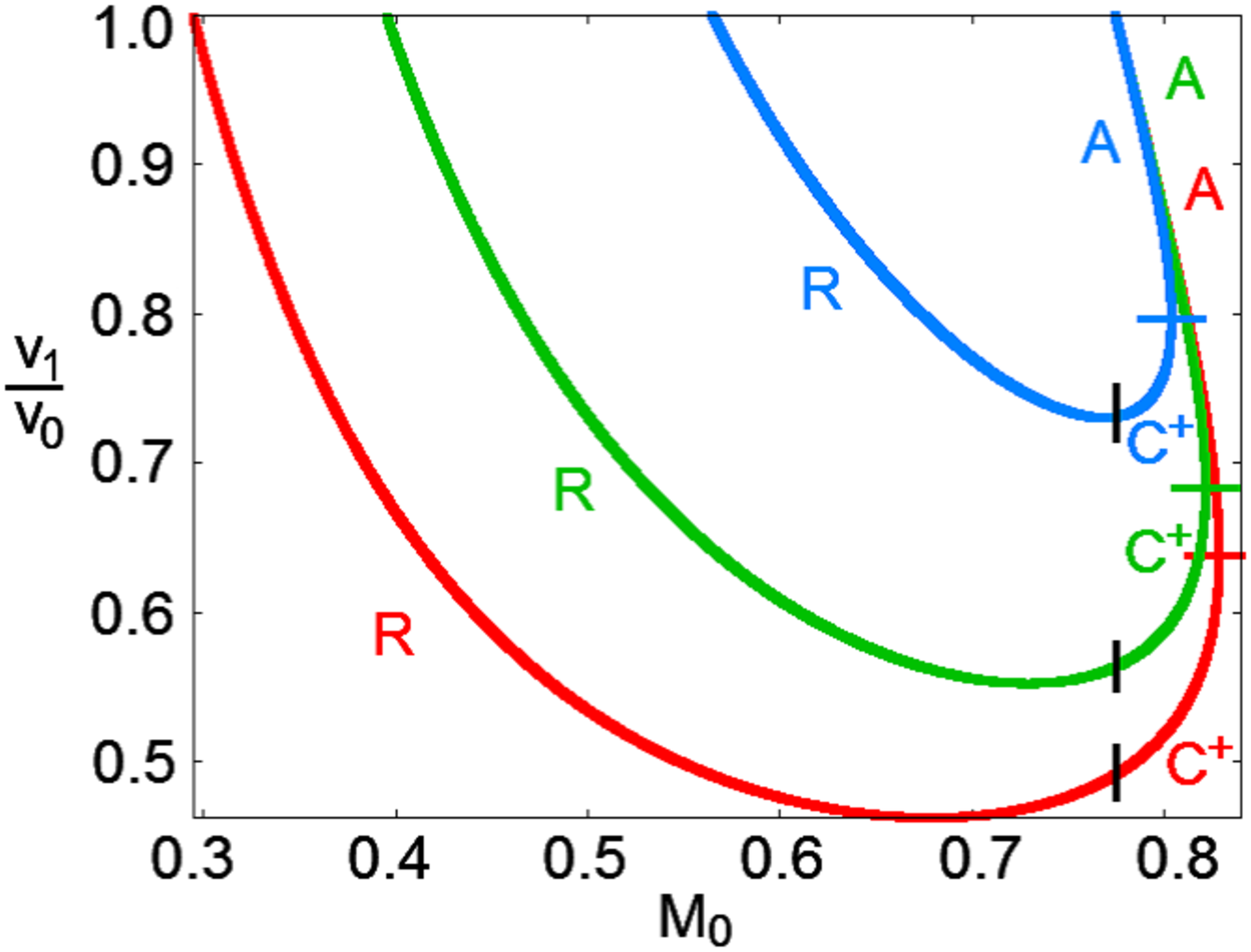}
\end{center}
\end{minipage} &
\begin{minipage}{0.45\hsize}
\begin{center}
\includegraphics[scale=0.26]{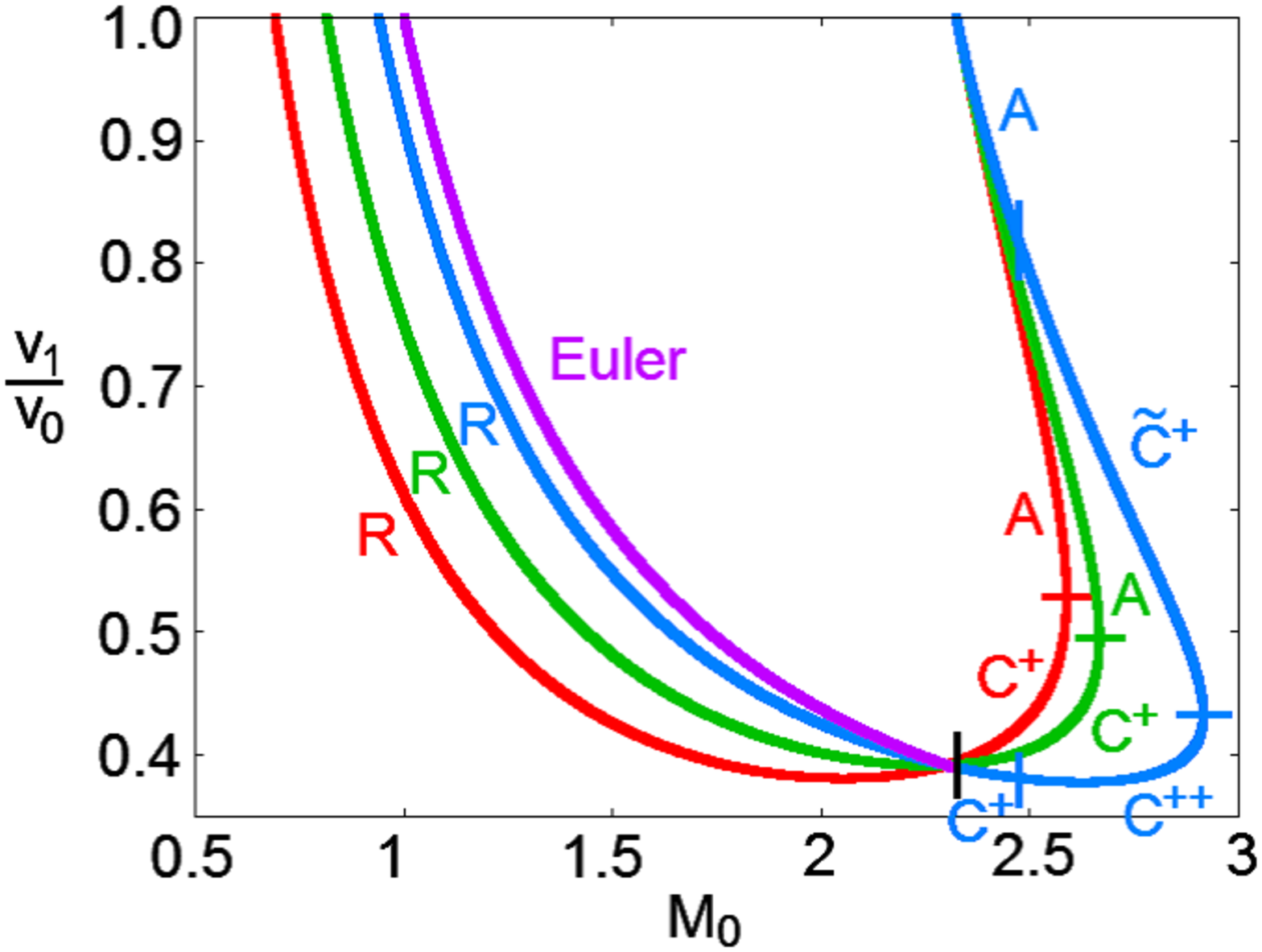}
\end{center}
\end{minipage} 
\end{tabular}
\caption{The slow loci for different combinations of the normal ($B$) and transverse ($A$) component of magnetic field.
The left panels: $B=1$ and $A = 3$ (red), $2$ (green) and $1$ (blue). The vertical black dashes indicate the points, at which 
$M_0 = \hat{c}_{A0}$ and switch-off shocks ($2,3 \rightarrow 4$ shocks) occur. Since $\hat{c}_{A0}$ is independent of $A$, the 
Mach numbers, $M_0$'s, at the points for all loci coincide with one another. These points mark the boundary between the regular slow 
shocks and non-regular intermediate shocks. The characters, $R$, $C^+$ and $A$, attached to each locus stand for the regular 
slow ($3 \rightarrow 4$), $2 \rightarrow 4$ intermediate and $2 \rightarrow 3$ intermediate shocks, respectively. The horizontal 
dash on each locus shows the point, at which the Mach number reaches its maximum on the locus and a $2 \rightarrow 3,4$ shock occurs.
This is the boundary between the $2 \rightarrow 4$ ($C^+$) and $2 \rightarrow 3$ ($A$) intermediate shocks. 
The right panels: $B=3$ and $A = 3$ (red), $2$ (green), $1$ (blue) and $0$ (purple, Euler shocks).
The vertical black dashes again give the boundary between the regular and non-regular shocks, at which switch-off shocks 
($2,3 \rightarrow 4$ shocks) occur. The characters, $R$, $C^+$ and $A$, have the same meaning as in the left panels whereas 
$C^{++}$ and $\tilde{C}^+$, which emerge only for small $A$'s, represent the $1\rightarrow 4$ and $1\rightarrow 3$ intermediate 
shocks, respectively. The horizontal dash on each locus marks again the point, at which the maximum Mach number is reached. 
On the other hand, the two vertical blue dashes on each blue locus indicate the points, at which $M_0 = \hat{c}_{f0}$. 
A $1,2\rightarrow 4$ shock occurs at the point closer to the vertical black dashes whereas a $1,2 \rightarrow 3$ shock emerges at 
the other point. A $1 \rightarrow 3,4$ shock occurs at the point indicated by the horizontal blue dash.
Note in passing that the locus vanishes at $B=0$.
}
\label{fig.locus2}
\end{figure}


The slow family is characterized by the feature that matter is compressed but the transverse magnetic field is reduced and in some cases 
reversed by the shock passage. 
The slow loci are shown in Fig.~\ref{fig.locus2} as a function of the upstream Mach number ($M_0$) for a number of combinations of 
the upstream normal ($B$) and transverse ($A$) components of magnetic field. It is evident that some loci are two-valued as a function
of $M_0$. There is a maximum Mach number in each locus, at which the downstream flow speed becomes equal to the the slow speed.
In some cases the minimum values appear in the transverse magnetic fields for given $A$ and $B$. 

The intermediate shocks are those that give negative downstream transverse magnetic fields. The shocks that nullify the transverse 
magnetic field are called the switch-off shock ($2,3 \rightarrow 4$ shock). It is apparent that the switch-off shocks are 
located at the boundary between
the regular slow shocks and the intermediate shocks, the fact which will be important later. Each locus is terminated at the point
that corresponds to a rotational discontinuity, which rotates magnetic field by $\upi$ radian, i.e. $\Bth = -A$.  
In the limit of $B \rightarrow 0$ or $A \rightarrow 0$, the slow branch vanishes, i.e. there is no solution that satisfies Eq.~(\ref{fast branch}) and the inequalities: $\hat{c}_{s0} <M_0 < \hat{c}_{f0}$ and $0 < \vh < 1$, except when $A =0$ and the upstream \Alfven \ speed 
is larger than the acoustic speed, i.e. $\hat{c}_{A}( := c_{A0}/a_0) > 1$, in which case the Euler shock branch takes its place. 
This branch is extended to the regime of $M_0 > \hat{c}_{f0}$ and connected smoothly to the fast-shock counterpart as mentioned earlier.



\section{Solving the MHD Riemann problem} \label{sec.method}
The basic for finding solutions of the MHD Riemann problem is essentially the same as that of \citet{T02}.
However, we have extended it substantially so that we could obtain all types of solutions for arbitrary initial conditions, 
including non-regular ones. In the following we briefly describe the basic ideas and the technical details will be published 
elsewhere \citep{TY12}.

First we consider the regular solutions that were treated by \citet{T02}. It is assumed that both normal and transverse components 
of magnetic field are non-vanishing. This implies that the degenerate case mentioned in \S~\ref{smp} is ignored.
As a definition of the regular solution, no intermediate shock is involved. It is further assumed that there is no switch-off rarefaction wave,
either. Since the switch-off rarefaction wave is regular, this is another genuine restriction on the solutions that can be handled.
Thanks to these assumptions, which types of waves are generated in what order are known a priori: the fast, \Alfven,\ and slow waves 
are fanning out in this order on both sides of the contact discontinuity. Since each regular wave forms a one-parameter family, 
the regular solutions considered now are specified by the seven parameters that are associated with the waves involved in the solution. 
At the contact discontinuity, the six quantities other than density are continuous. Hence the six parameters out of seven that correspond
to the regular waves other than contact discontinuity are determined by the requirement that the six quantities other than density
be continuous at the contact discontinuity. Then the last parameter that specifies the contact discontinuity is automatically obtained.
As pointed out by \citet{T02}, further simplification arises from the observation that only the rotational discontinuities rotate 
transverse magnetic fields. In fact, the sum of rotation angles at the left- and right-going rotational discontinuities is equal 
to the angle that are made by the transverse components of the magnetic fields on both sides of the initial discontinuity. Hence we
have only to solve one of the rotation angles. 

We thus arrive at the reduced problem, in which the remaining five parameters should be determined. The system of equations is 
expressed symbolically as 
\begin{equation}
\label{System of Riemann problem}
\bm{F}_L^s( \psi _L^s ;\bm{u}_L^r(\psi ^r ;\bm{u}_L^f(\psi _L^f;\bm{u}_L) ) ) -\bm{F}_R^s( \psi _R^s ;\bm{u}_R^r(\psi ^r ;\bm{u}_R^f(\psi _R^f;\bm{u}_R) ) ) = \bm{0} ,
\end{equation}
where $\psi _L^f$, $\psi _L^s$, $\psi _R^f$, $\psi _R^s$ and $\psi^r$ are the five parameters that specify the jumps at the left-going 
fast and slow waves, right-going fast and slow waves, and rotational discontinuities, respectively. The initial left and right states are
given by $\bm{u}_{L,R} = {}^t(\rho _{L,R}, p_{L,R}, \bm{v}_{L,R}, \bm{B}_{t L,R})$. The downstream states of the left- and right-going
fast waves are expressed as $\bm{u}_{L,R}^f$, which are functions of the upstream states, $\bm{u}_{L,R}$, and the parameters, $\psi _{L,R}^f$.
Similarly, $\bm{u}_{L,R}^r$ represent the downstream states of the left- and right-going rotational discontinuities. The five downstream
quantities,  $p$, $\bm{v}$ and $|\bm{B}_t|$, of the left- and right-going slow waves, which should be continuous at the contact discontinuity, 
are denoted by $\bm{F}_{L,R}^s$. These equations are numerically solved by the Newton-Raphson method in our code.

As mentioned already, if the switch-off rarefaction is involved in the solution, the above method cannot be applied as it is, since
we do not know a priori which waves emerge in the solution. \citet{T02,T03} neglected such possibilities by assuming that both normal 
and transverse components of magnetic fields are non-vanishing. This is also the case for the code developed by \citet{ATJweb} and 
available online, which appears to be based on \citet{T02,T03}. Hence their codes cannot handle all the regular solutions in addition 
to the non-regular ones, which they did not take into account from the beginning.

On the contrary, our code can treat all these cases: vanishing normal or transverse component of magnetic field with a possible 
existence of the switch-off rarefaction waves as well as the emergence of intermediate shocks including the switch-on and switch-off 
shocks. All the intermediate shocks other than switch-on shocks are parametrized as slow shocks because they are parts of the slow loci.
The switch-on shocks and switch-off rarefactions, on the other hand, are treated as a part of the fast shock and rarefaction, 
respectively. As mentioned already, since we do not know a priori which waves are generated, we need to try all possible combinations.
For example, if the transverse magnetic fields are non-vanishing initially, we first attempt to find solutions that do not include
a fast wave on one side of the contact discontinuity, i.e. the solution with either a $"1 \rightarrow 3"$, $"1 \rightarrow 4"$, 
$"1\rightarrow 3,4"$, $"1,2 \rightarrow 3"$, $"1,2 \rightarrow 4"$ or $"1,2 \rightarrow 3,4"$ intermediate shock being included.
We then try to obtain a solution that does include a fast wave but not a rotational discontinuity, i.e. the solution with either a 
$"2 \rightarrow 3"$, $"2 \rightarrow 4"$ or $"2 \rightarrow 3,4"$ intermediate shock or a switch-off shock. Finally, regular solutions
are sought. The technical details of these procedures will be described in a separate paper~\citep{TY12}.

\section{Results} \label{sec.results}
In this section, we apply our new code to two specific initial conditions and their neighbors. One of them is first picked up by
\citet{BW88} and is known to have a non-regular solution with a compound wave. This is one of the initial conditions in the MHD 
Riemann problem that are best known to developers of numerical MHD codes. As demonstrated by \citet{ATJweb} there is also a regular
solution to this problem. The controversy in the literature is why the former is almost always obtained in the numerical solutions.
After confirming the above regular and non-regular solutions by our code, we show that there are actually uncountably many non-regular
solutions to this problem. By studying the properties of these solutions as well as exploring the neighboring solutions, we obtain new
insights into the controversy.

In the second application, we present the initial condition, for which there seems to be no regular solution. In fact the non-regular 
solution we found is a limit of a sequence of regular solutions and include a switch-off shock wave. We investigate the neighboring
solutions in detail and reveal how the solution is approached by various sequences of solutions. The result forces us to rethink the 
interpretations of the regular waves and switch-off shocks.

\subsection{Brio \& Wu problem: uncountably many non-regular solutions}
We present here the solutions for the initial condition in the MHD Riemann problem solved numerically by \citet{BW88}.
The left and right states in the initial condition are given by
\begin{eqnarray}
(\rho_L,\ p_L,\ v_{xL},\ v_{yL},\ v_{zL},\ B_{yL},\ B_{zL}) =& (1,\ 1,\ 0,\ 0,\ 0,\ 1,\ 0), \\
(\rho_R,\ p_R,\ v_{xR},\ v_{yR},\ v_{zR},\ B_{yR},\ B_{zR}) =& (0.125,\ 0.1,\ 0,\ 0,\ 0,\ -1,\ 0).
\end{eqnarray}
The normal component of magnetic field is $B_n = 0.75$ and we take for the adiabatic index $\gamma = 5/3$ instead of $\gamma = 2$, 
the value adopted by \citet{BW88}. This is because the Riemann invariants are singular at $\gamma = 2$~\citep[e.g.][]{TY12} and cannot be 
used for check as usually done. Note, however, there is no problem to find solutions by our code even in this case. In fact, 
we confirmed that the solutions are not different qualitatively between $\gamma = 5/3$ and $\gamma = 2$. This is a coplanar problem,
for which both regular and non-regular solutions are expected. It is also noted that this can be regarded as a Riemann problem of the 
reduced MHD system.

\begin{figure}
\begin{tabular}{cc}
\begin{minipage}{0.45\hsize}
\begin{center}
\includegraphics[scale=0.26]{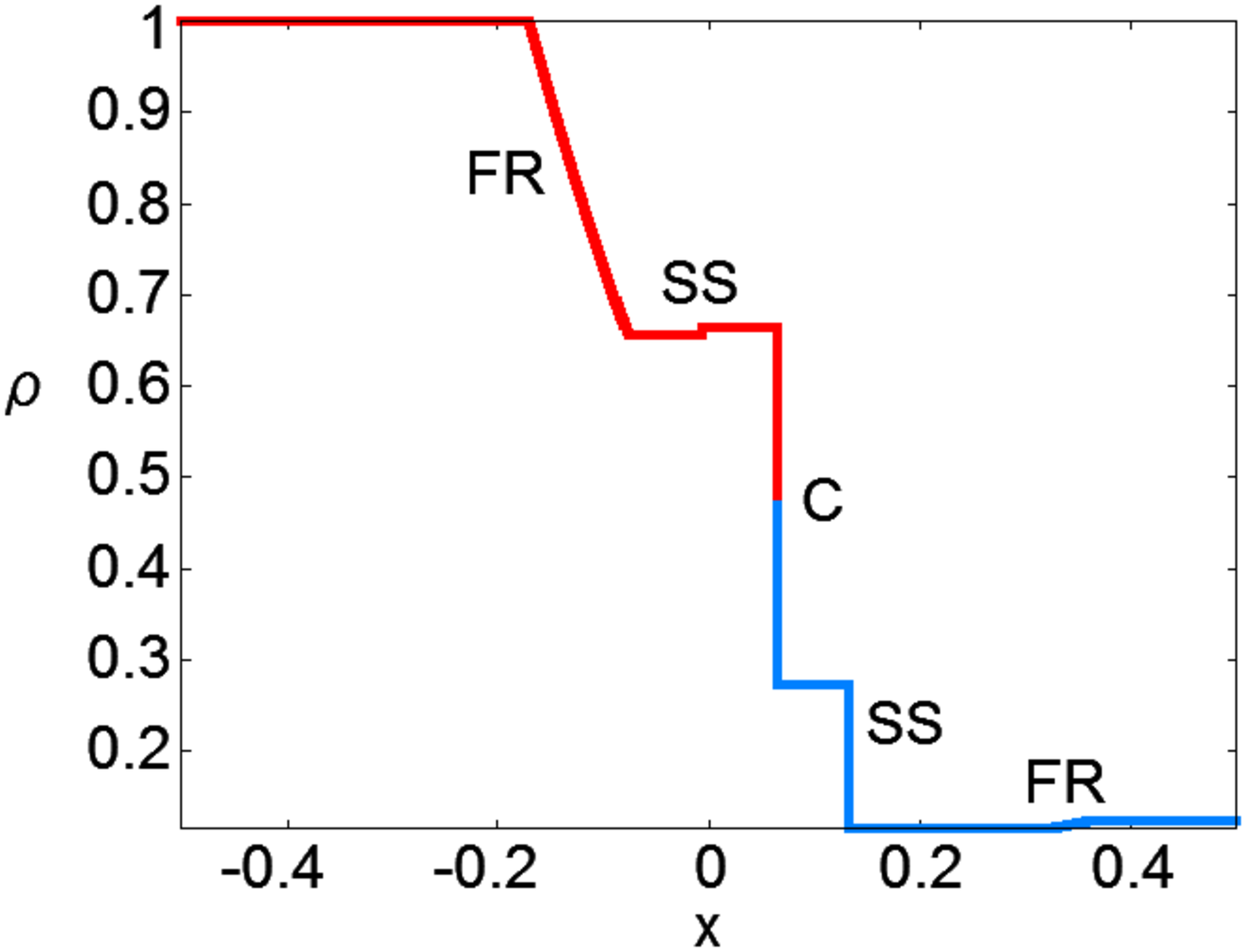}
\end{center}
\end{minipage} &
\begin{minipage}{0.45\hsize}
\begin{center}
\includegraphics[scale=0.26]{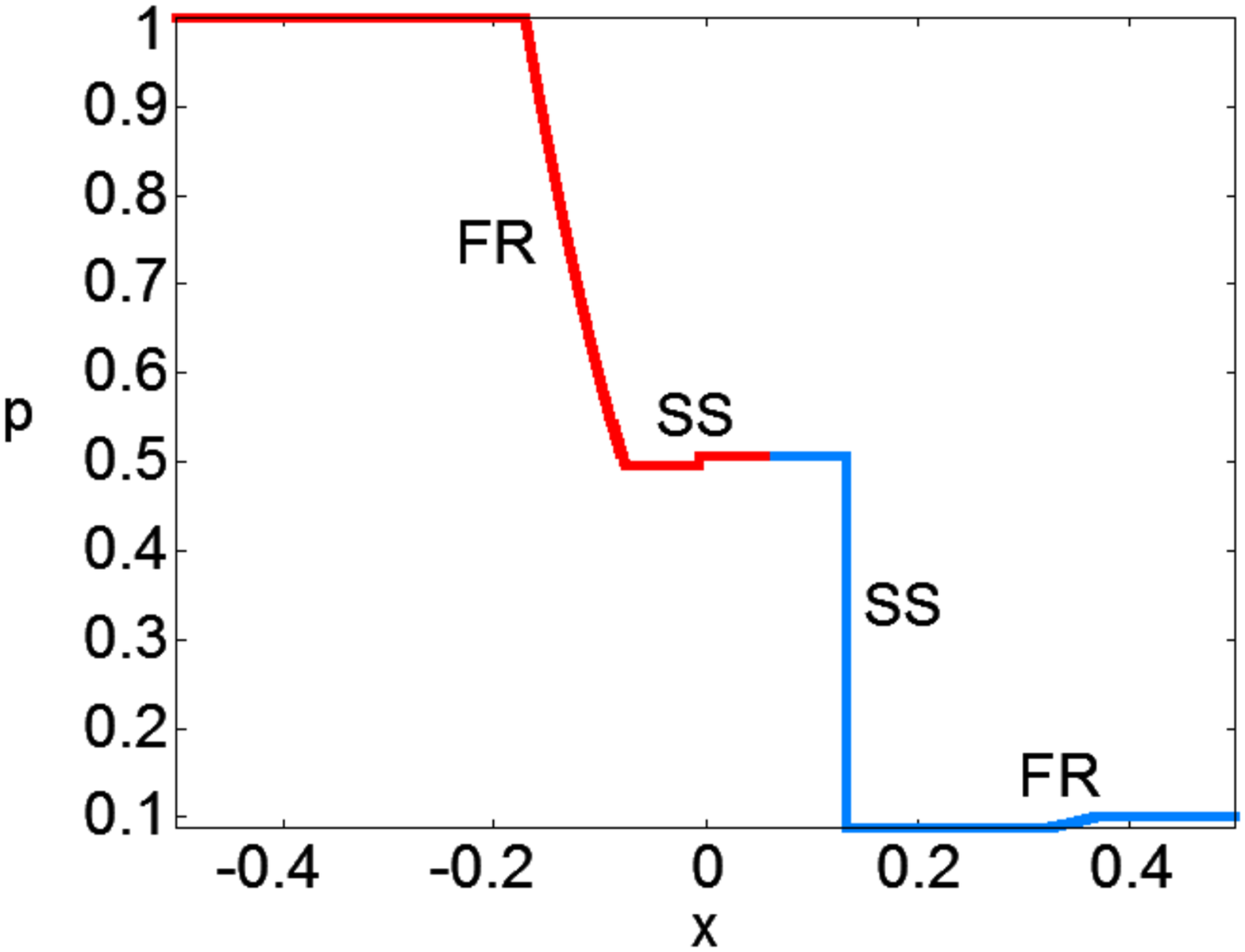}
\end{center}
\end{minipage} \\
\begin{minipage}{0.45\hsize}
\begin{center}
\includegraphics[scale=0.26]{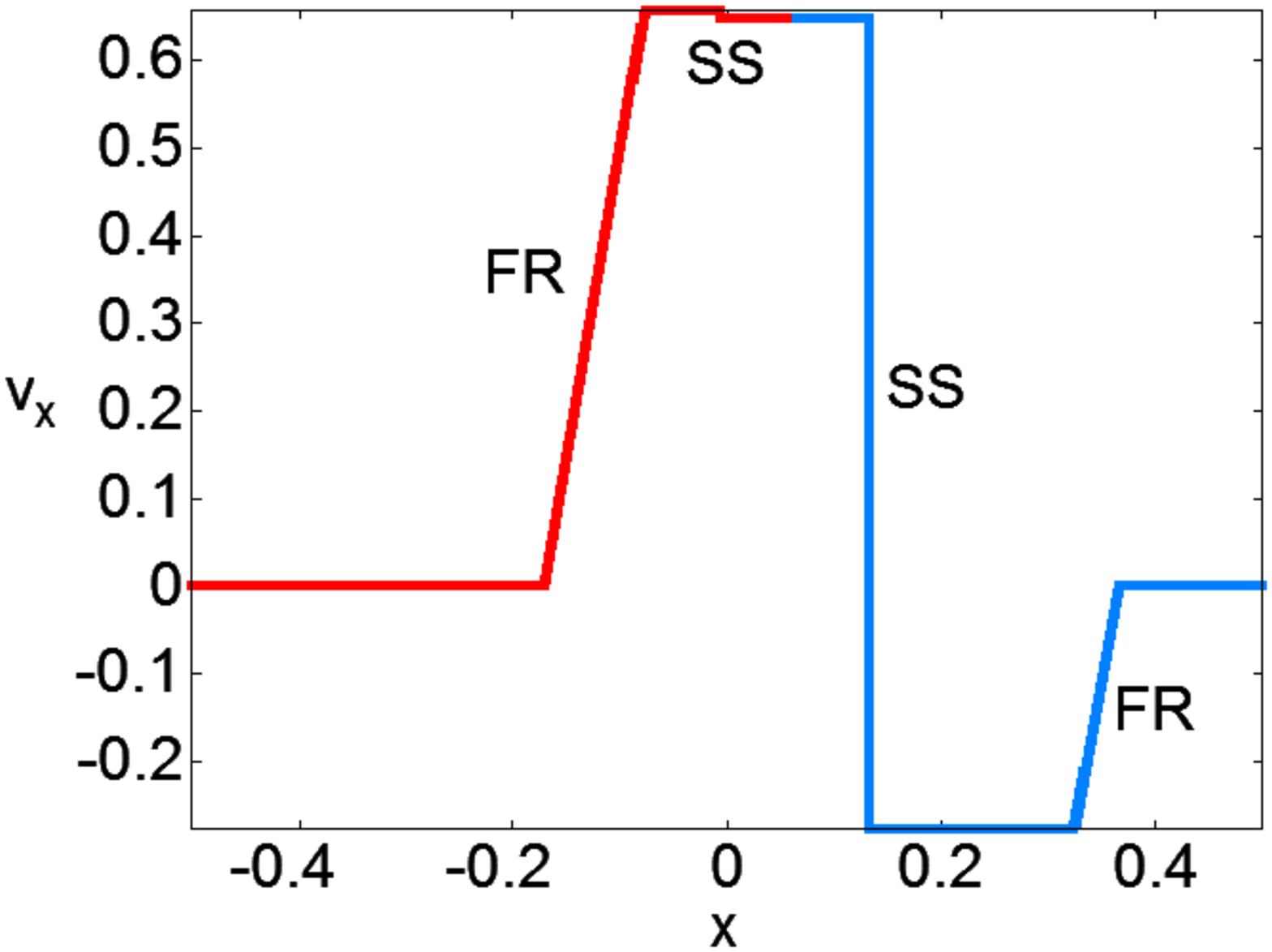}
\end{center}
\end{minipage} &
\begin{minipage}{0.45\hsize}
\begin{center}
\includegraphics[scale=0.26]{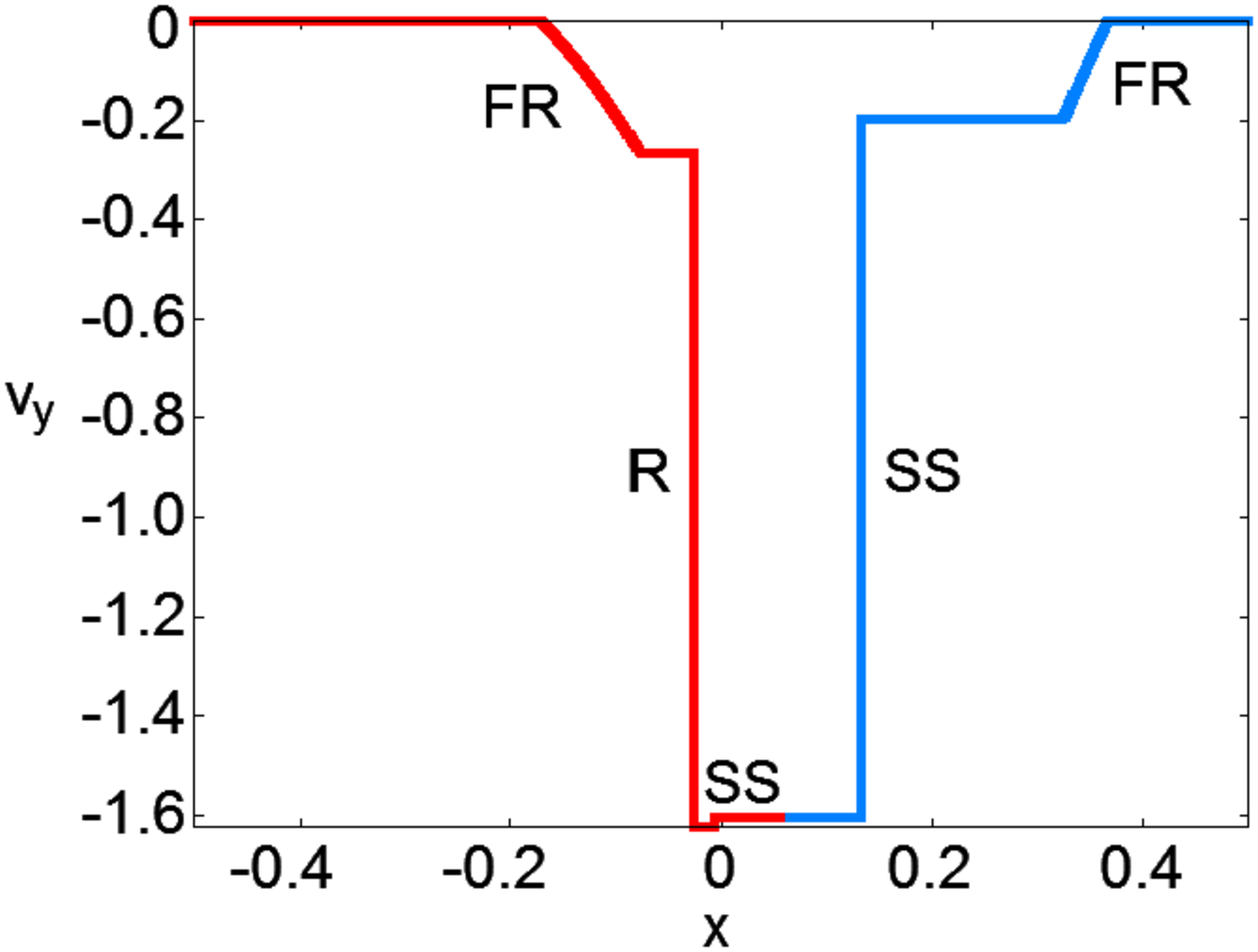}
\end{center}
\end{minipage} \\
\begin{minipage}{0.45\hsize}
\begin{center}
\includegraphics[scale=0.26]{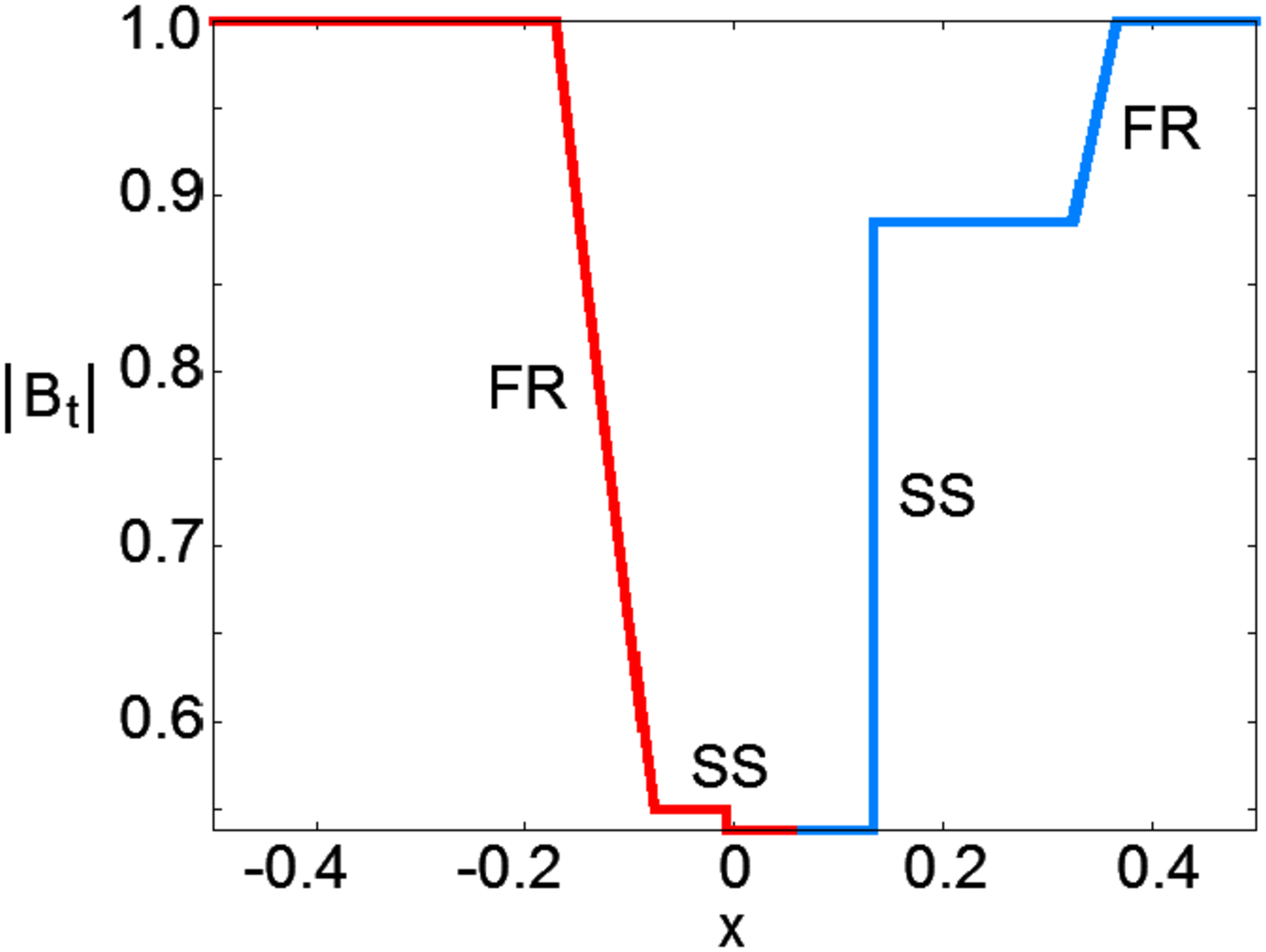}
\end{center}
\end{minipage} &
\begin{minipage}{0.4\hsize}
\begin{center}
\includegraphics[scale=0.26]{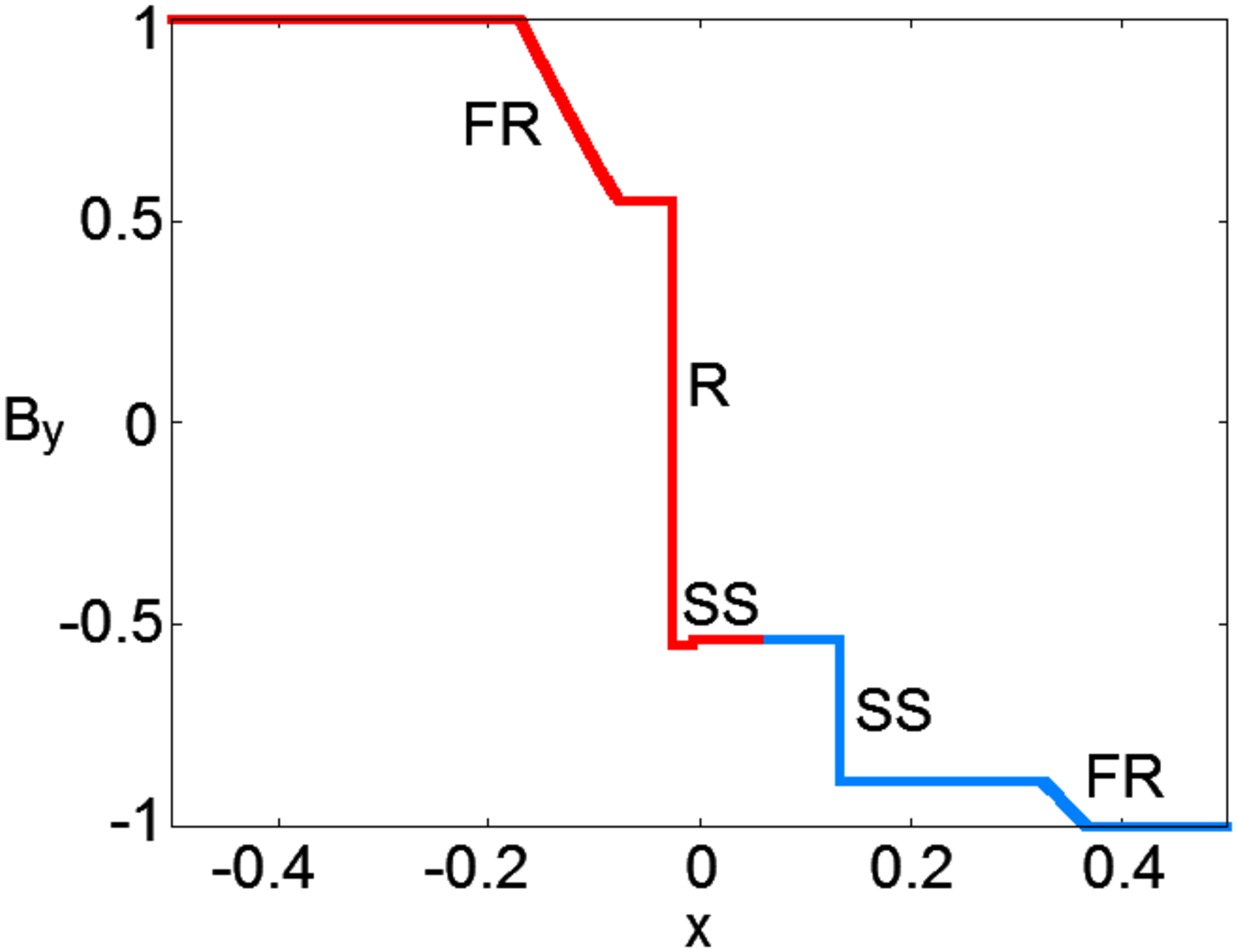}
\end{center}
\end{minipage} 
\end{tabular}
\caption{The regular solution for the \citet{BW88} problem. The waves in the red and blue portions that are separated by the contact 
discontinuity are left- and right-going, respectively. The designations FR, R, SS and C represent the fast rarefaction wave, rotational
discontinuity, slow shock and contact discontinuity, respectively.}
\label{BW regular}
\end{figure}

We begin with the regular solution. In Fig.~\ref{BW regular} we show various quantities in the solution at $t=0.1$. $v_z$ and $B_z$ are 
identically zero and omitted in the figure. The solution consists of a fast rarefaction wave (denoted by FR in the figure), rotational 
discontinuity (R) and slow shock (SS) going leftwards and a fast rarefaction wave, slow shock and contact discontinuity running in the
right direction in addition to a contact discontinuity (C). No right-going rotational discontinuity exist because the left-going one 
has rotated the magnetic field by $\upi $ radian already. This is the same solution as the one obtained by \citet{ATJweb}. 
 
\begin{figure}
\begin{tabular}{cc}
\begin{minipage}{0.45\hsize}
\begin{center}
\includegraphics[scale=0.26]{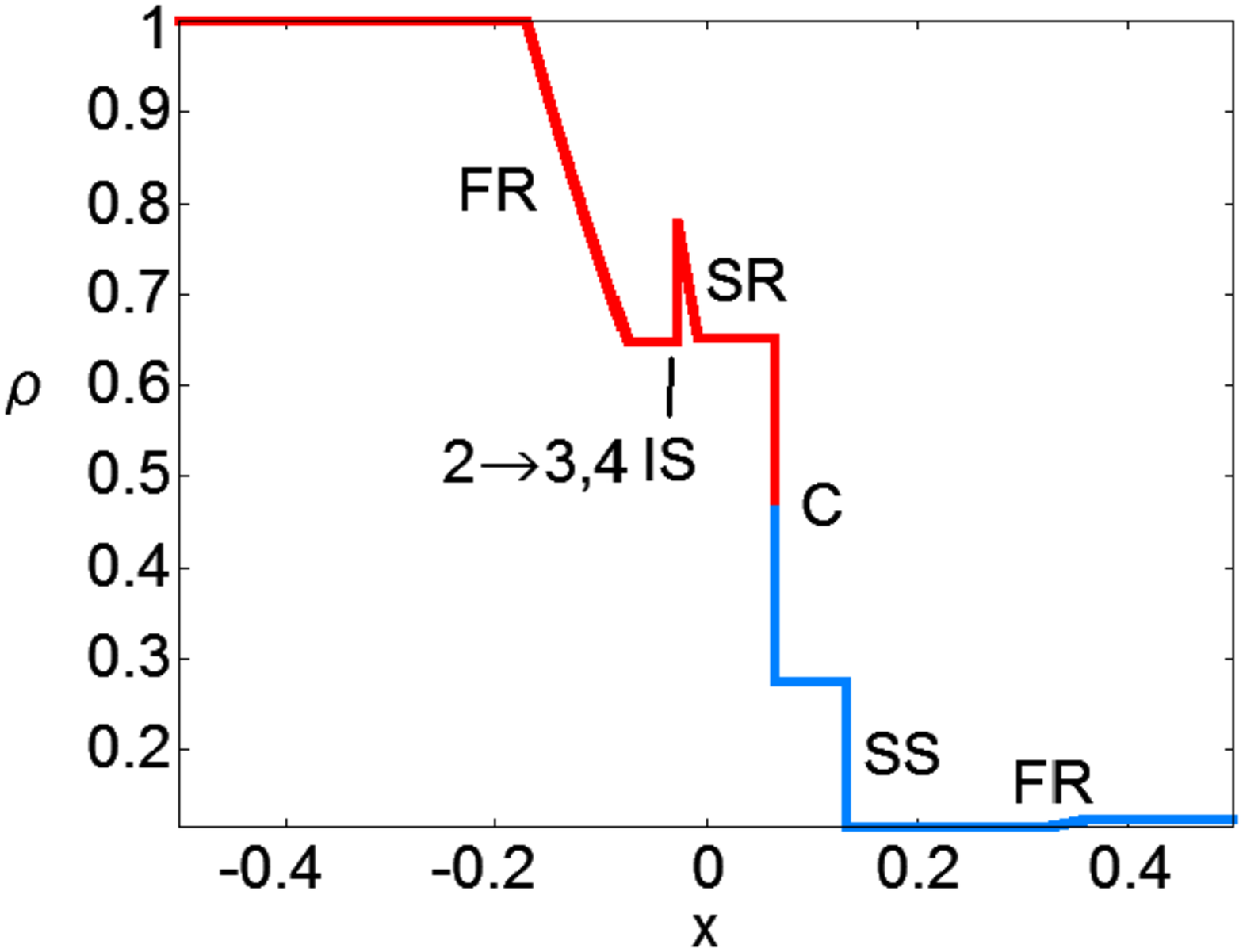}
\end{center}
\end{minipage} &
\begin{minipage}{0.45\hsize}
\begin{center}
\includegraphics[scale=0.26]{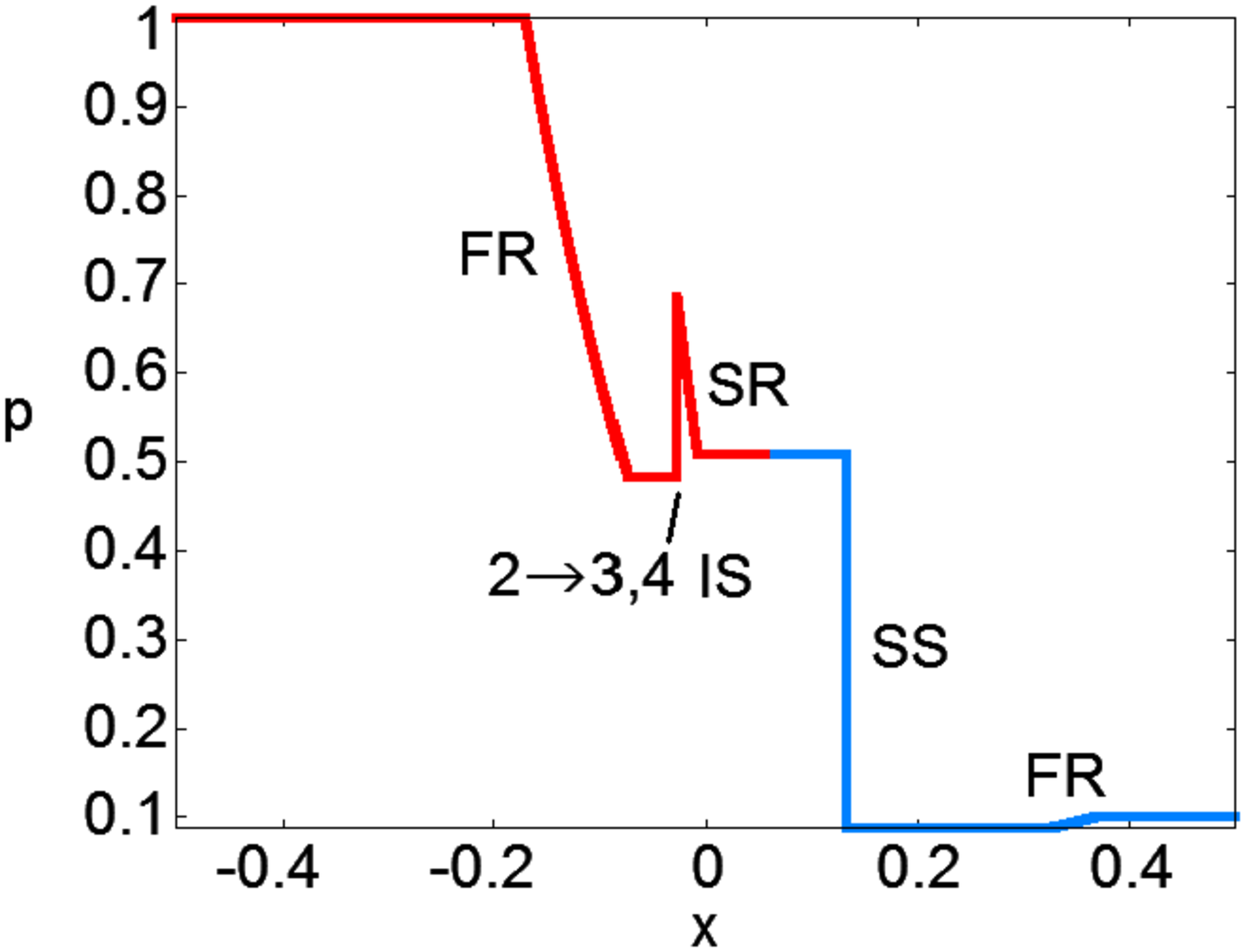}
\end{center}
\end{minipage} \\
\begin{minipage}{0.45\hsize}
\begin{center}
\includegraphics[scale=0.26]{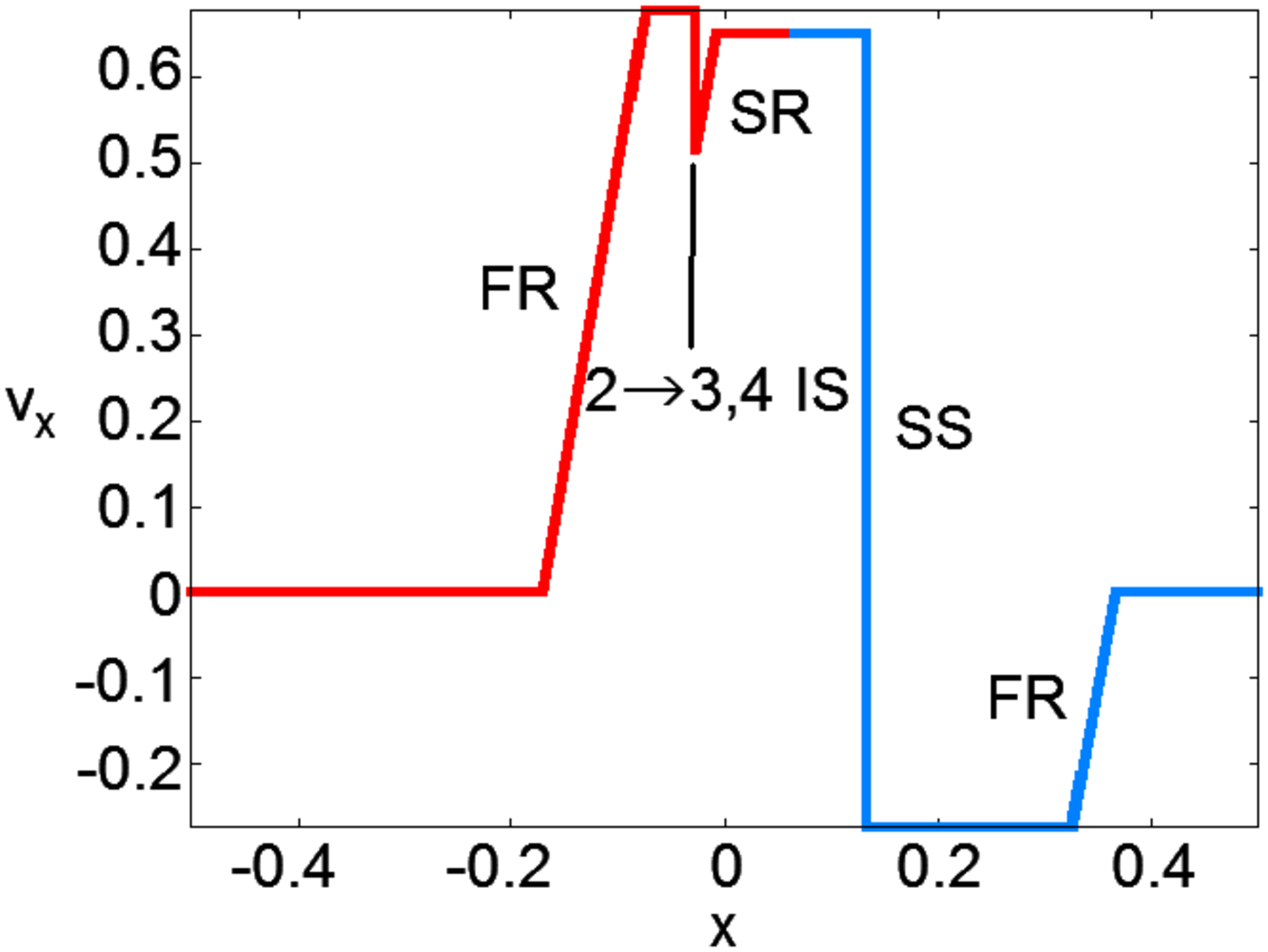}
\end{center}
\end{minipage} &
\begin{minipage}{0.45\hsize}
\begin{center}
\includegraphics[scale=0.26]{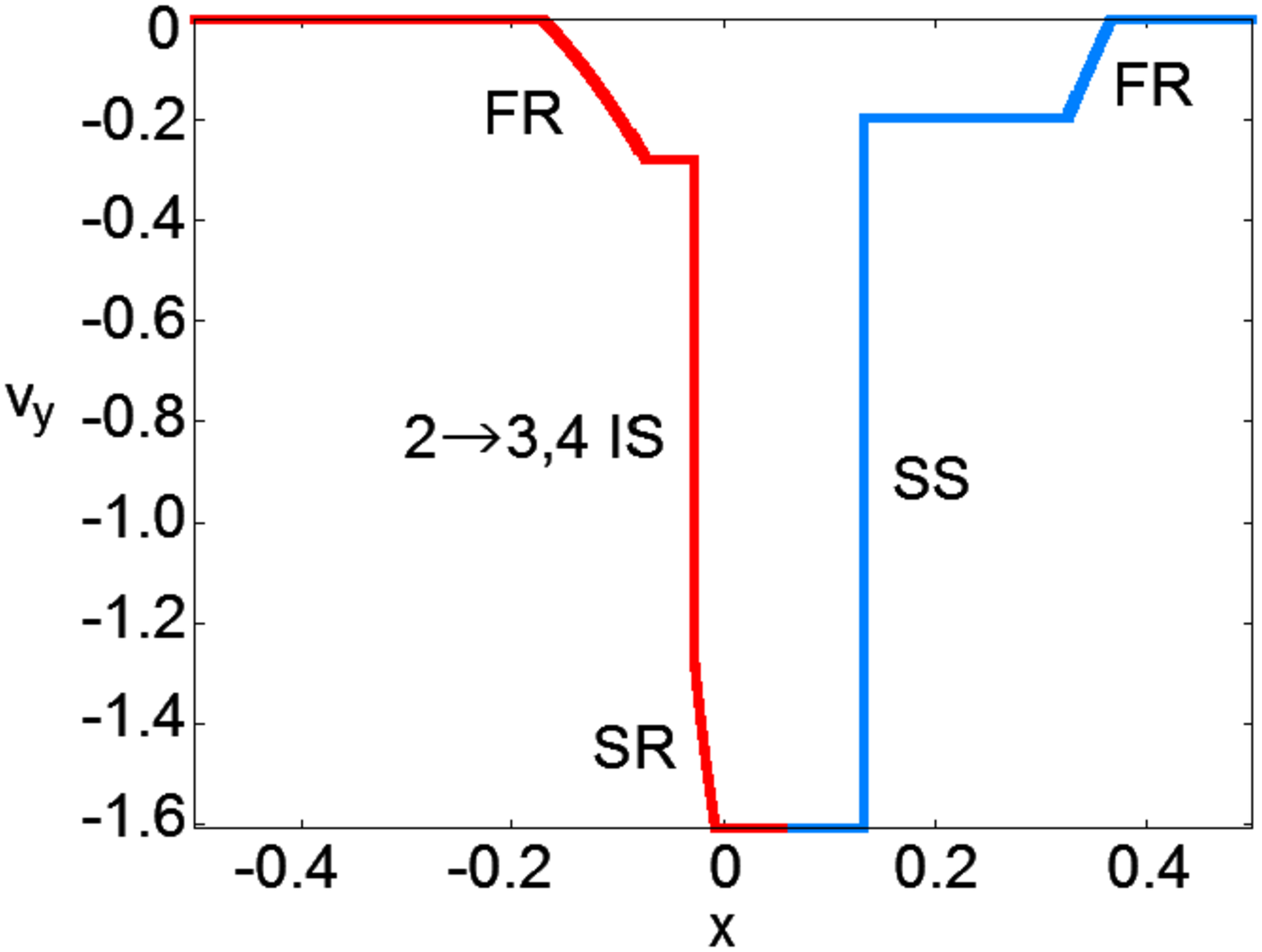}
\end{center}
\end{minipage} \\
\begin{minipage}{0.45\hsize}
\begin{center}
\includegraphics[scale=0.26]{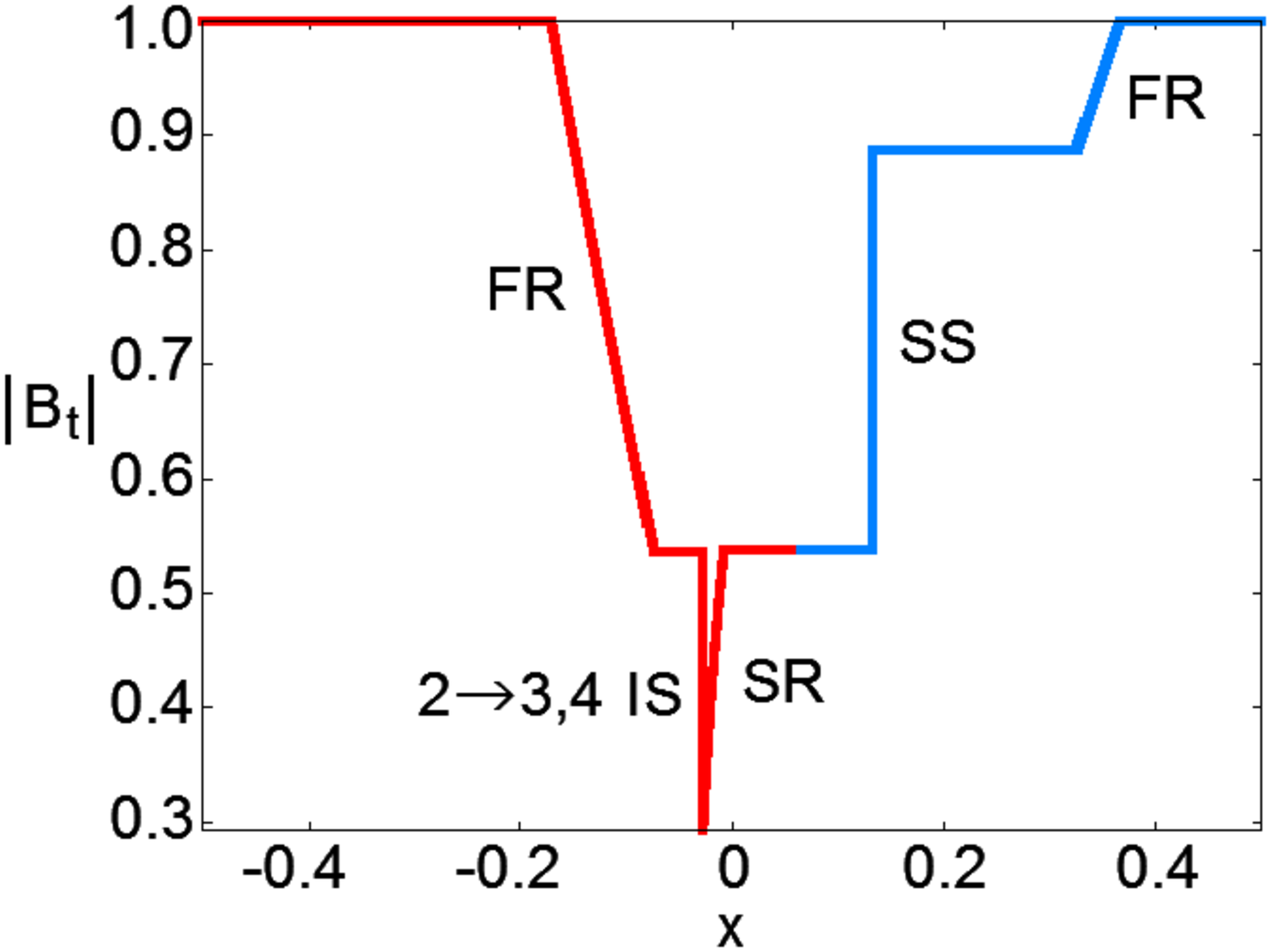}
\end{center}
\end{minipage} &
\begin{minipage}{0.45\hsize}
\begin{center}
\includegraphics[scale=0.26]{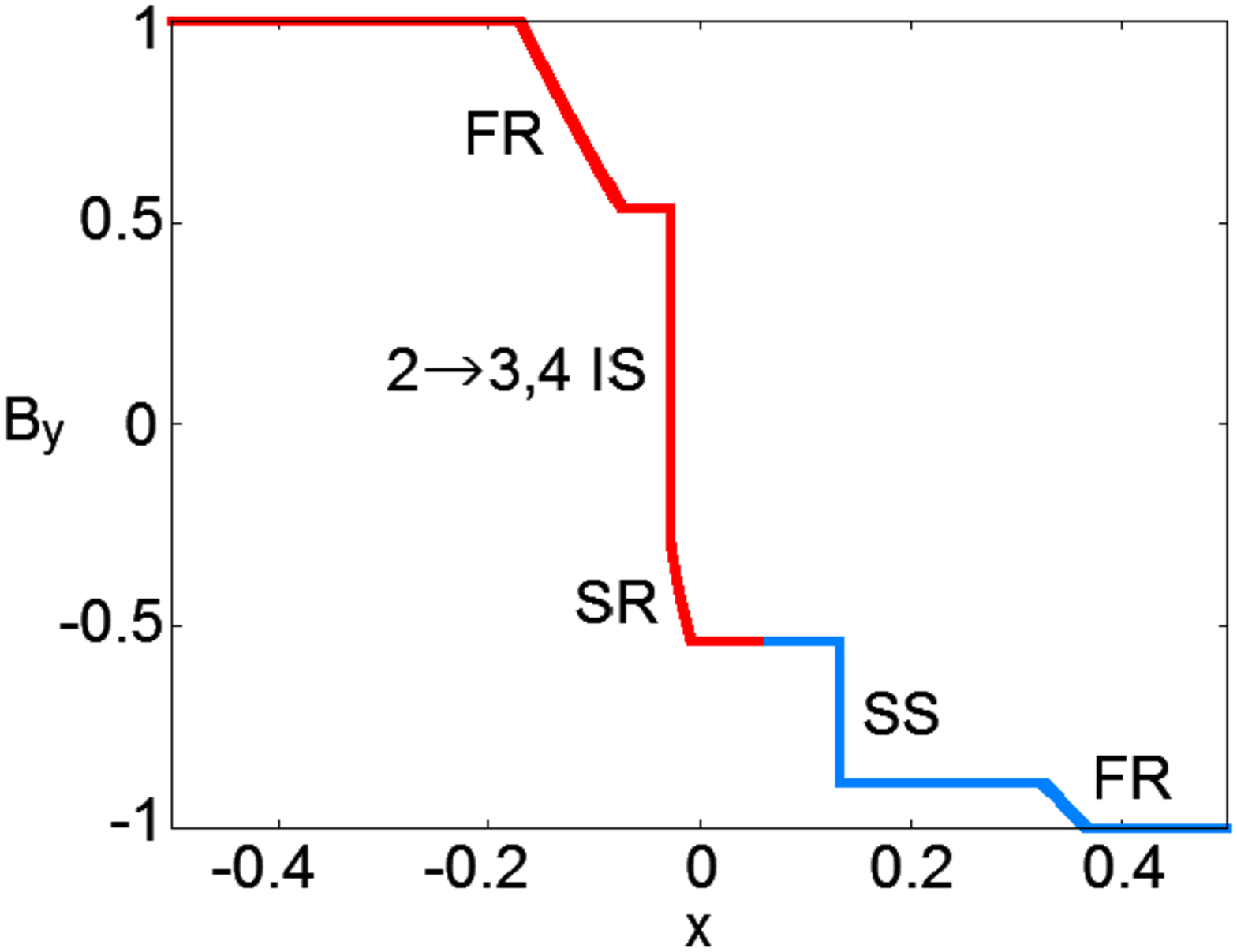}
\end{center}
\end{minipage} 
\end{tabular}
\caption{The non-regular solution for the Brio \& Wu problem that contains a compound wave.
The notation is the same as in Fig.~\ref{BW regular} except for the designations SR and IS,
which stand for the slow rarefaction wave and intermediate shock, respectively.}
\label{BW compound}
\end{figure}

Next we present a non-regular solution, various quantities in which are displayed at $t = 0.1$ in Fig.~\ref{BW compound}. As is evident, 
this solution contains a compound wave (denoted by $2\rightarrow 3,4$ IS and SR in the figure) and is exactly the solution \citet{BW88} obtained in their numerical
simulations. The compound wave is a $2 \rightarrow 3,4$ intermediate shock, to which a slow rarefaction wave is attached. Other waves that
constitute the solution are a left-going fast rarefaction wave and a contact discontinuity, slow shock and fast rarefaction wave going
rightward. Note that the compound wave is responsible for changing the direction of magnetic field. It should be also pointed out that 
the strengths of the waves other than the compound wave are different between the regular and non-regular solutions. It is the merit of our
code that these non-regular solutions can be handled unlike the one developed and made available for public use by \citet{ATJweb}, which 
can treat only regular solutions.

\begin{figure}
\begin{tabular}{cc}
\begin{minipage}{0.45\hsize}
\begin{center}
\includegraphics[scale=0.26]{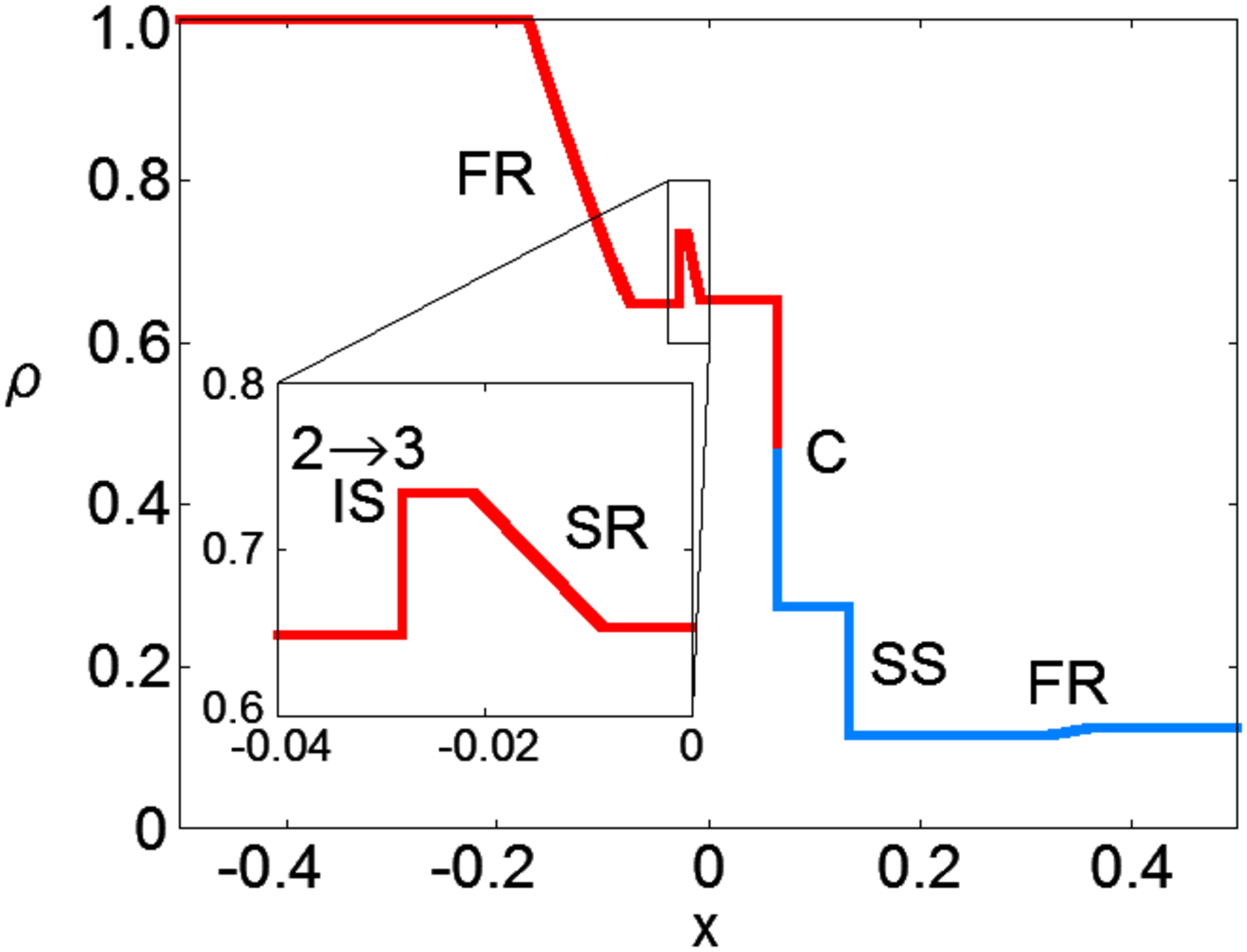}
\end{center}
\end{minipage} &
\begin{minipage}{0.45\hsize}
\begin{center}
\includegraphics[scale=0.26]{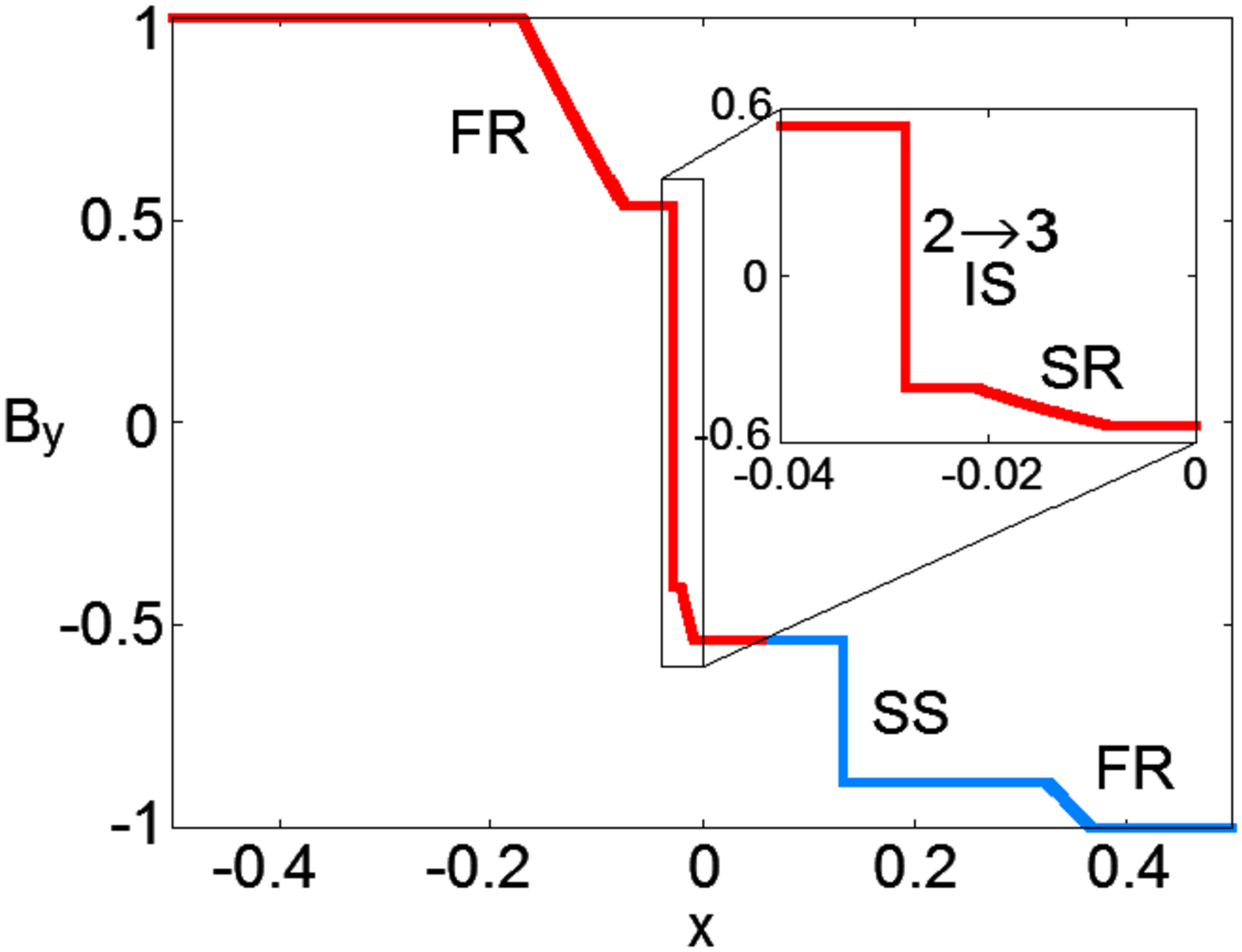}
\end{center}
\end{minipage} \\
\begin{minipage}{0.45\hsize}
\begin{center}
\includegraphics[scale=0.26]{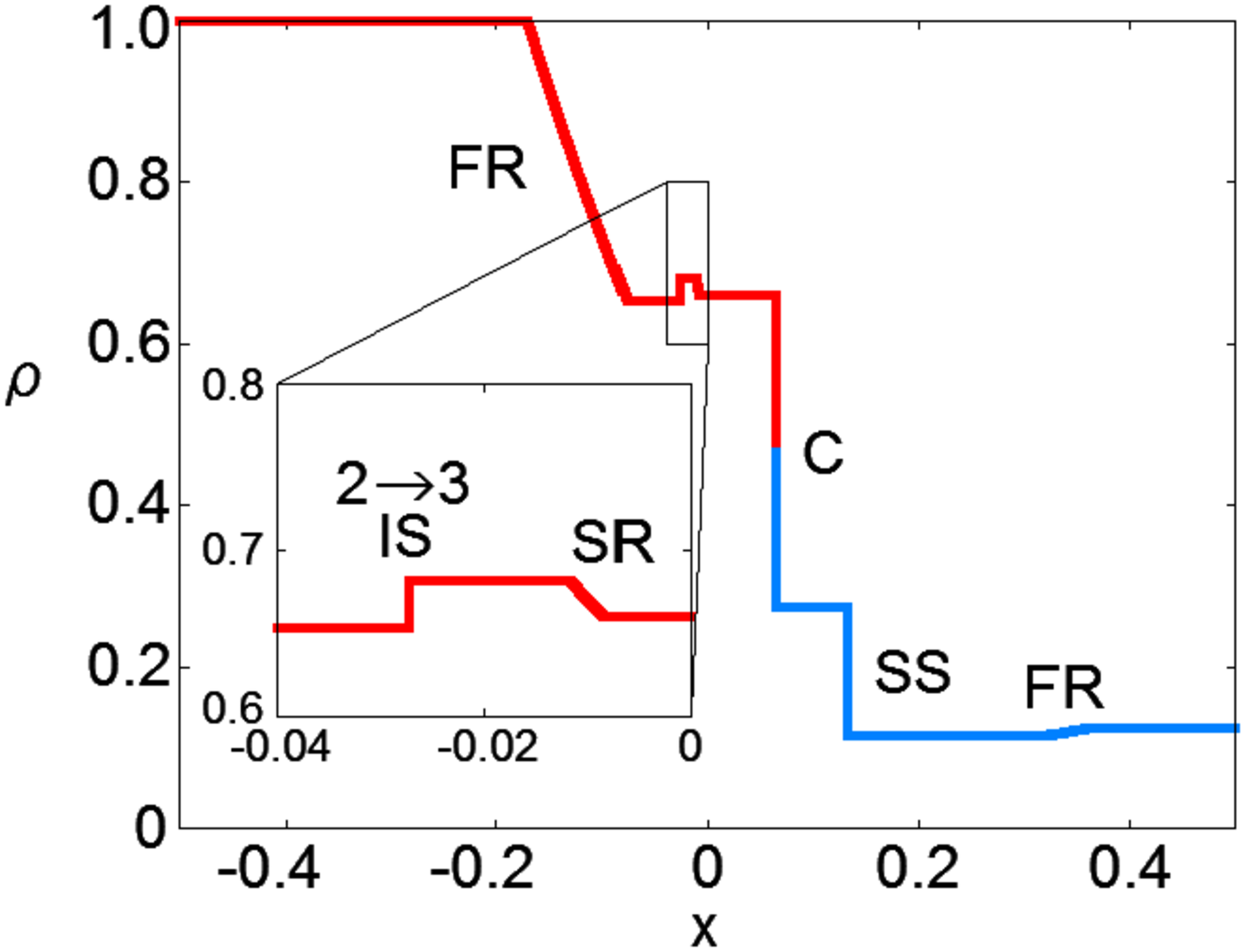}
\end{center}
\end{minipage} &
\begin{minipage}{0.45\hsize}
\begin{center}
\includegraphics[scale=0.26]{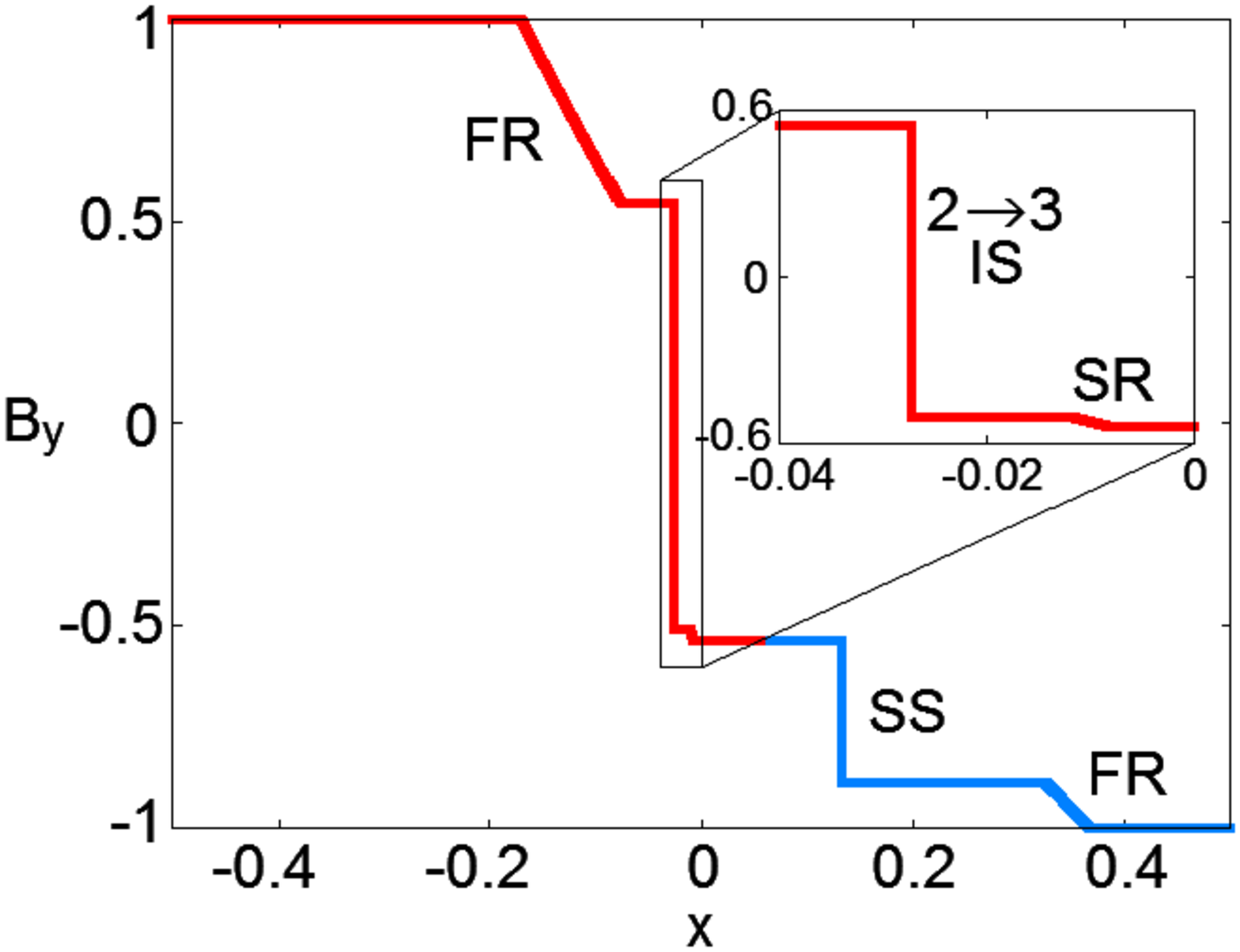}
\end{center}
\end{minipage} \\
\begin{minipage}{0.45\hsize}
\begin{center}
\includegraphics[scale=0.26]{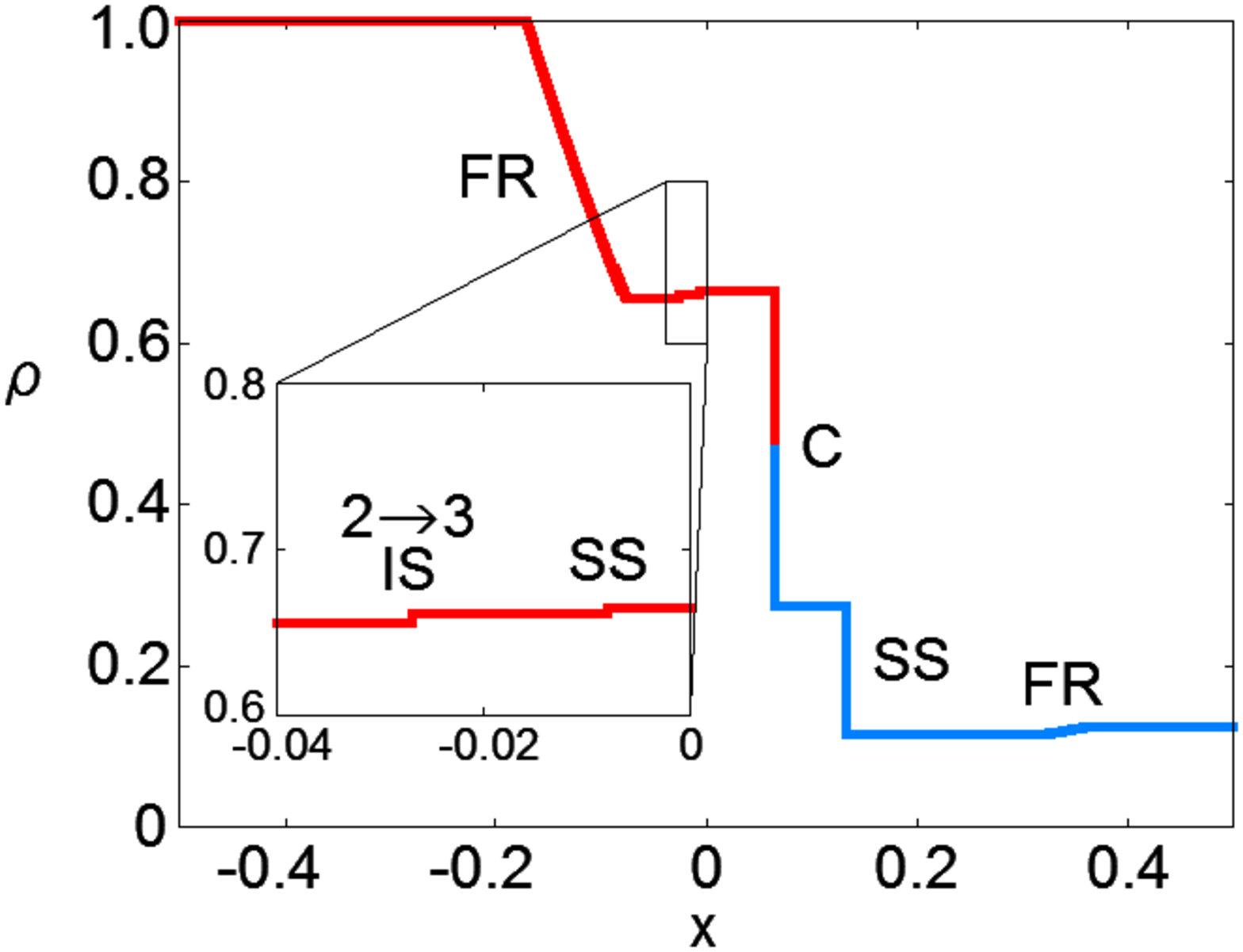}
\end{center}
\end{minipage} &
\begin{minipage}{0.45\hsize}
\begin{center}
\includegraphics[scale=0.26]{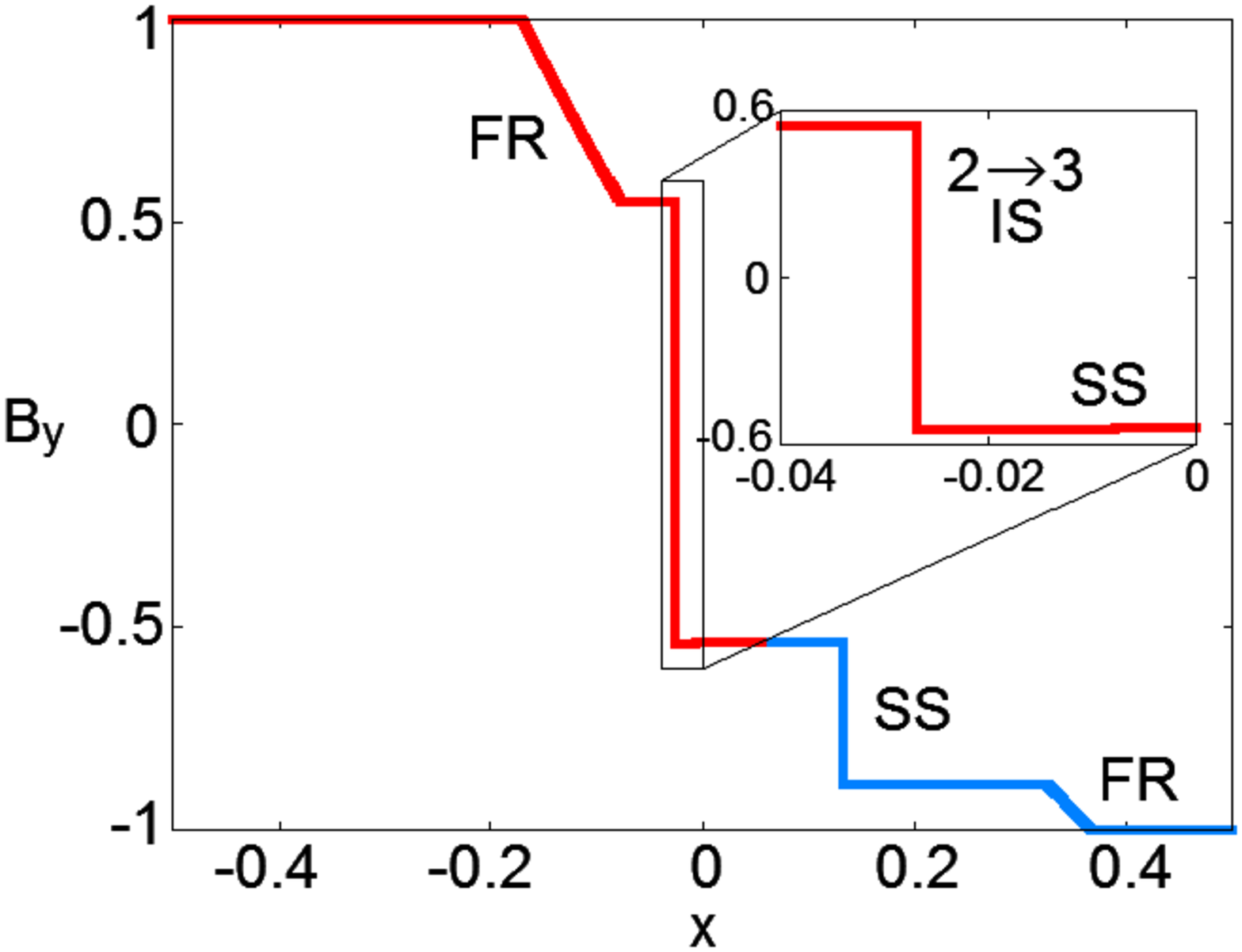}
\end{center}
\end{minipage} 
\end{tabular}
\caption{Some of the uncountably many non-regular solutions for the Brio \& Wu problem that contain a 
$2 \rightarrow 3$ intermediate shock. The notation is the same as in Fig.~\ref{BW compound}. The insets
are the close-ups of indicated regions.}
\label{BW others}
\end{figure}

Another advantage of our code is that it attempts to find all solutions, trying all combinations of constituent waves including the 
intermediate shocks. In its application to the current problem, we found that other non-regular solutions exist, which involve various  
$2 \rightarrow 3$ intermediate shocks. Some of such solutions are shown at $t=0.1$ in Fig.~\ref{BW others}. The solution given in the top
panel resembles the non-regular solution with the compound wave presented above. A closer look (see the inset), however, reveals 
a difference: a $2 \rightarrow 3$ intermediate shock is closely followed by a slow rarefaction wave but they are detached. The separation 
of the two waves are more apparent in the middle panel, in which the strengths of the intermediate shock and rarefaction wave are also 
weaker than in the top panel. In the bottom panel, the solution does not include a slow rarefaction wave but a slow shock instead. The 
$2 \rightarrow 3$ intermediate shock is even weaker in this case. It is interesting that the strengths of other waves than the intermediate 
shock do not change very much. It should be emphasized here that these are solutions to the same initial condition.

As a matter of fact, these solutions form a one-parameter family, with the Brio \& Wu solution with the compound wave at one end and 
the regular solution with a rotational discontinuity at the other end point of the sequence. 
The parameter that characterizes the solution family is a magnitude of jump in the transverse magnetic field across the 
$2 \rightarrow 3$ intermediate shock. Since the sequence of solutions is obtained by gradually changing this parameter, we can 
deduce that there are uncountably many solutions in fact. 
As already mentioned, of these non-regular solutions and the unique regular solution it is the Brio \& Wu non-regular solution that 
numerical simulations almost always obtain. From the point of view of the evolutionary conditions, the regular solution is the only 
solution that contains evolutionary shocks alone. Both $2 \rightarrow 3,4$ and $2 \rightarrow 3$ shocks are non-evolutionary. Hence 
numerical simulations seem to prefer the non-evolutionary solution to the evolutionary solution. This is the original controversy.
Now we have added another question: why is the Brio \& Wu solution is singled out from the uncountably many non-regular solutions with 
non-evolutionary shocks? 


 
\begin{figure}
\begin{tabular}{cc}
\begin{minipage}{0.45\hsize}
\begin{center}
\includegraphics[scale=0.26]{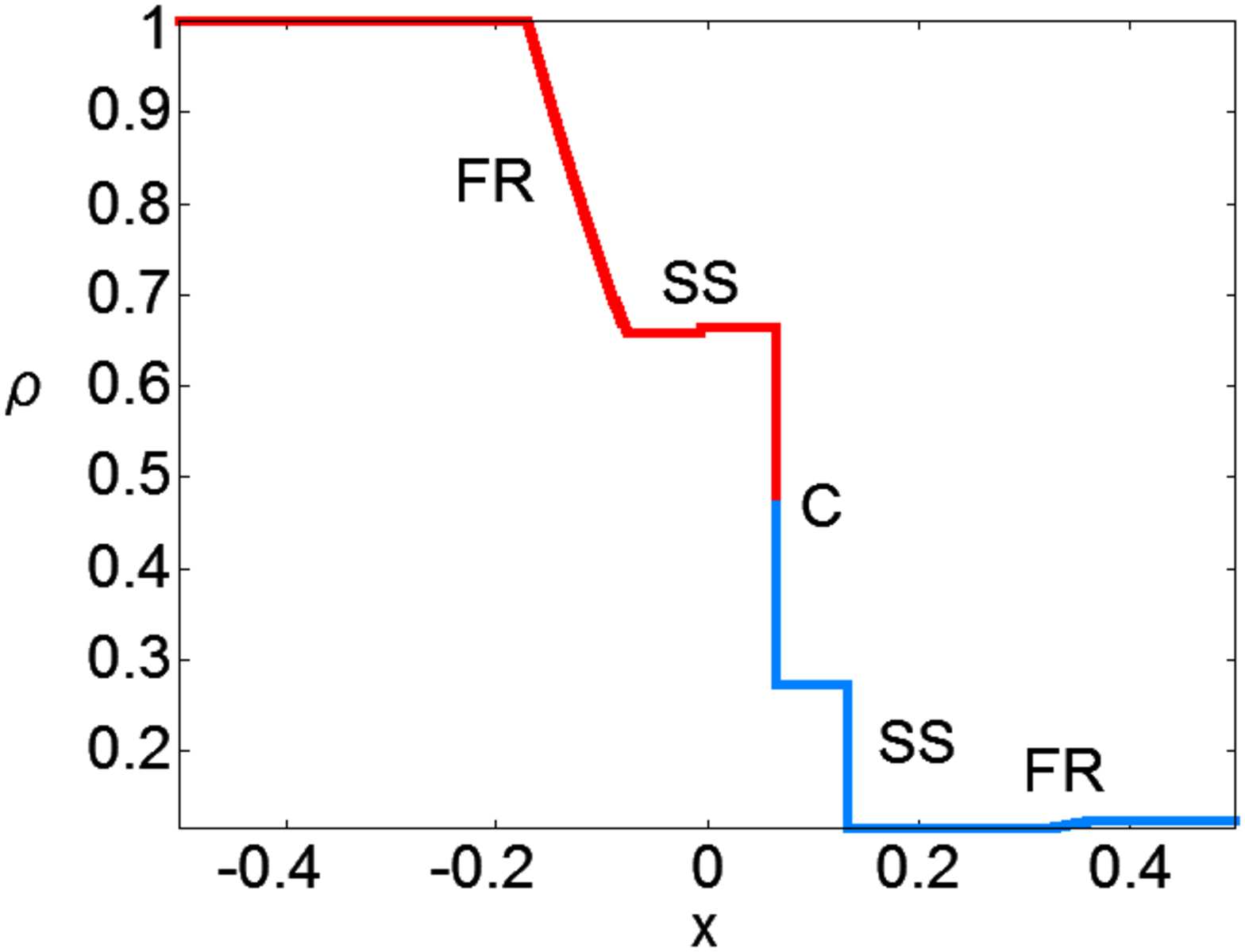}
\end{center}
\end{minipage} &
\begin{minipage}{0.45\hsize}
\begin{center}
\includegraphics[scale=0.26]{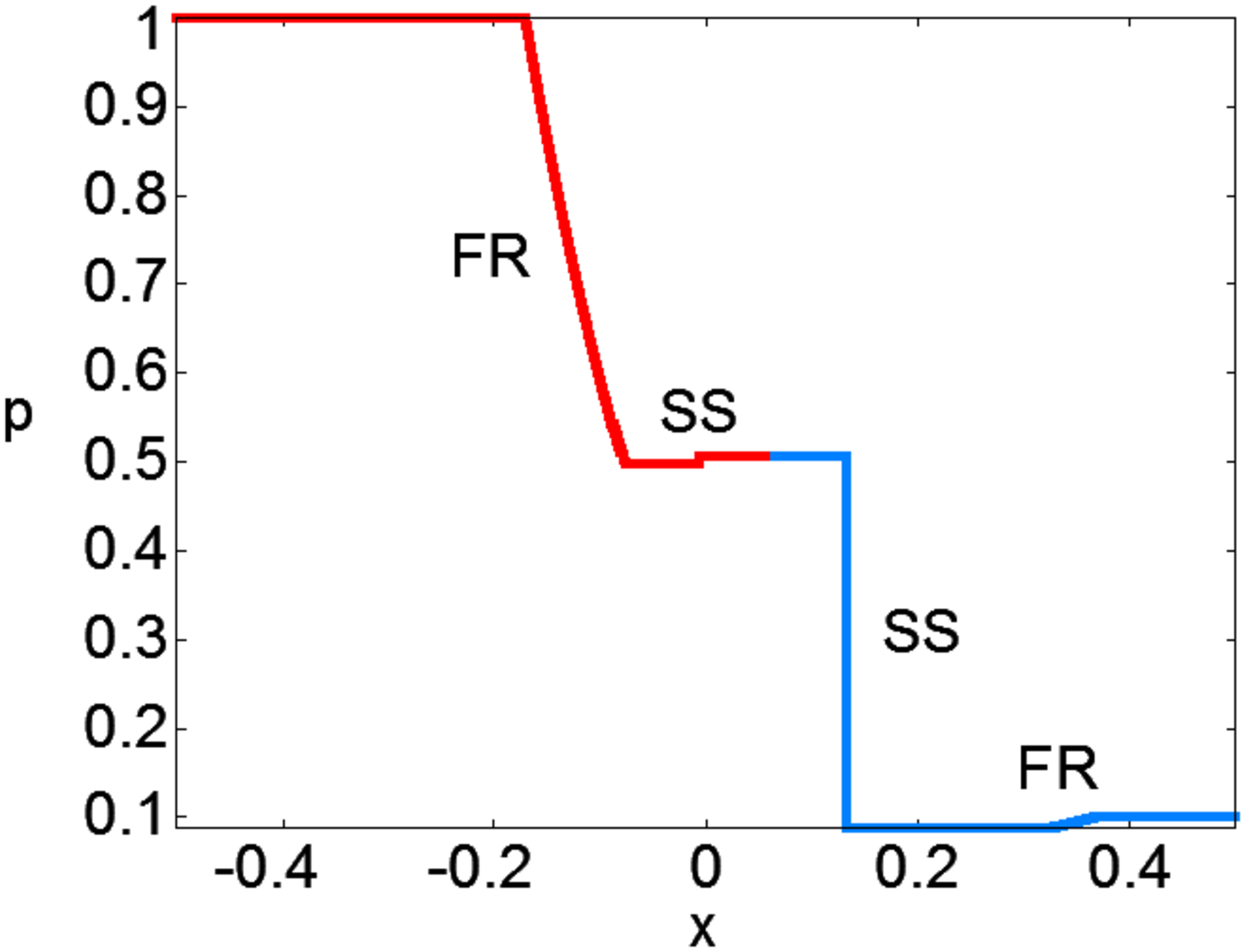}
\end{center}
\end{minipage} \\
\begin{minipage}{0.45\hsize}
\begin{center}
\includegraphics[scale=0.26]{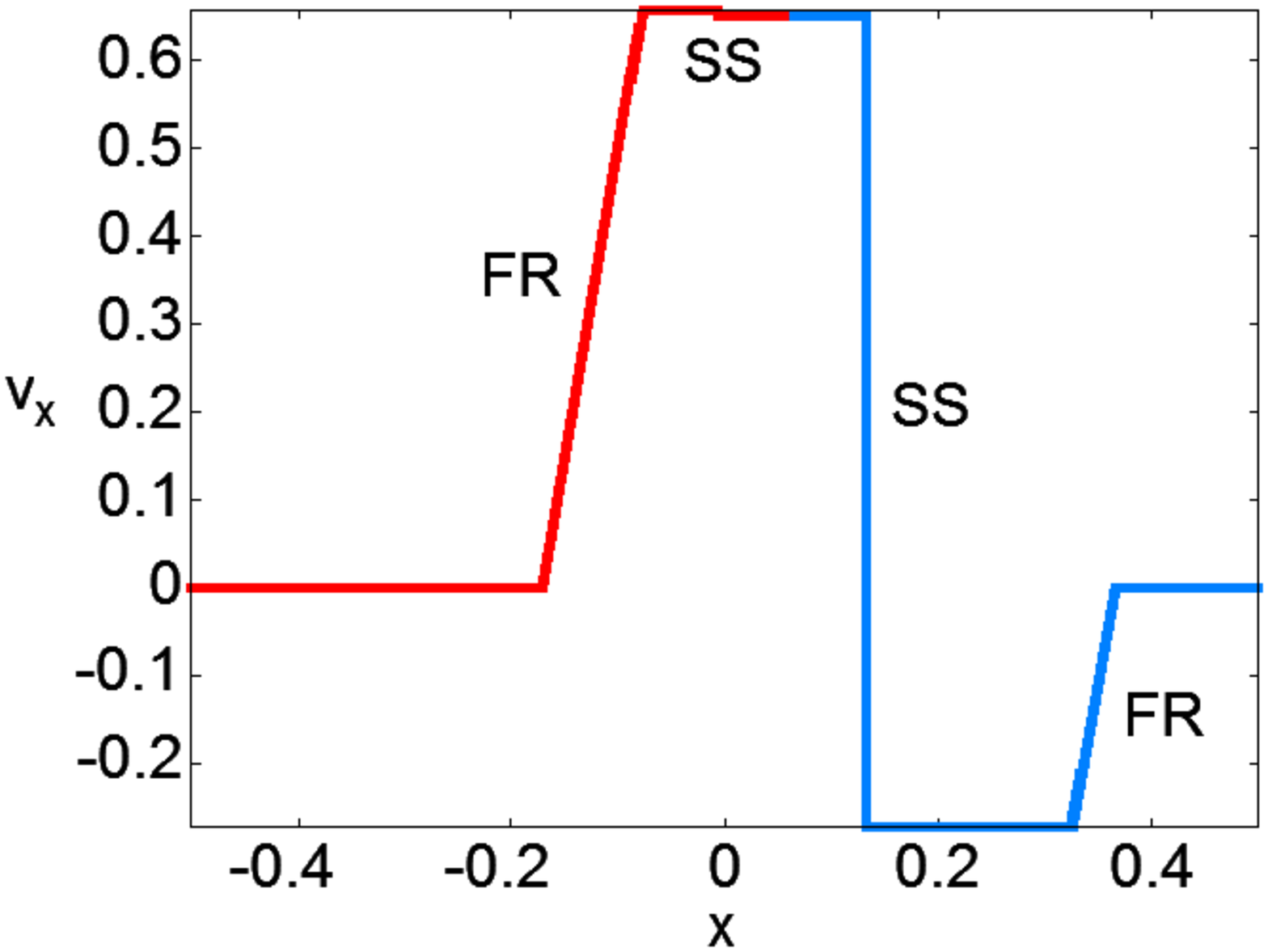}
\end{center}
\end{minipage} &
\begin{minipage}{0.45\hsize}
\begin{center}
\includegraphics[scale=0.26]{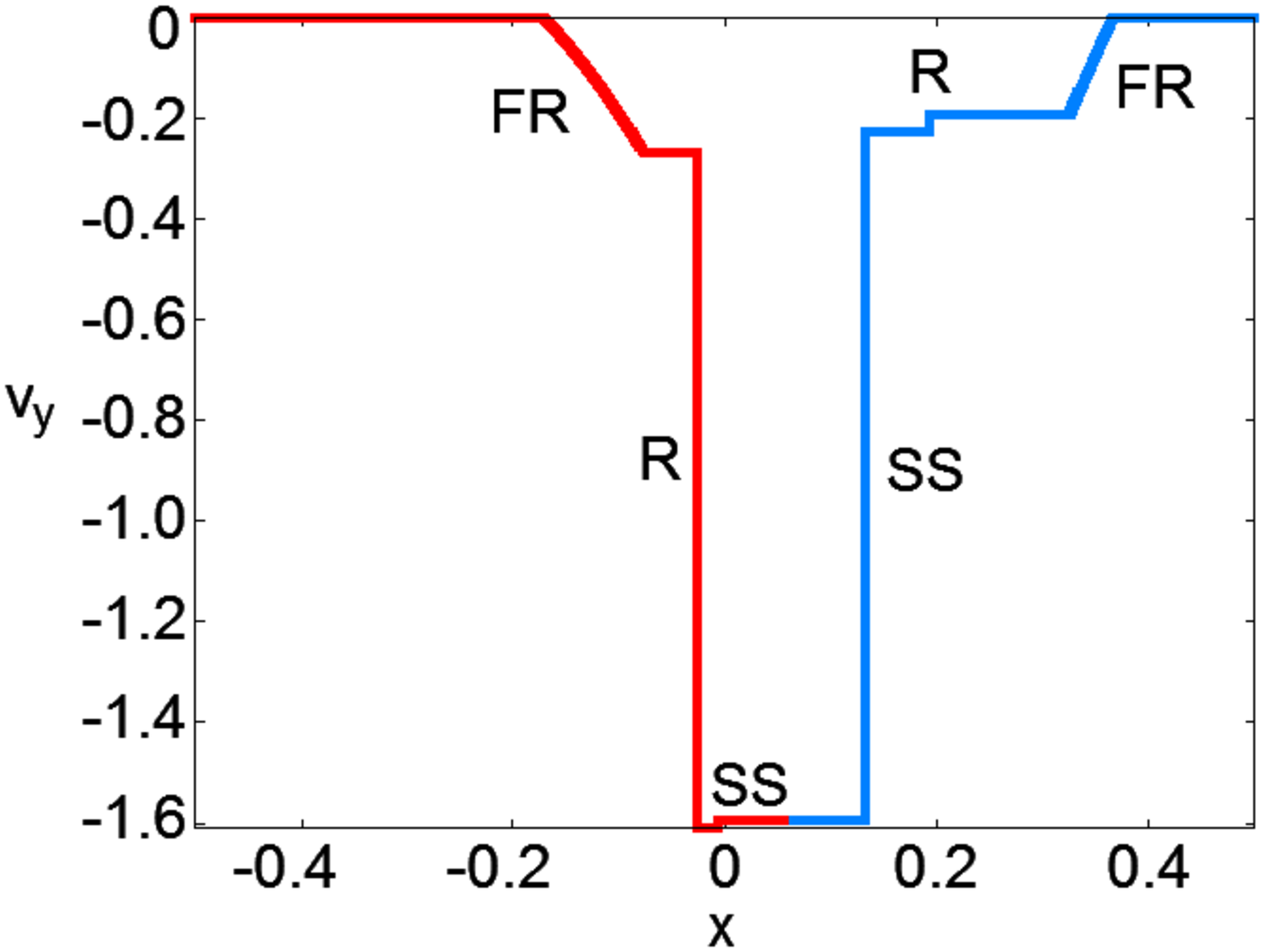}
\end{center}
\end{minipage} \\
\begin{minipage}{0.45\hsize}
\begin{center}
\includegraphics[scale=0.26]{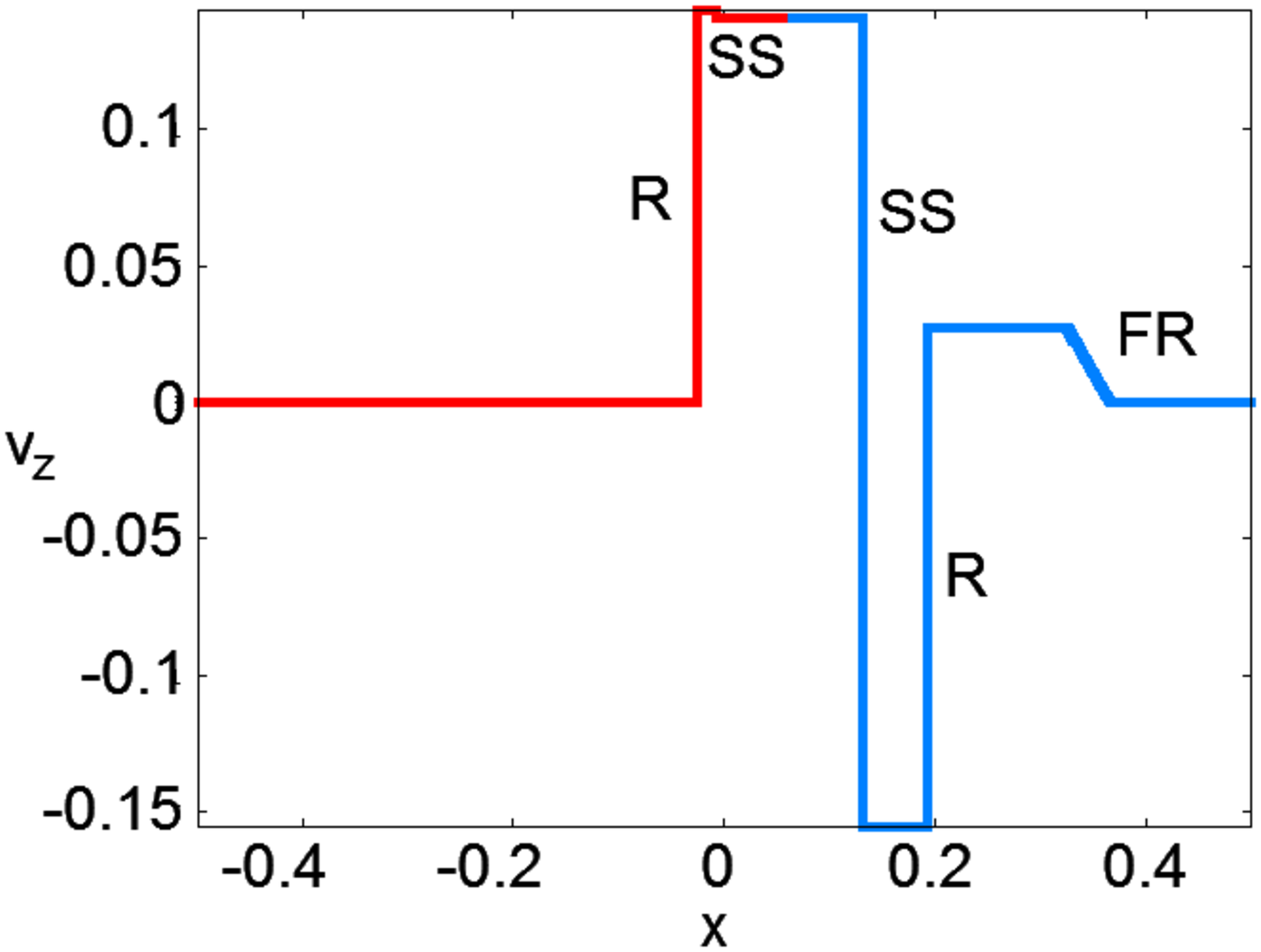}
\end{center}
\end{minipage} &
\begin{minipage}{0.45\hsize}
\begin{center}
\includegraphics[scale=0.26]{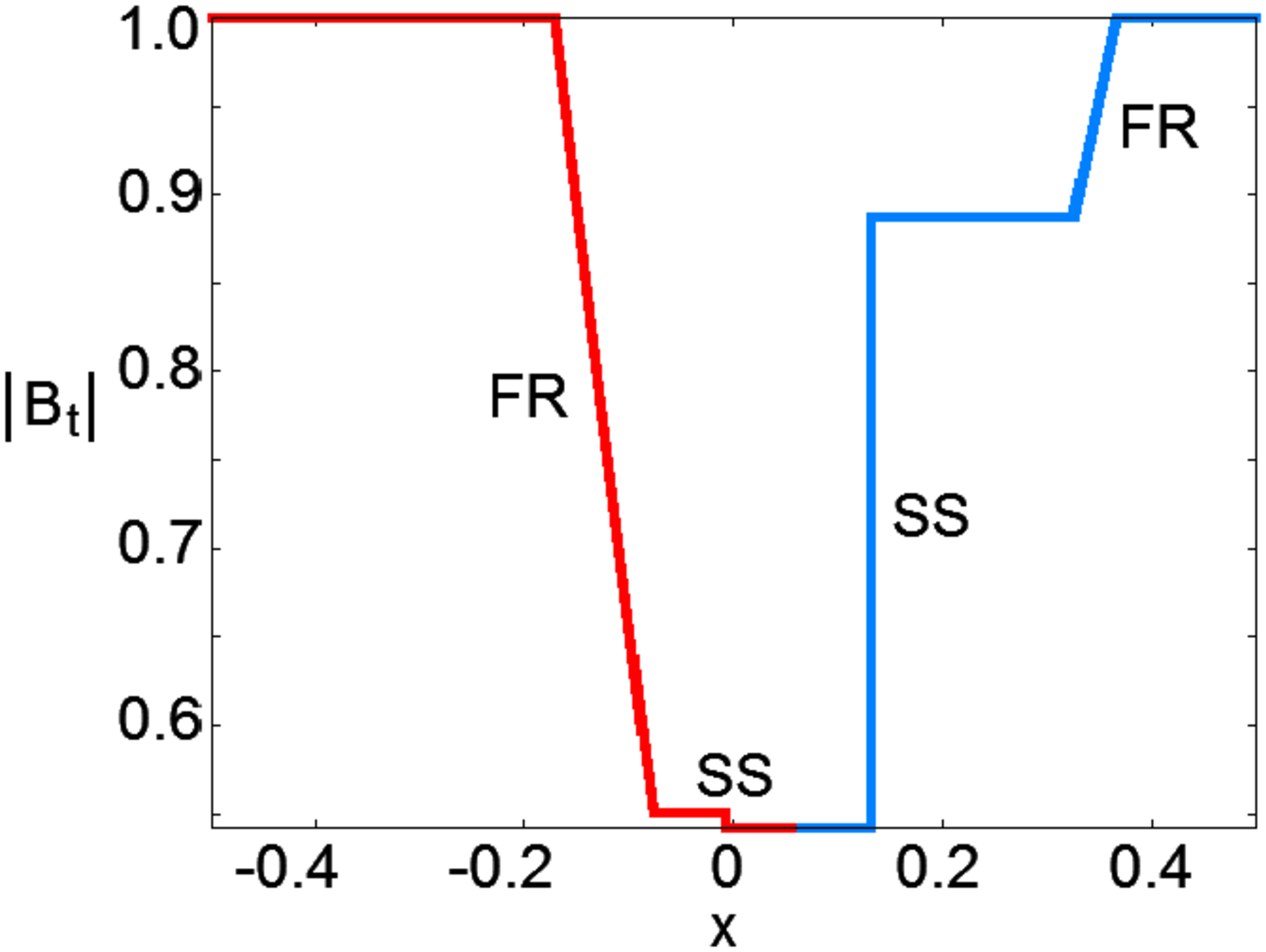}
\end{center}
\end{minipage} \\
\begin{minipage}{0.45\hsize}
\begin{center}
\includegraphics[scale=0.26]{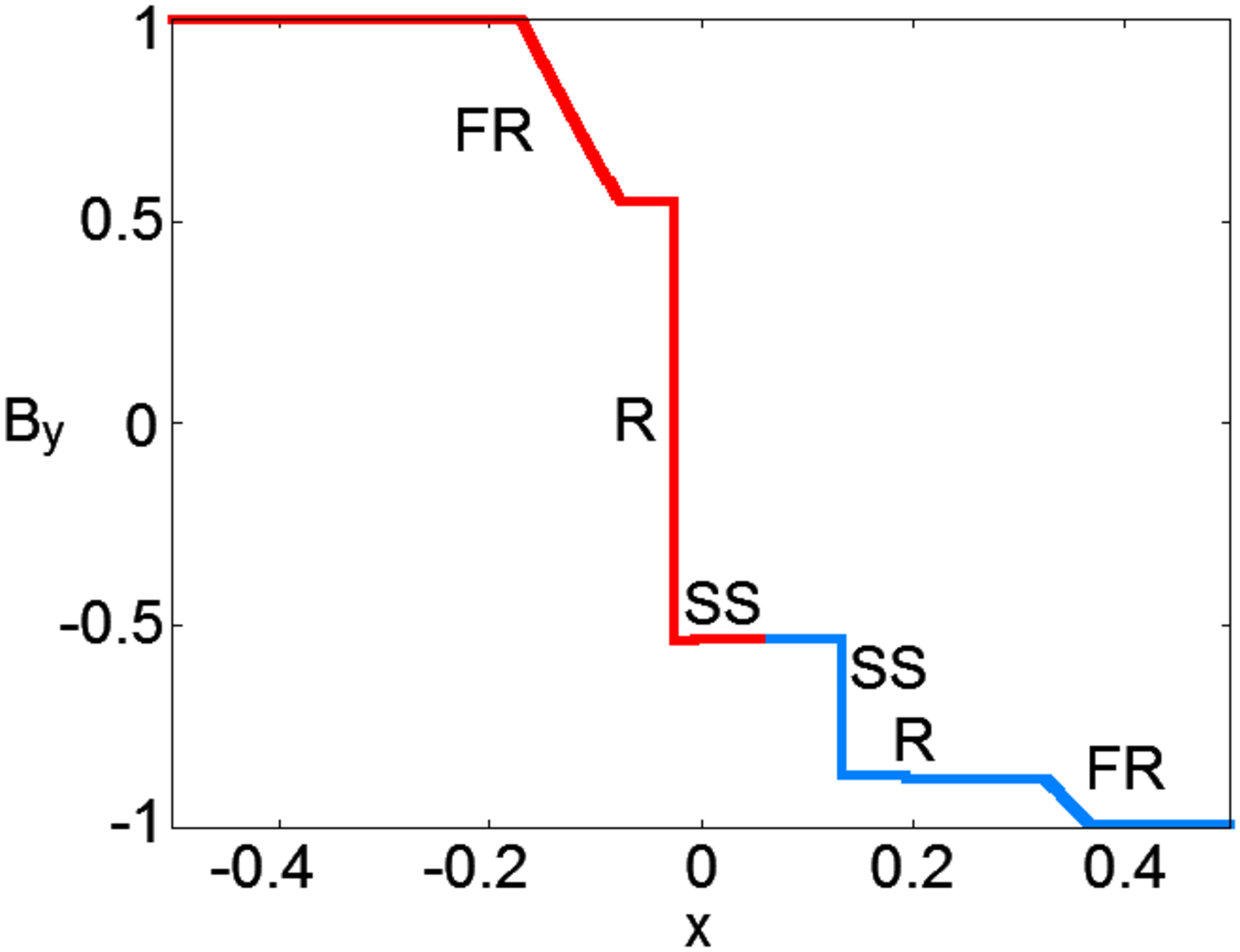}
\end{center}
\end{minipage} &
\begin{minipage}{0.45\hsize}
\begin{center}
\includegraphics[scale=0.26]{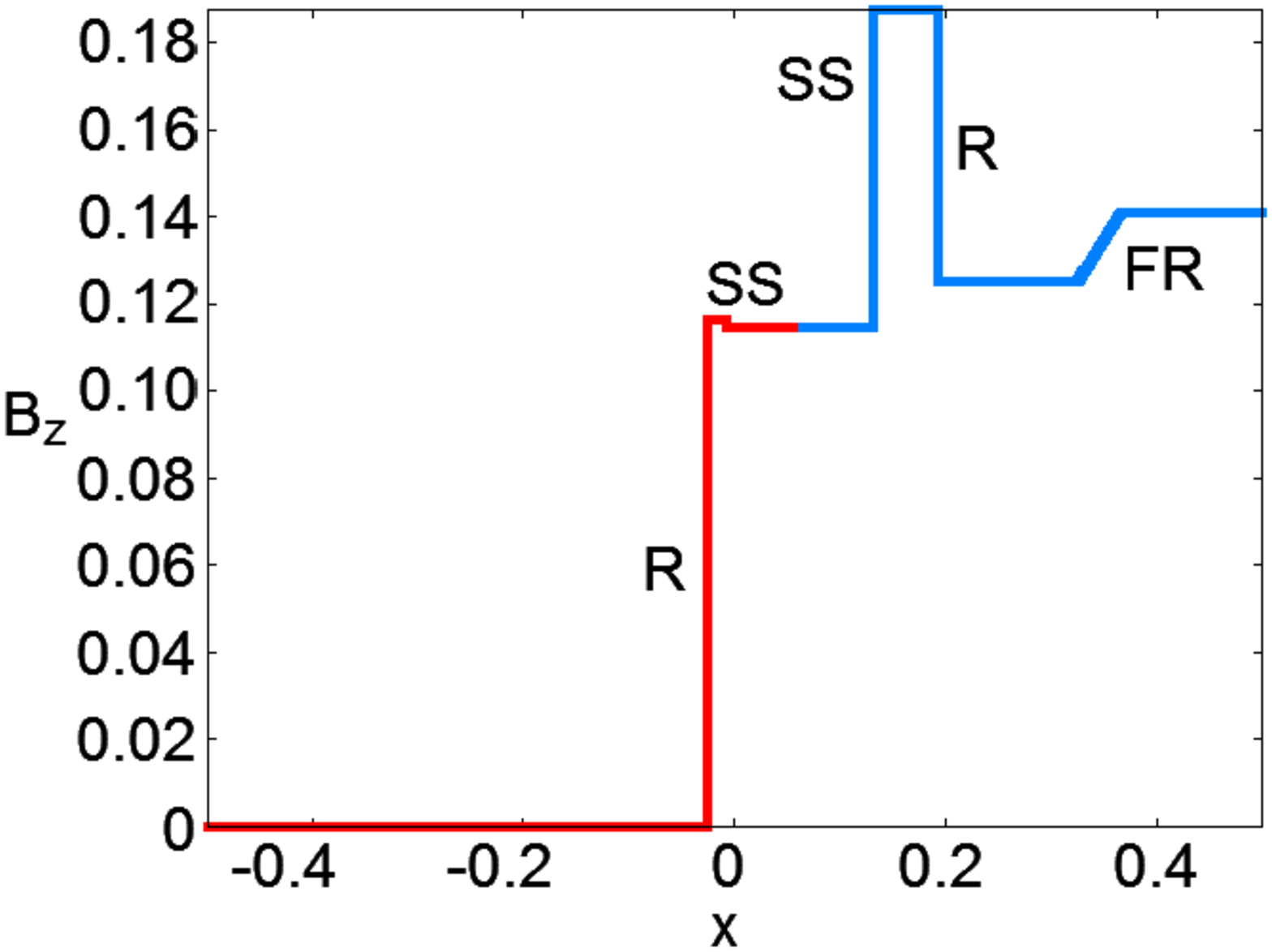}
\end{center}
\end{minipage} 
\end{tabular}
\caption{A non-coplanar neighboring solutions to the Brio \& Wu problem. The initial magnitude of the transverse 
magnetic field  of the right state is the same as the original value. The notation is the same as in Fig.~\ref{BW regular}.
}
\label{BW_neighbors_non}
\end{figure}

Since the evolutionary conditions are related with the neighboring solutions, we consider the initial conditions that are close to the original 
one. Both non-coplanar and coplanar perturbations are discussed in turn separately. As a non-coplanar initial configuration we 
take the following: 
\begin{eqnarray}
(\rho_L,\ p_L,\ v_{xL},\ v_{yL},\ v_{zL},\ B_{yL},\ B_{zL}) =& (1,\ 1,\ 0,\ 0,\ 0,\ 1,\ 0), \\
(\rho_R,\ p_R,\ v_{xR},\ v_{yR},\ v_{zR},\ B_{yR},\ B_{zR}) =& (0.125,\ 0.1,\ 0,\ 0,\ 0,\ \cos 3.0,\ \sin 3.0).
\end{eqnarray}
The transverse magnetic field in the right state is rotated slightly with other quantities being unchanged. In accord to the 
evolutionary condition, we found only a regular solution, which is shown in Fig.~\ref{BW_neighbors_non}. The solution includes 
left- and right-going rotational discontinuities, which are responsible for the rotation of magnetic fields. As is evident from 
the comparison (see Fig.~\ref{BW regular}), this solution is indeed close to the regular solution to the original problem and 
can be regarded as the perturbed solution. 
On the other hand, no neighboring non-regular solution exists in this case. 
\citet{Wu88a, Wu88b} observed in his dissipative MHD simulations, however, 
a new type of non-coplanar shock-like structures 
that do not satisfy the Rankine-Hugoniot relations. They concluded that this time-dependent "intermediate shocks" 
are the neighboring states to the ordinary intermediate shocks. It is noted, however, that the time-dependent non-coplanar structures of 
these new "intermediate shocks" depend upon the dissipation coefficients as shown in \citet{Wu88a, Wu88b} and his succeeding one 
\citep{Wu90} and may not be realized in ideal MHD.

It is of great interest to investigate coplanar perturbations, since the coplanarity is maintained even in numerical
simulations unless it is broken explicitly. If the symmetry is retained indeed, we are essentially dealing with the reduced MHD
system, in which the \Alfven \ characteristics do not exist. It should be noted then that the evolutionary conditions will
be modified as pointed out by \citet{FK01}. For example, the $1 \rightarrow 3$ and $2 \rightarrow 4$ shocks become
evolutionary and the rotational discontinuities become non-evolutionary. On the other hand, the $2 \rightarrow 3$ shock remains non-evolutionary with no converging characteristic. 
It is noteworthy that the $2 \rightarrow 3,4$ shock is evolutionary, since the slow characteristics are marginally converging and the 
linear independence of the eigenfunctions of outgoing waves and the initial jumps is recovered. As a consequence of these changes 
the Brio \& Wu solution should be regarded as a regular solution in the reduced system whereas the other solutions with 
$2 \rightarrow 3$ shocks are still non-regular even in this system. \citet{FK97,FK01} claimed that this is the reason why 
numerical simulations almost always obtain the Brio \& Wu solution. The problem may be a bit more complicated, however, as will 
be demonstrated shortly. 

\begin{figure}
\begin{tabular}{cc}
\begin{minipage}{0.45\hsize}
\begin{center}
\includegraphics[scale=0.26]{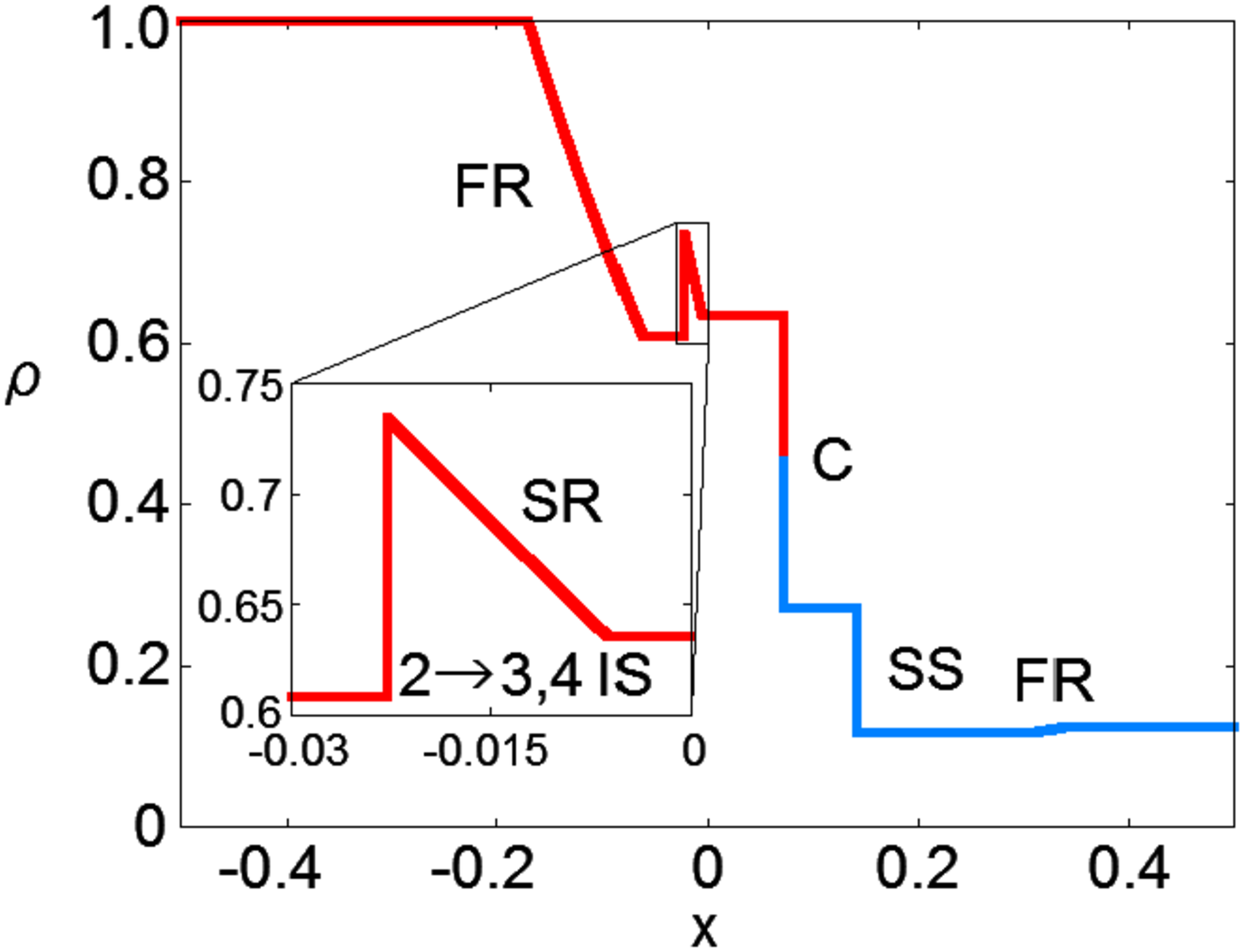}
\end{center}
\end{minipage} &
\begin{minipage}{0.45\hsize}
\begin{center}
\includegraphics[scale=0.26]{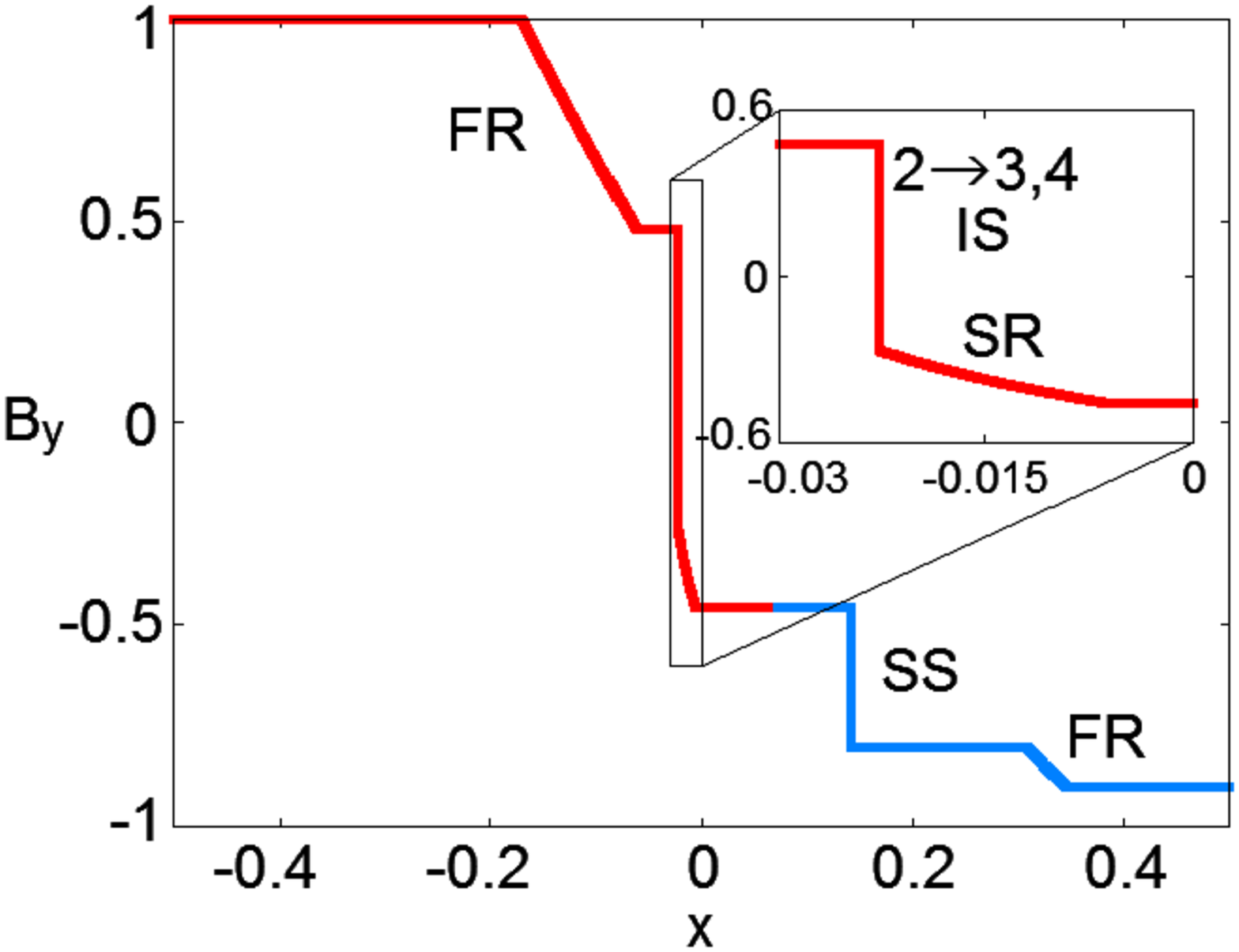}
\end{center}
\end{minipage} \\
\begin{minipage}{0.45\hsize}
\begin{center}
\includegraphics[scale=0.26]{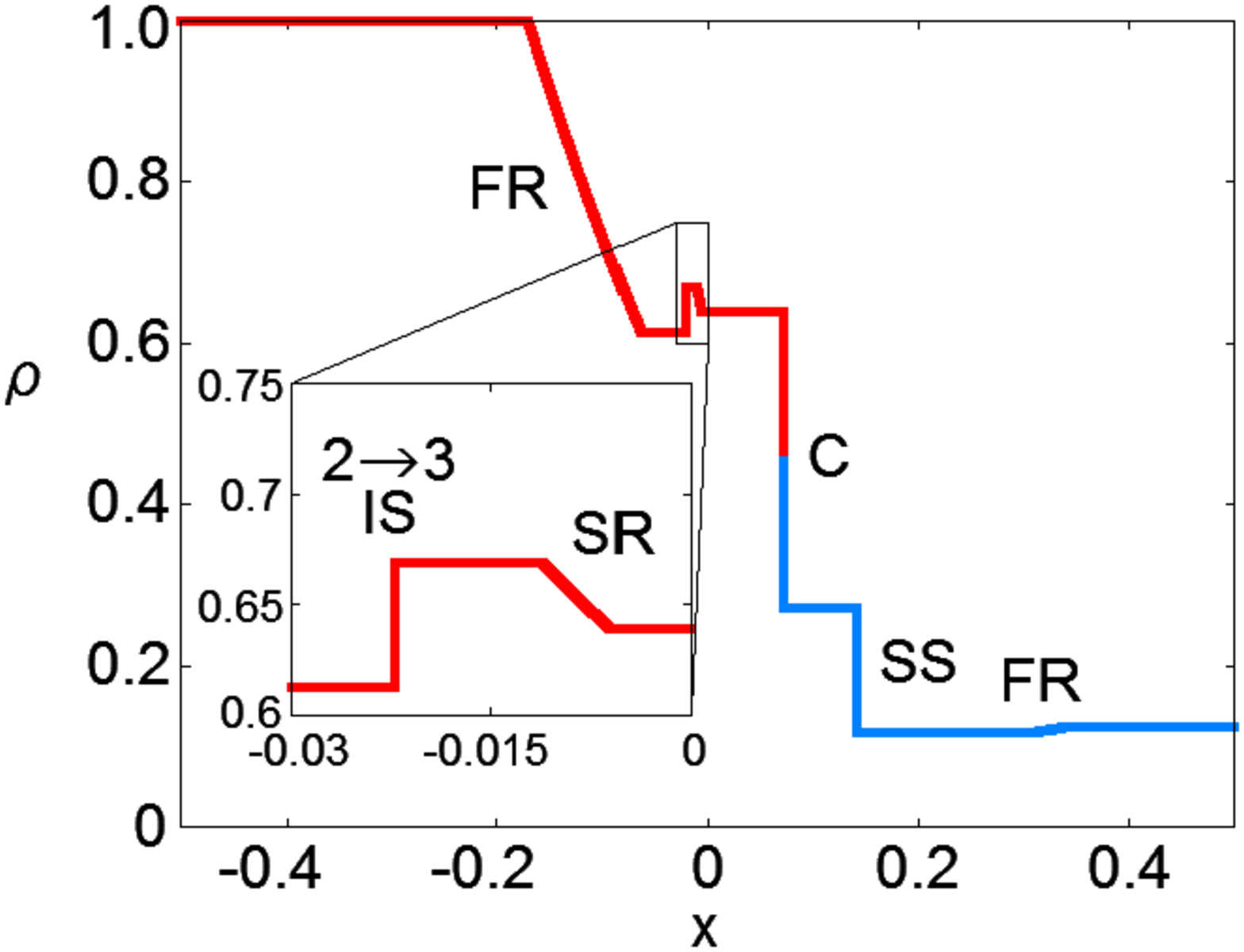}
\end{center}
\end{minipage} &
\begin{minipage}{0.45\hsize}
\begin{center}
\includegraphics[scale=0.26]{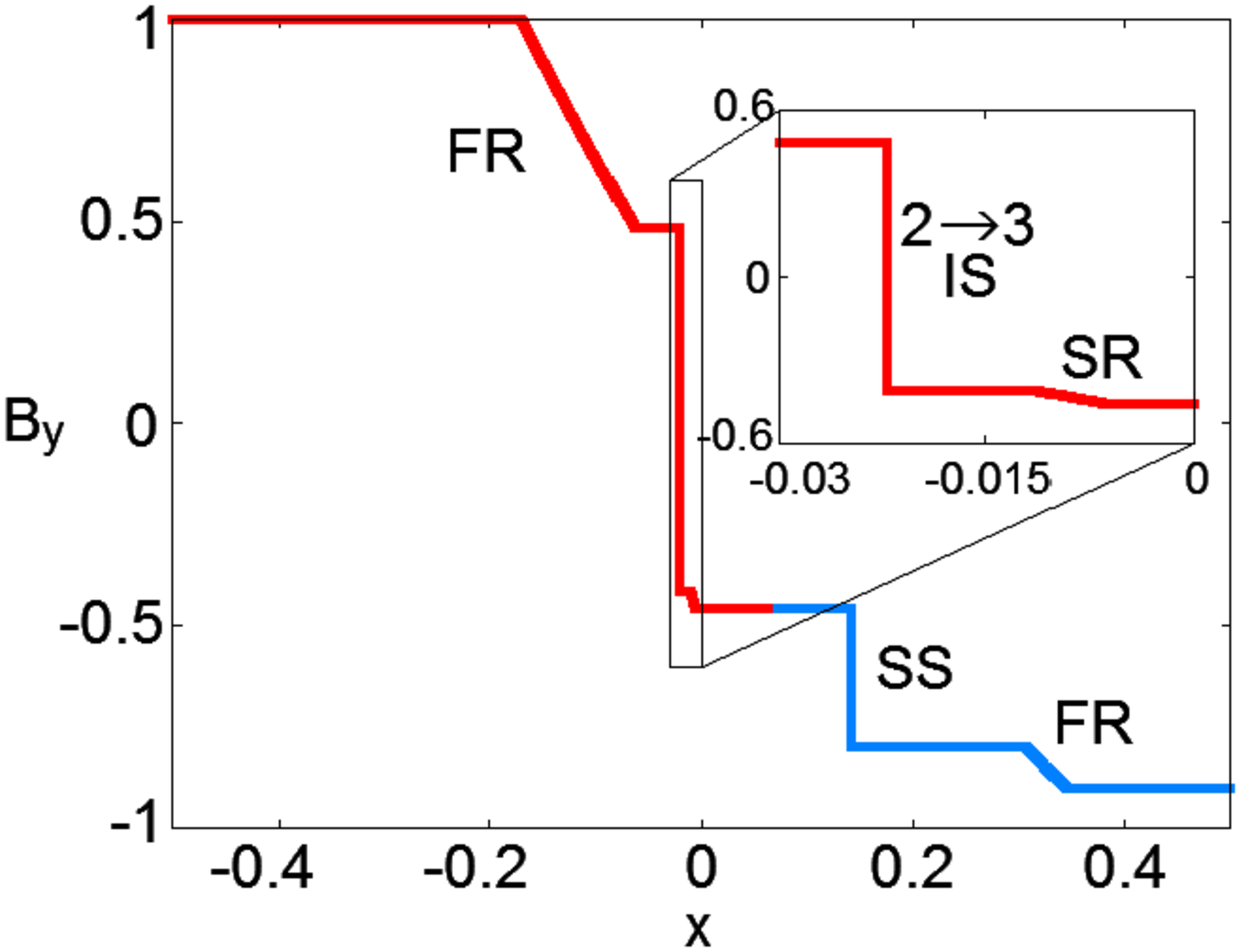}
\end{center}
\end{minipage} \\
\begin{minipage}{0.45\hsize}
\begin{center}
\includegraphics[scale=0.26]{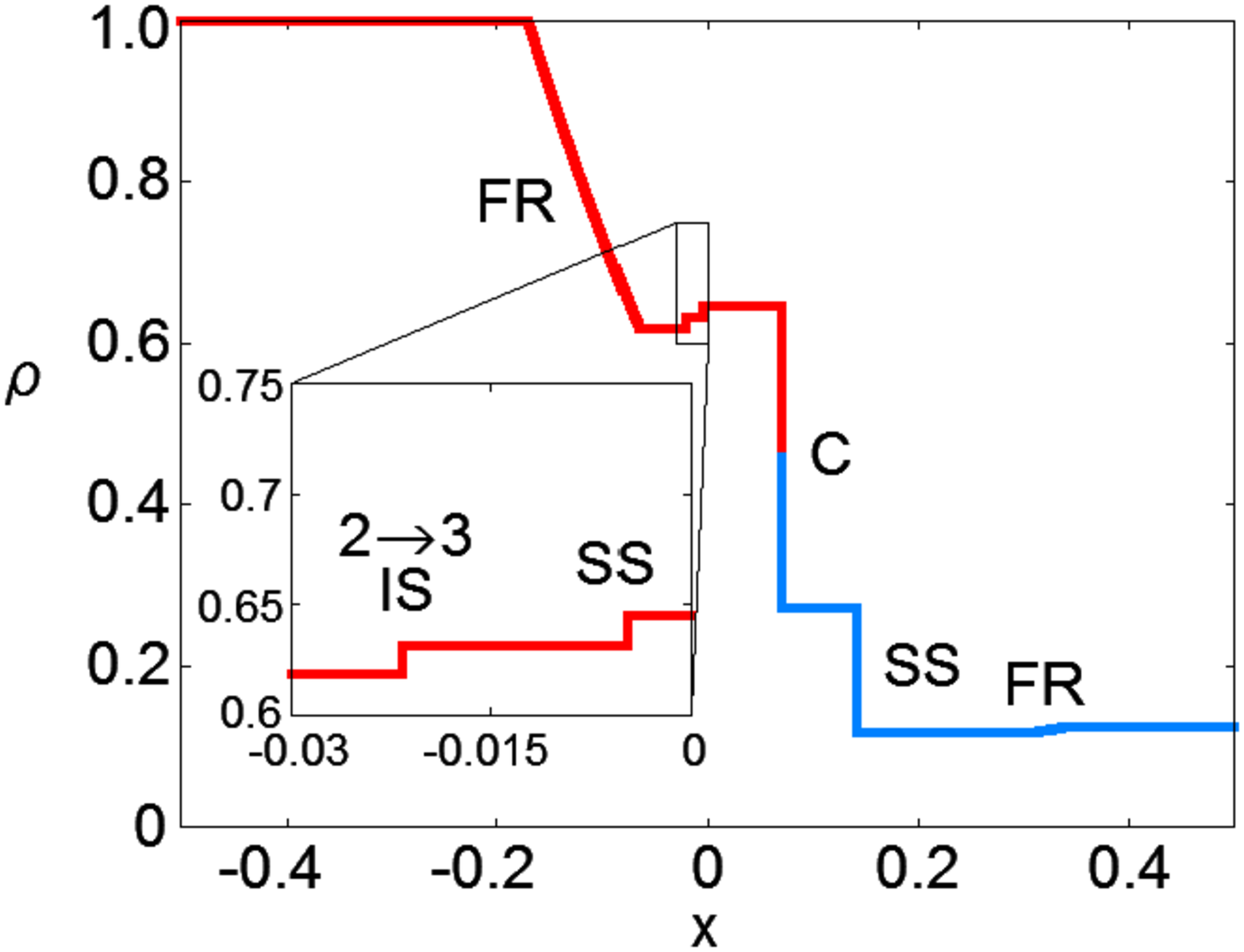}
\end{center}
\end{minipage} &
\begin{minipage}{0.45\hsize}
\begin{center}
\includegraphics[scale=0.26]{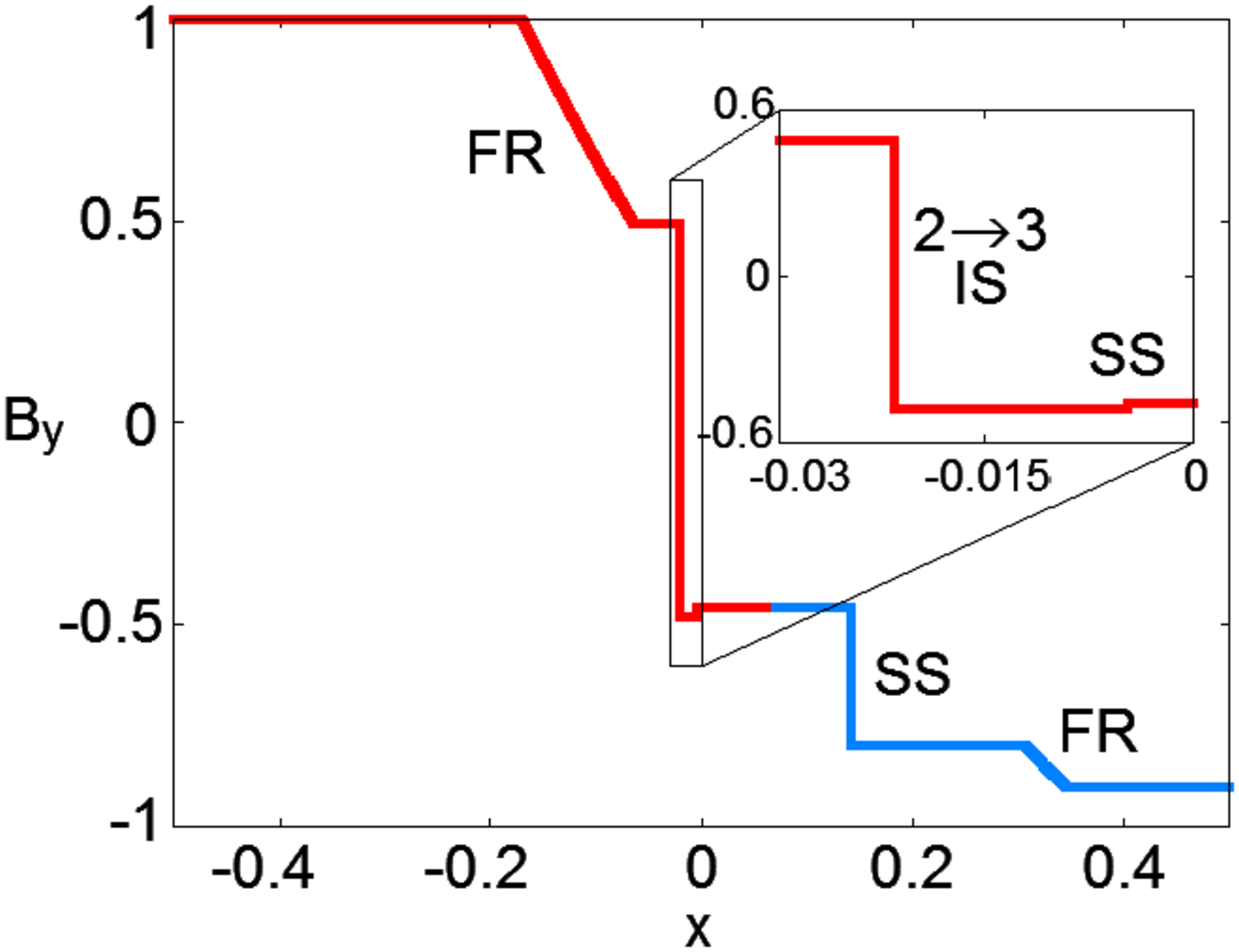}
\end{center}
\end{minipage} \\
\begin{minipage}{0.45\hsize}
\begin{center}
\includegraphics[scale=0.26]{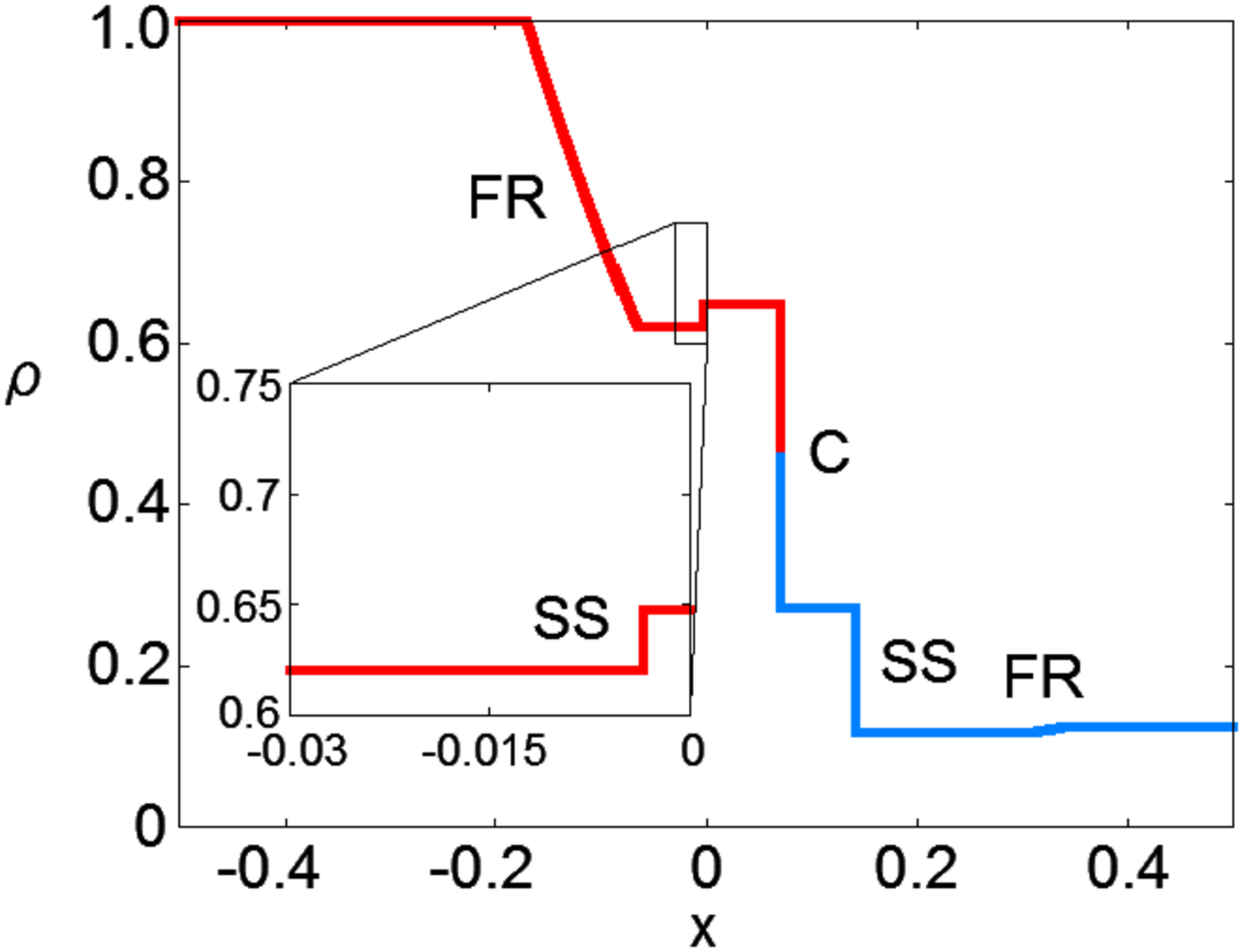}
\end{center}
\end{minipage} &
\begin{minipage}{0.45\hsize}
\begin{center}
\includegraphics[scale=0.26]{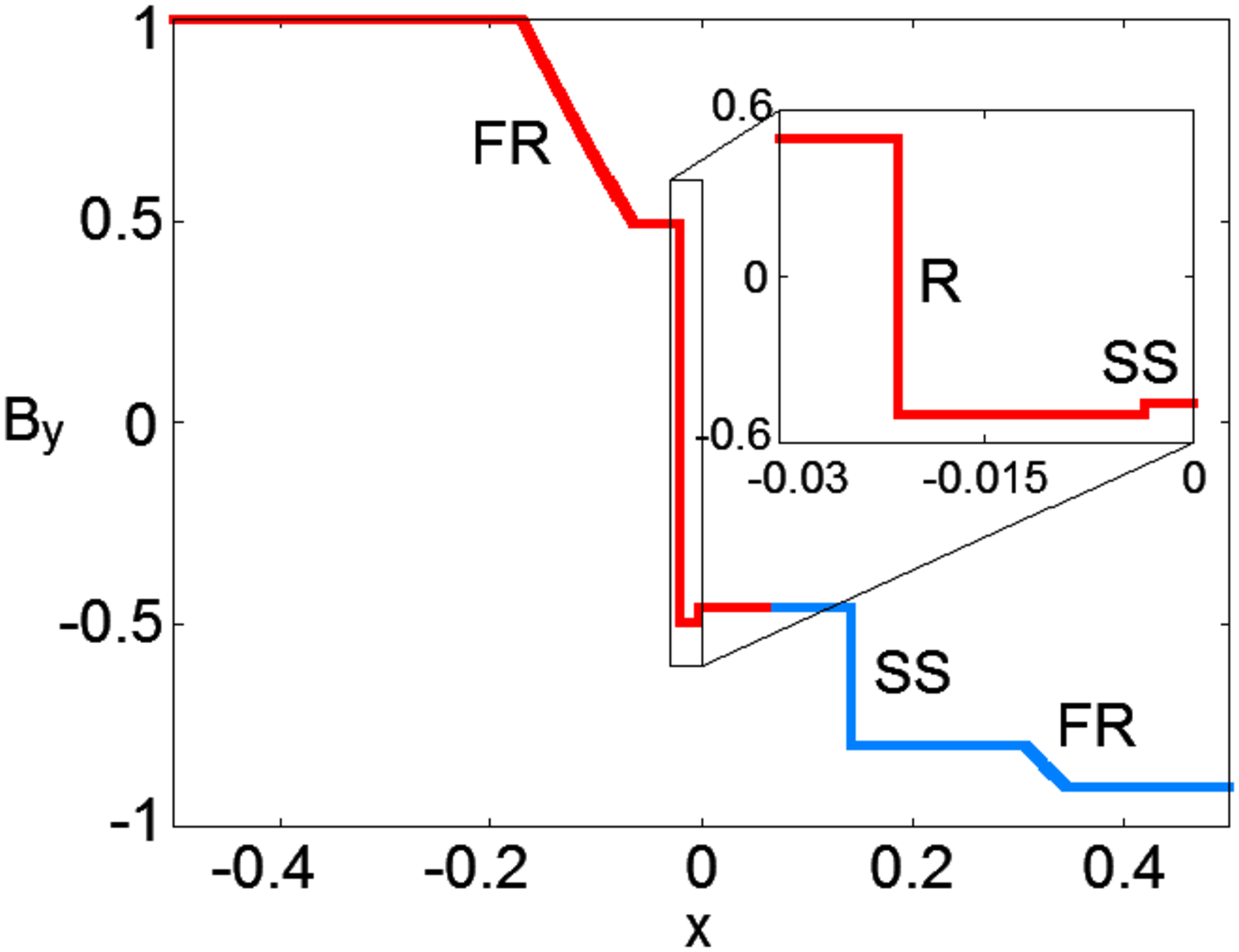}
\end{center}
\end{minipage} 
\end{tabular}
\caption{Some coplanar neighboring solutions to the Brio \& Wu problem. The initial magnitude of the transverse 
magnetic field  of the right state is reduced from the original value. The notation is the same as in Fig.~\ref{BW compound}. 
The insets are the close-ups of indicated regions.
}
\label{BW_neighbors_small}
\end{figure}

\begin{figure}
\begin{tabular}{cc}
\begin{minipage}{0.45\hsize}
\begin{center}
\includegraphics[scale=0.26]{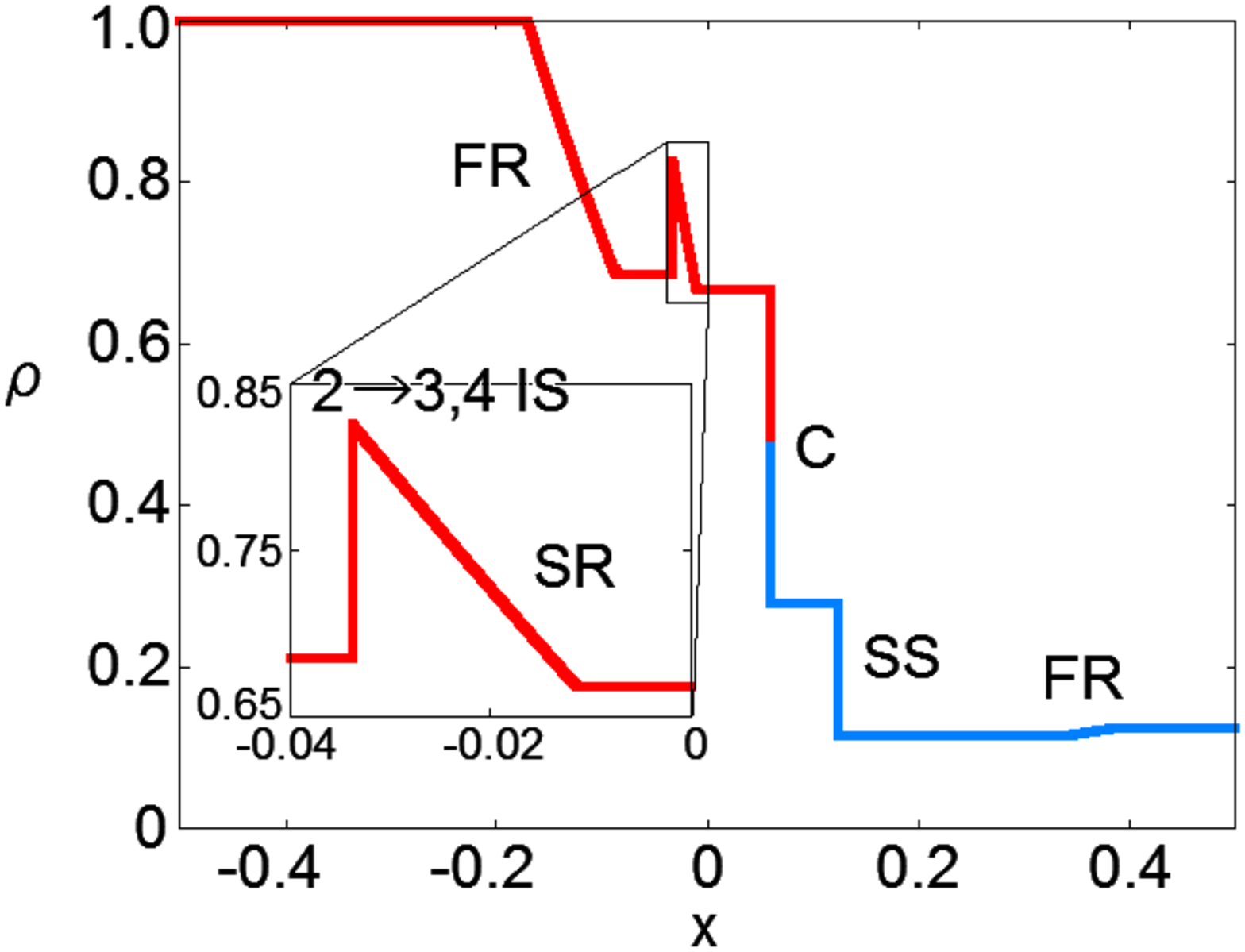}
\end{center}
\end{minipage} &
\begin{minipage}{0.45\hsize}
\begin{center}
\includegraphics[scale=0.26]{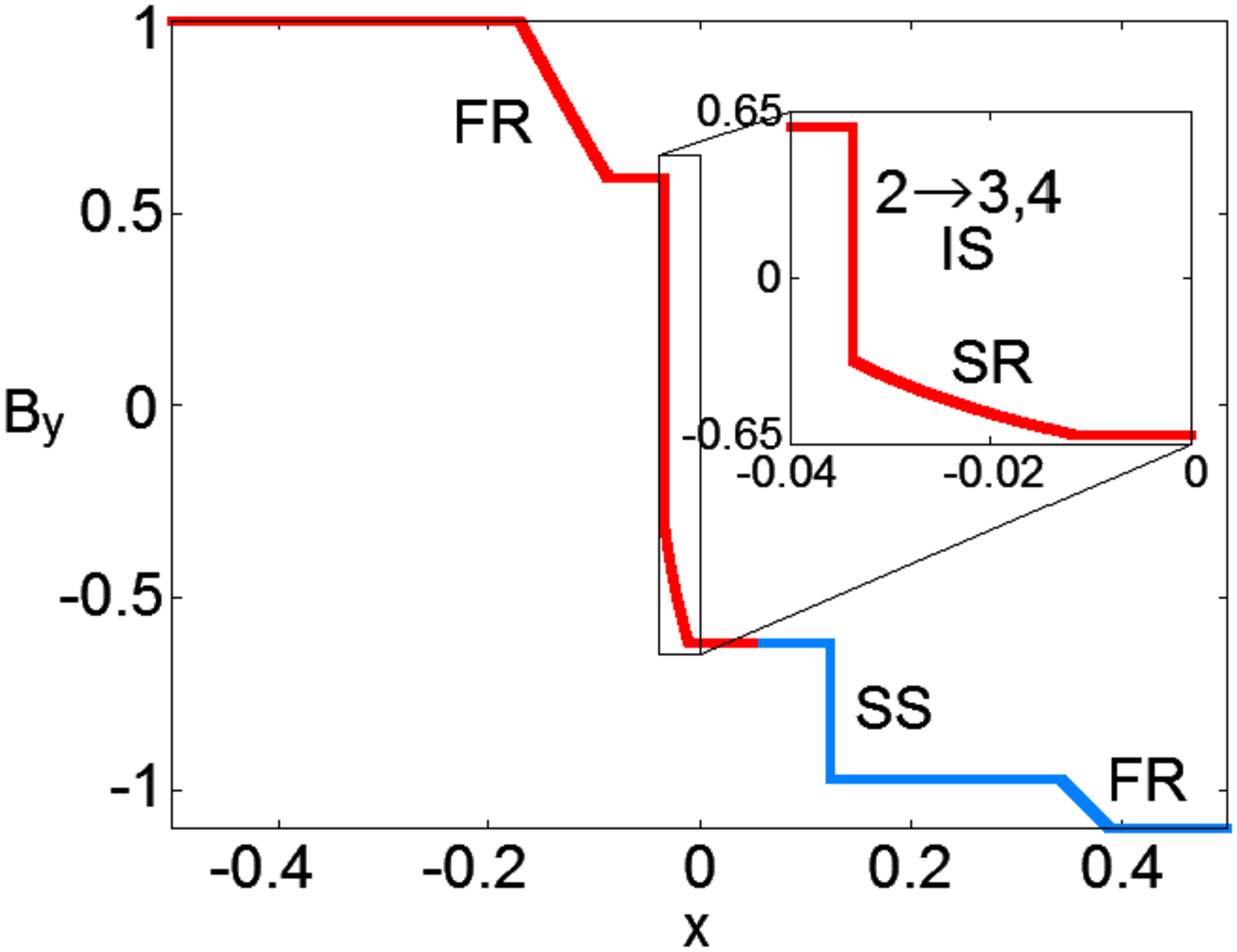}
\end{center}
\end{minipage} \\
\begin{minipage}{0.45\hsize}
\begin{center}
\includegraphics[scale=0.26]{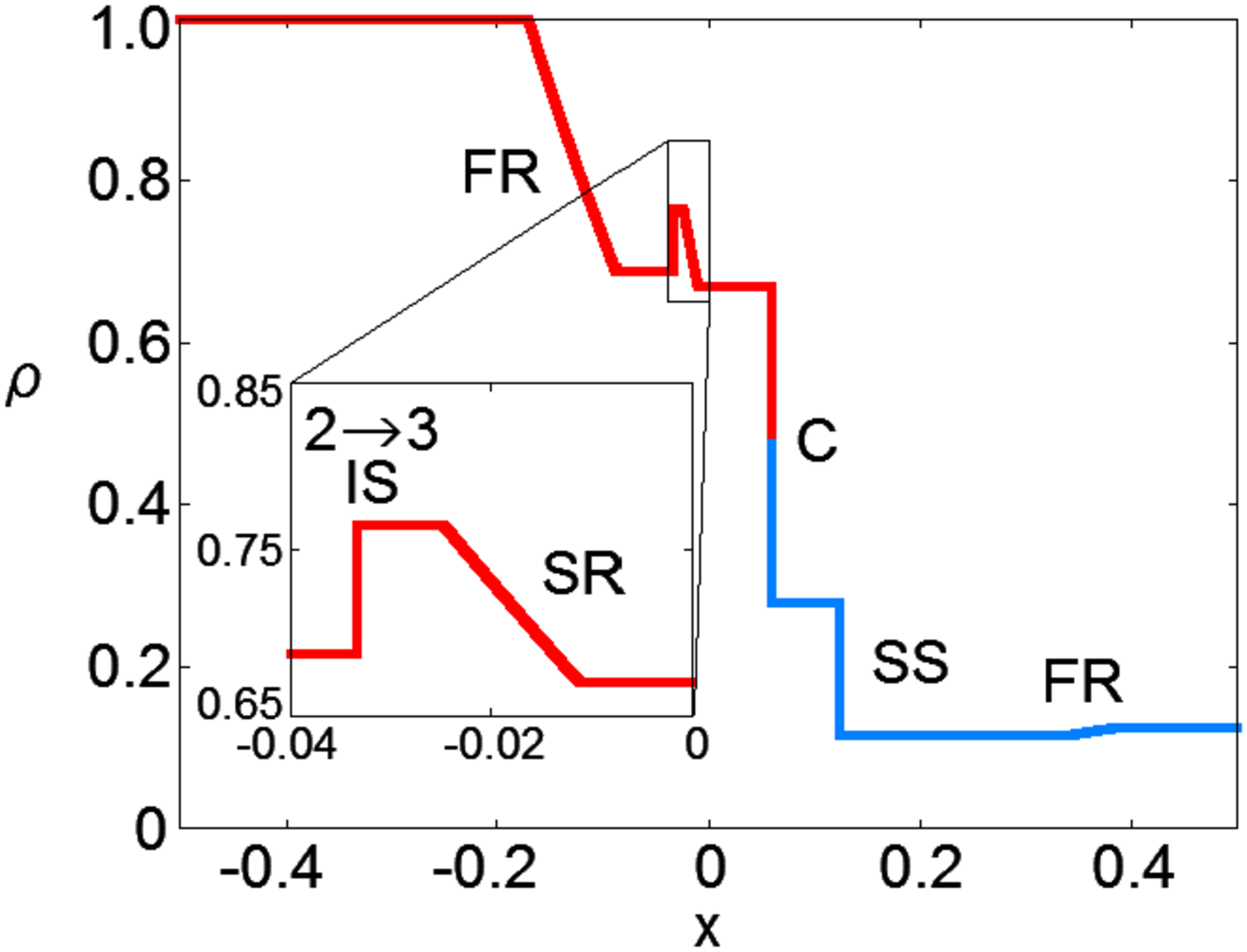}
\end{center}
\end{minipage} &
\begin{minipage}{0.45\hsize}
\begin{center}
\includegraphics[scale=0.26]{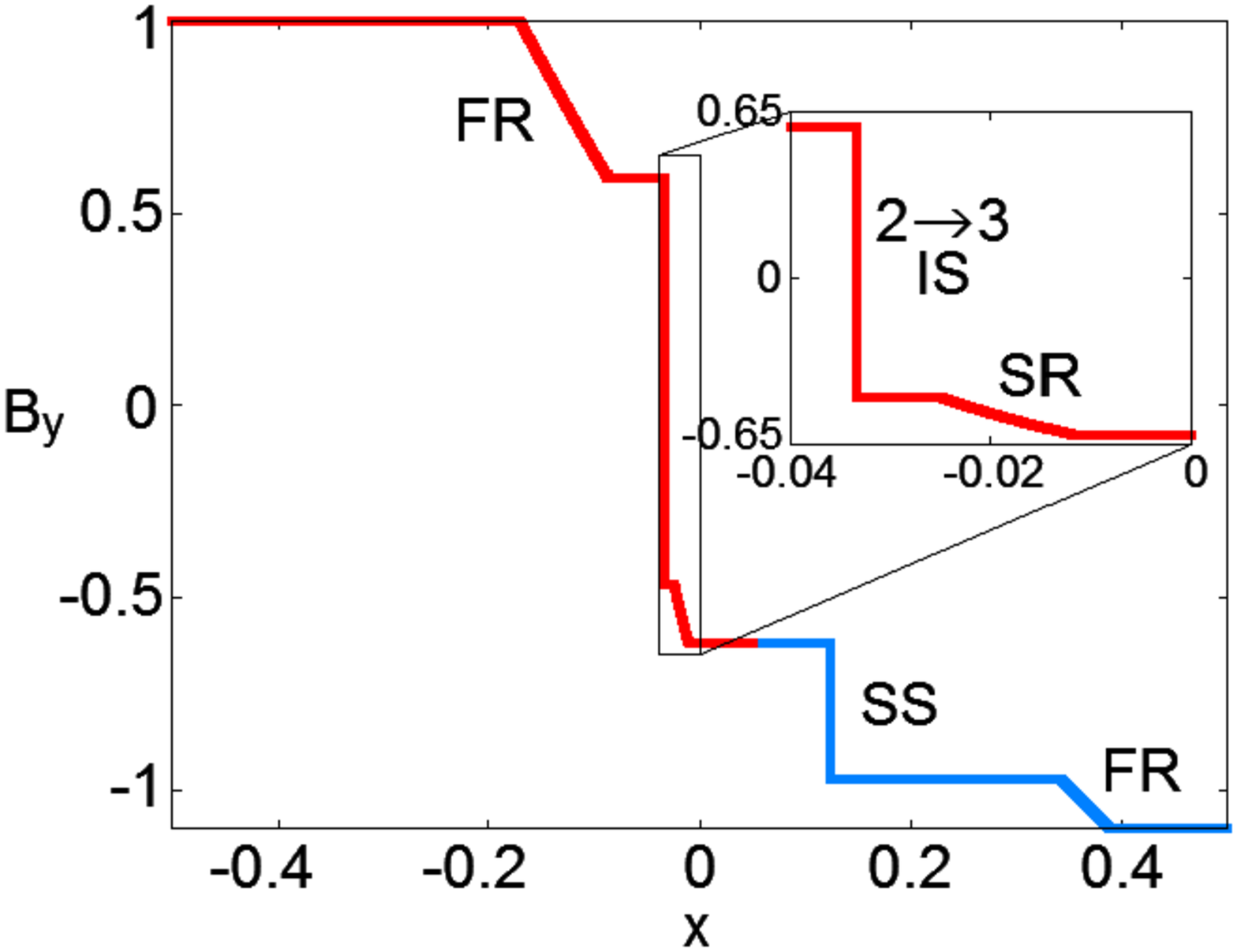}
\end{center}
\end{minipage} \\
\begin{minipage}{0.45\hsize}
\begin{center}
\includegraphics[scale=0.26]{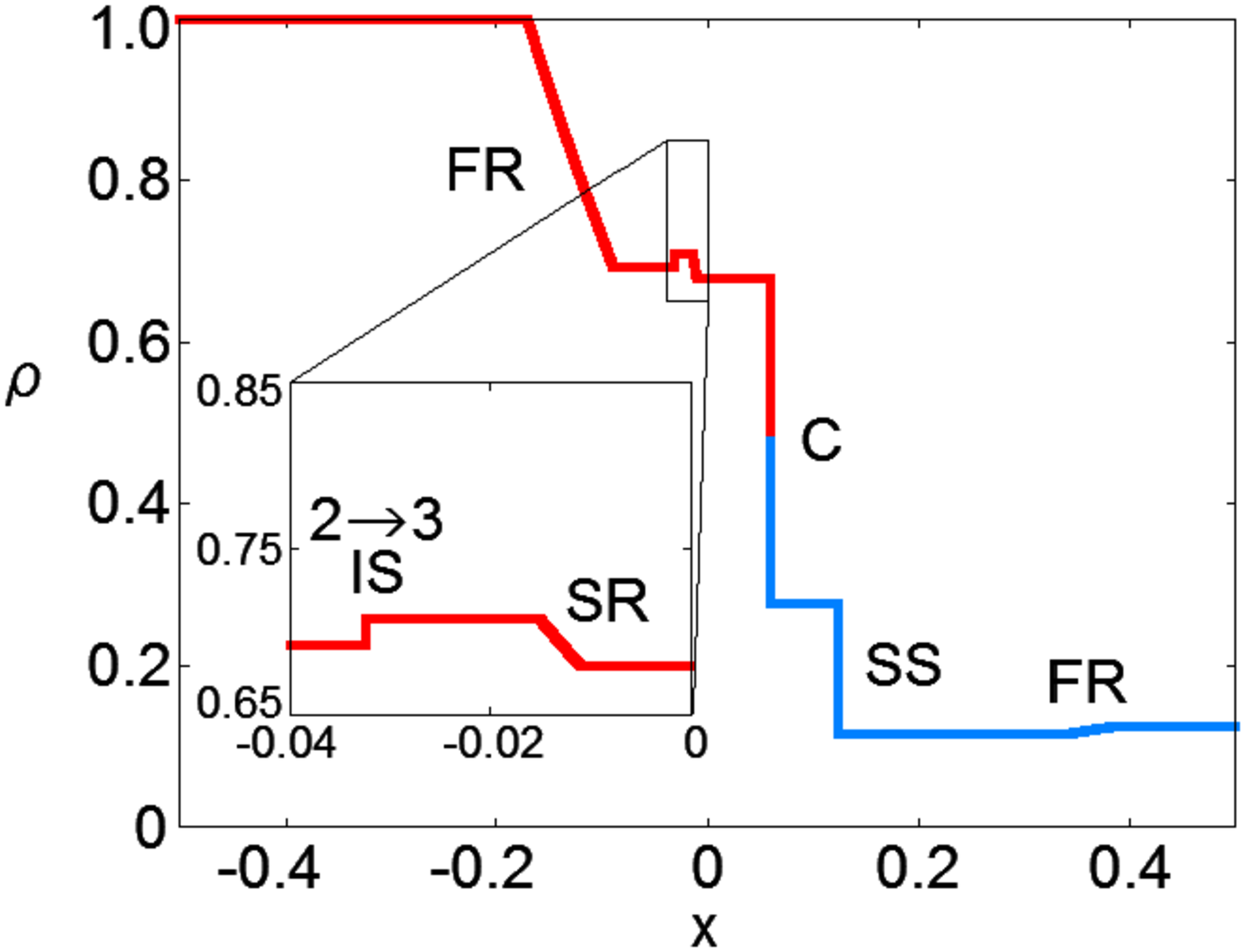}
\end{center}
\end{minipage} &
\begin{minipage}{0.45\hsize}
\begin{center}
\includegraphics[scale=0.26]{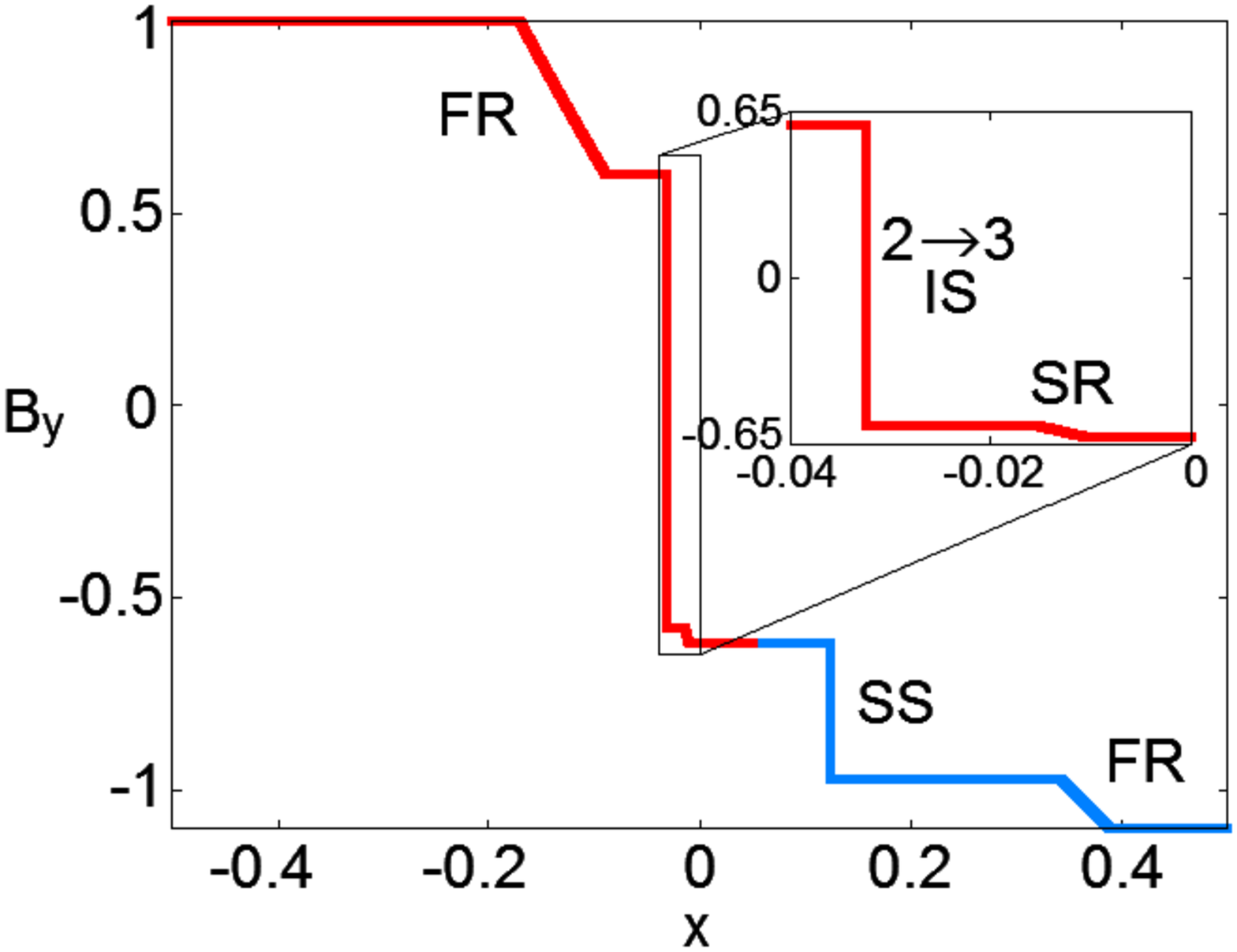}
\end{center}
\end{minipage} \\
\begin{minipage}{0.45\hsize}
\begin{center}
\includegraphics[scale=0.26]{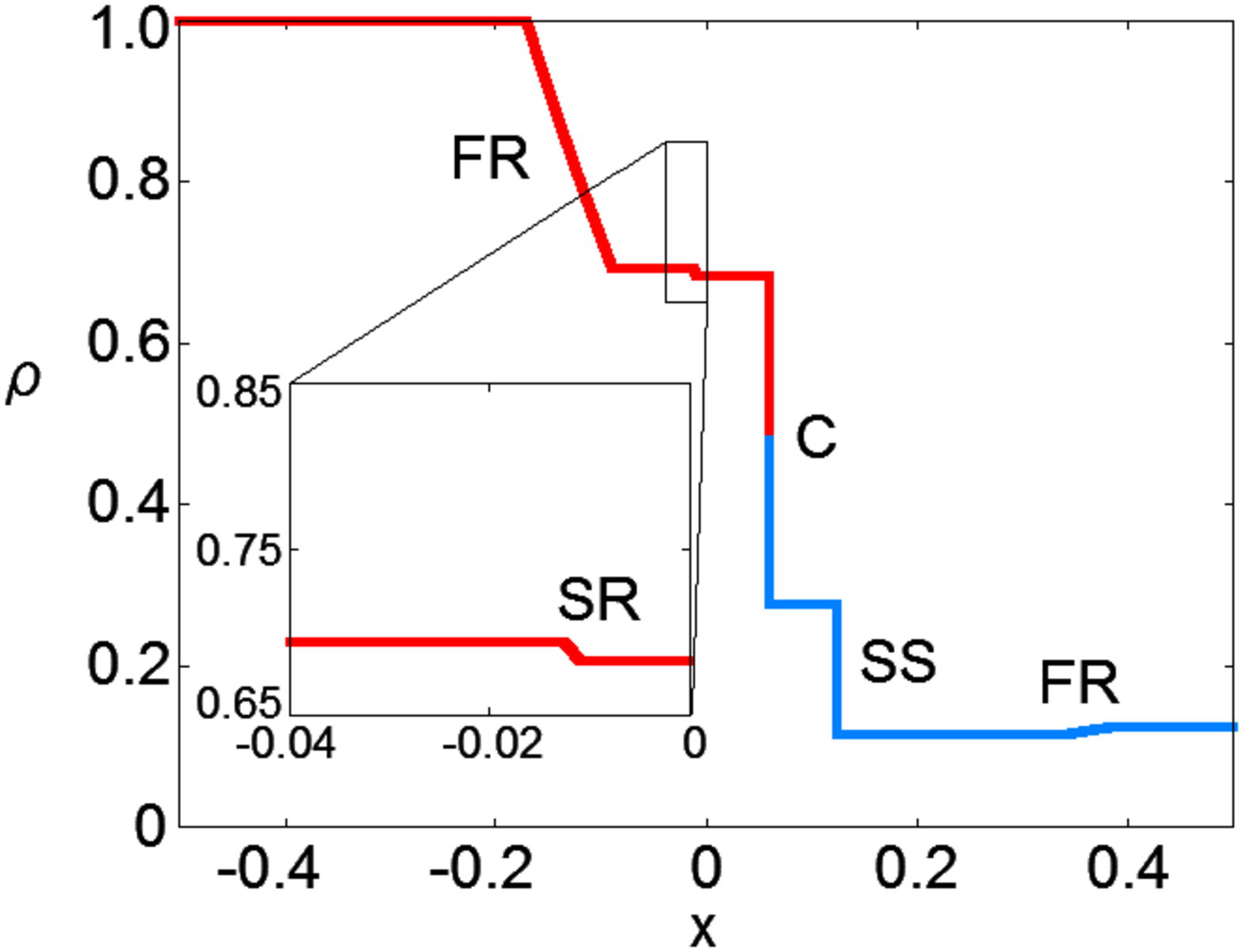}
\end{center}
\end{minipage} &
\begin{minipage}{0.45\hsize}
\begin{center}
\includegraphics[scale=0.26]{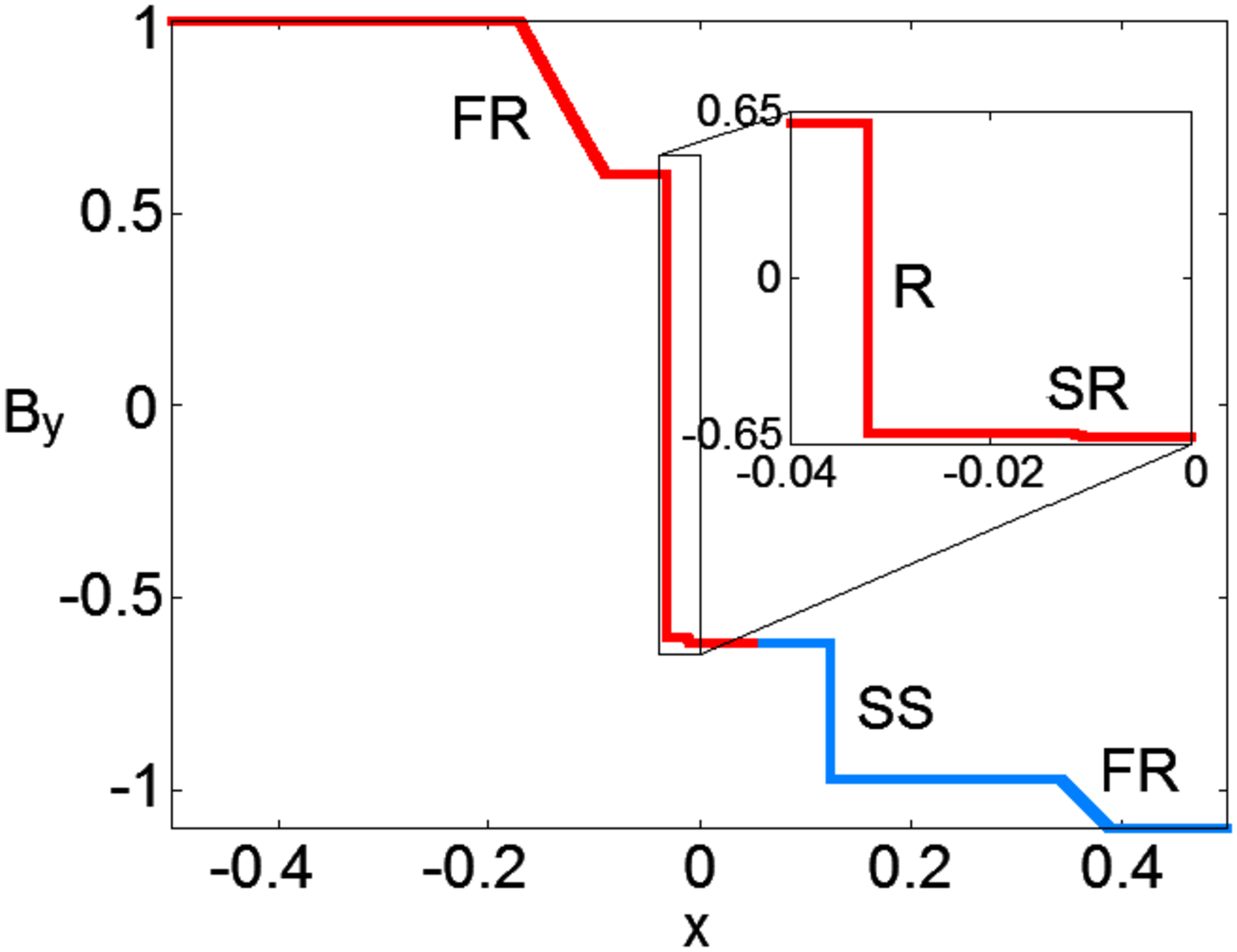}
\end{center}
\end{minipage} 
\end{tabular}
\caption{
Some coplanar neighboring solutions to the Brio \& Wu problem. The initial magnitude of the transverse 
magnetic field  of the right state is increased from the original value. The notation is the same as in Fig.~\ref{BW compound}. 
The insets are the close-ups of indicated regions.
}
\label{BW_neighbors_large}
\end{figure}

In Fig.~\ref{BW_neighbors_small} we present the solutions we found with our code for the following initial
condition:
\begin{eqnarray}
(\rho_L,\ p_L,\ v_{xL},\ v_{yL},\ v_{zL},\ B_{yL},\ B_{zL}) =& (1,\ 1,\ 0,\ 0,\ 0,\ 1,\ 0), \\
(\rho_R,\ p_R,\ v_{xR},\ v_{yR},\ v_{zR},\ B_{yR},\ B_{zR}) =& (0.125,\ 0.1,\ 0,\ 0,\ 0,\ -0.95,\ 0).
\end{eqnarray}
This time we reduced the transverse magnetic field in the right state slightly, retaining the coplanarity. We then found essentially 
the same types of solutions as in the original problem: (1) the solution with a rotational discontinuity\footnote{The rotational 
discontinuity may be an inappropriate term in the reduced system, in which the \Alfven \ characteristic ceases to exist. We use it,
however, to make clear the correspondence to the full system.}, (2) the solution with the compound wave consisting of a 
$2 \rightarrow 3,4$ shock and slow rarefaction wave, and (3) the solutions with a $2 \rightarrow 3$ shock. As mentioned, the second
solution is regular and the first and third are now non-regular in the reduced MHD system. The three types of solutions actually form 
a one-parameter solution family with the first and second solutions being the end points. As shown in Fig.~\ref{BW_neighbors_large} this 
is also the case for the following initial condition:
\begin{eqnarray}
(\rho_L,\ p_L,\ v_{xL},\ v_{yL},\ v_{zL},\ B_{yL},\ B_{zL}) =& (1,\ 1,\ 0,\ 0,\ 0,\ 1,\ 0), \\
(\rho_R,\ p_R,\ v_{xR},\ v_{yR},\ v_{zR},\ B_{yR},\ B_{zR}) =& (0.125,\ 0.1,\ 0,\ 0,\ 0,\ -1.05,\ 0),
\end{eqnarray}
in which the transverse magnetic field in the right state is slightly amplified.

These results show clearly that the Brio \& Wu solution, or the regular solution with the compound wave, has a unique neighboring 
solution of the same type as expected from the evolutionary conditions in the reduced system. In the case of the non-regular 
solutions either with the $2 \rightarrow 3$ shock or with the rotational discontinuity, however, the perturbed solutions are 
not uniquely determined, since there are too many outgoing characteristics. And the existence of uncountably many non-regular solutions
is a manifestation of this under-determination. Unlike the over-determination and, as a consequence, the non-existence
of neighboring solution, the meaning of the non-uniqueness of perturbed solutions is rather obscure. The results of numerical simulations indicate,
however, the evolutionary conditions in the reduced system appear to work in selecting the solution.

\subsection{An initial condition with no regular solution \label{sec5.2}}
In the continuous modifications of the Brio \& Wu problem, we ran into an interesting initial condition, for which
there seems to be no regular solution. If true, this is a counter-example to the conventional expectation that
a unique regular solution always exists. This initial condition is truly a special point unlike the Brio \& Wu 
problem. In fact, the regular and non-regular solutions of the Brio \& Wu problem is shared by the neighboring
coplanar initial conditions as demonstrated above. This is not the case for the problem considered here.
  
\begin{figure}
\begin{tabular}{cc}
\begin{minipage}{0.45\hsize}
\begin{center}
\includegraphics[scale=0.26]{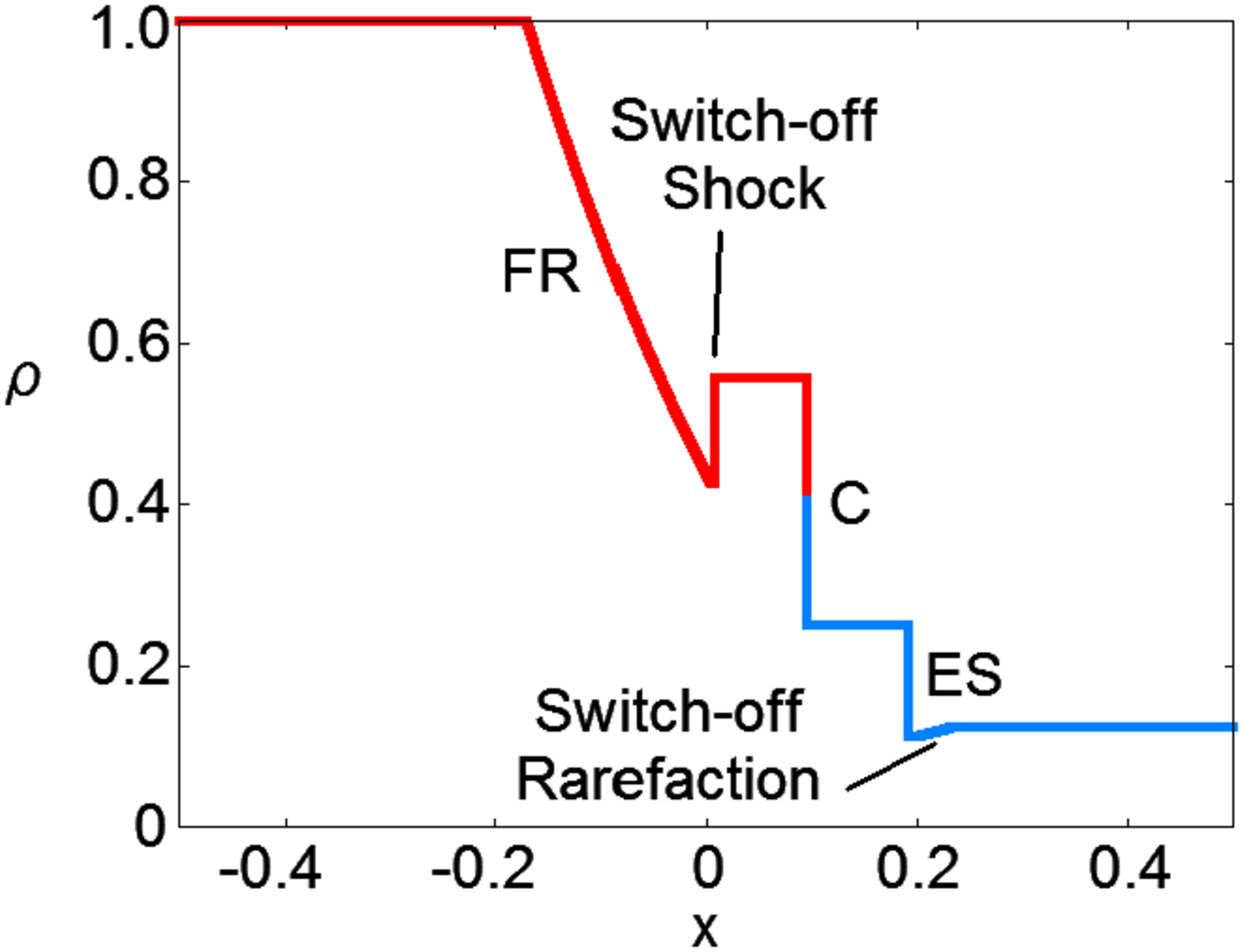}
\end{center}
\end{minipage} &
\begin{minipage}{0.45\hsize}
\begin{center}
\includegraphics[scale=0.26]{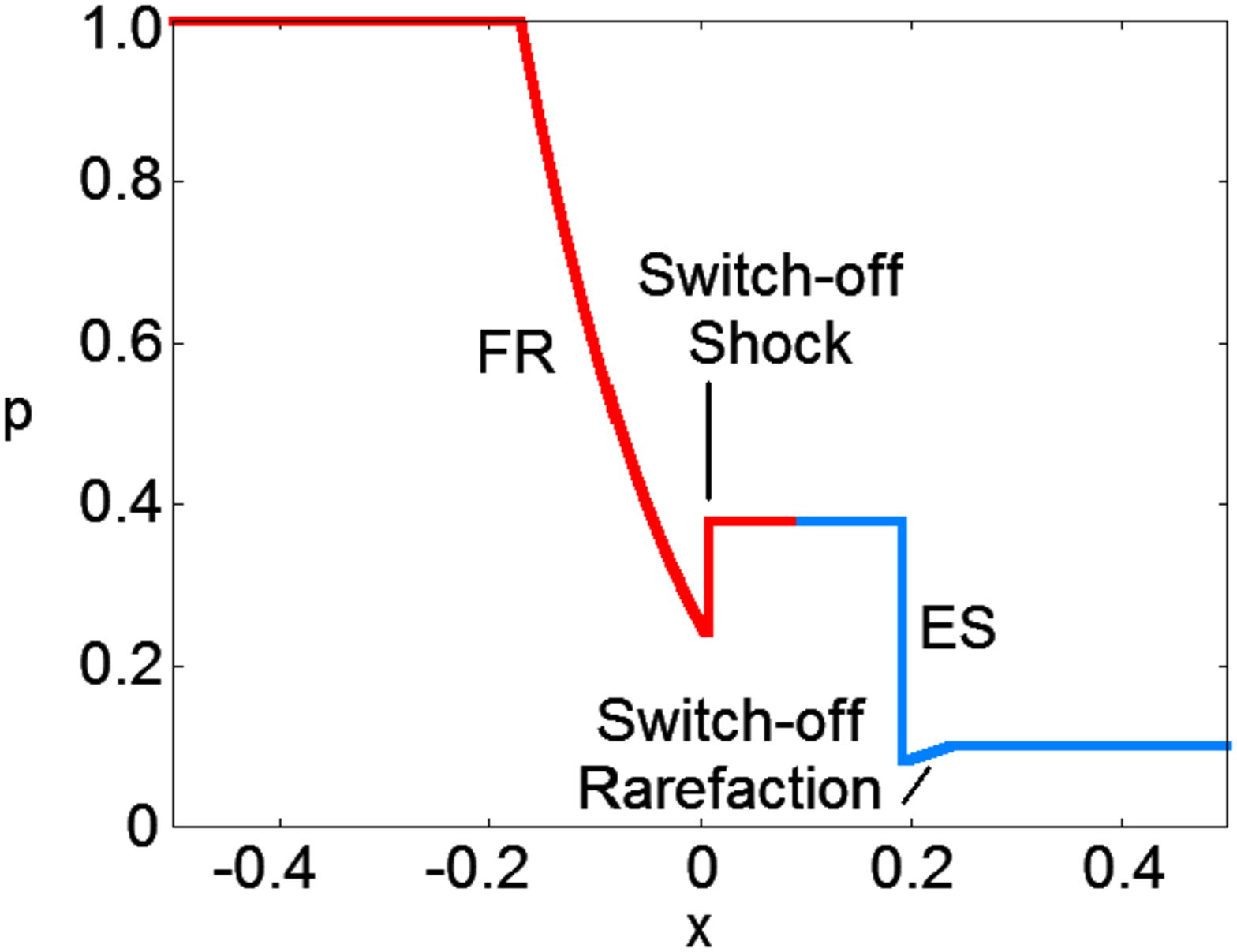}
\end{center}
\end{minipage} \\
\begin{minipage}{0.45\hsize}
\begin{center}
\includegraphics[scale=0.26]{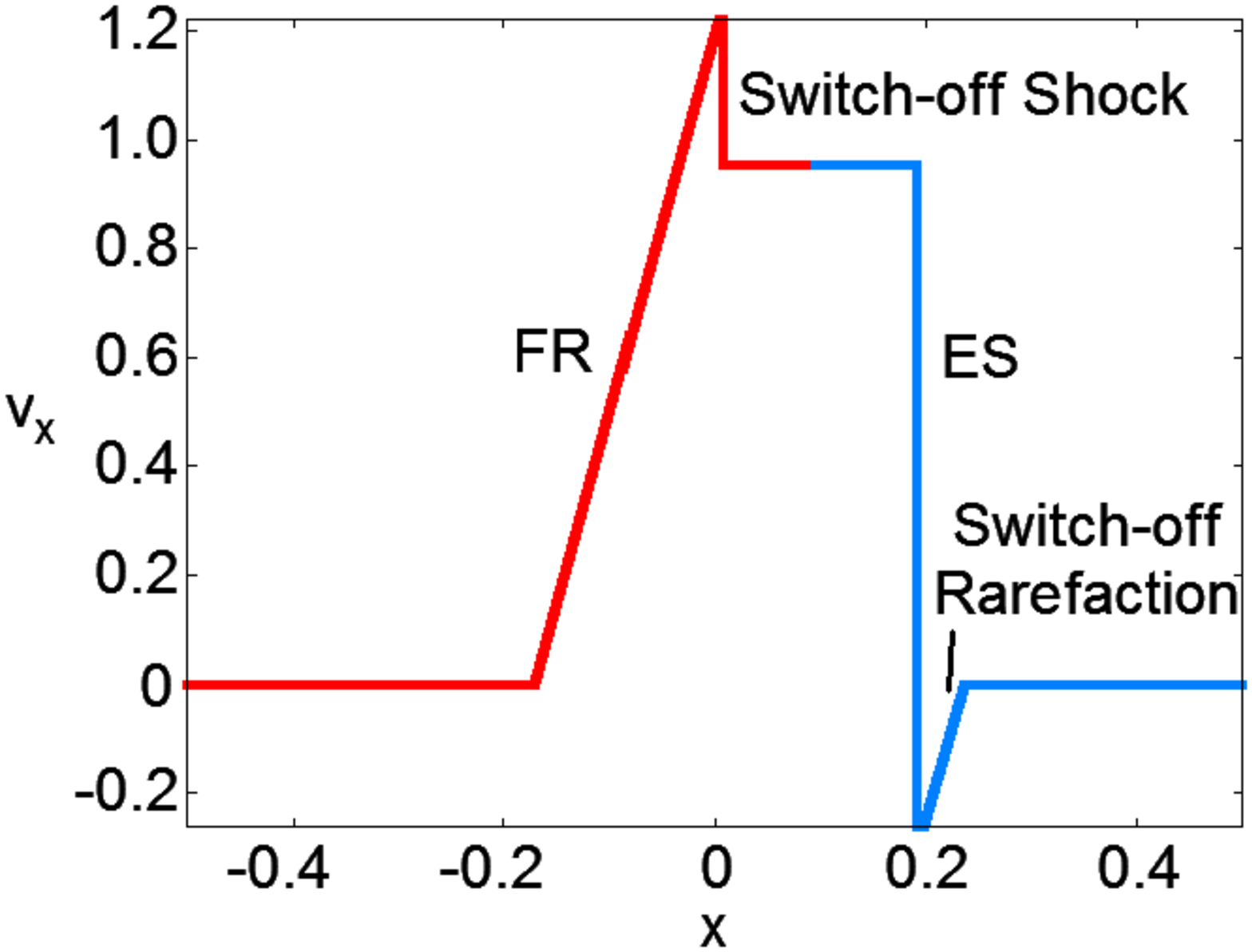}
\end{center}
\end{minipage} &
\begin{minipage}{0.45\hsize}
\begin{center}
\includegraphics[scale=0.26]{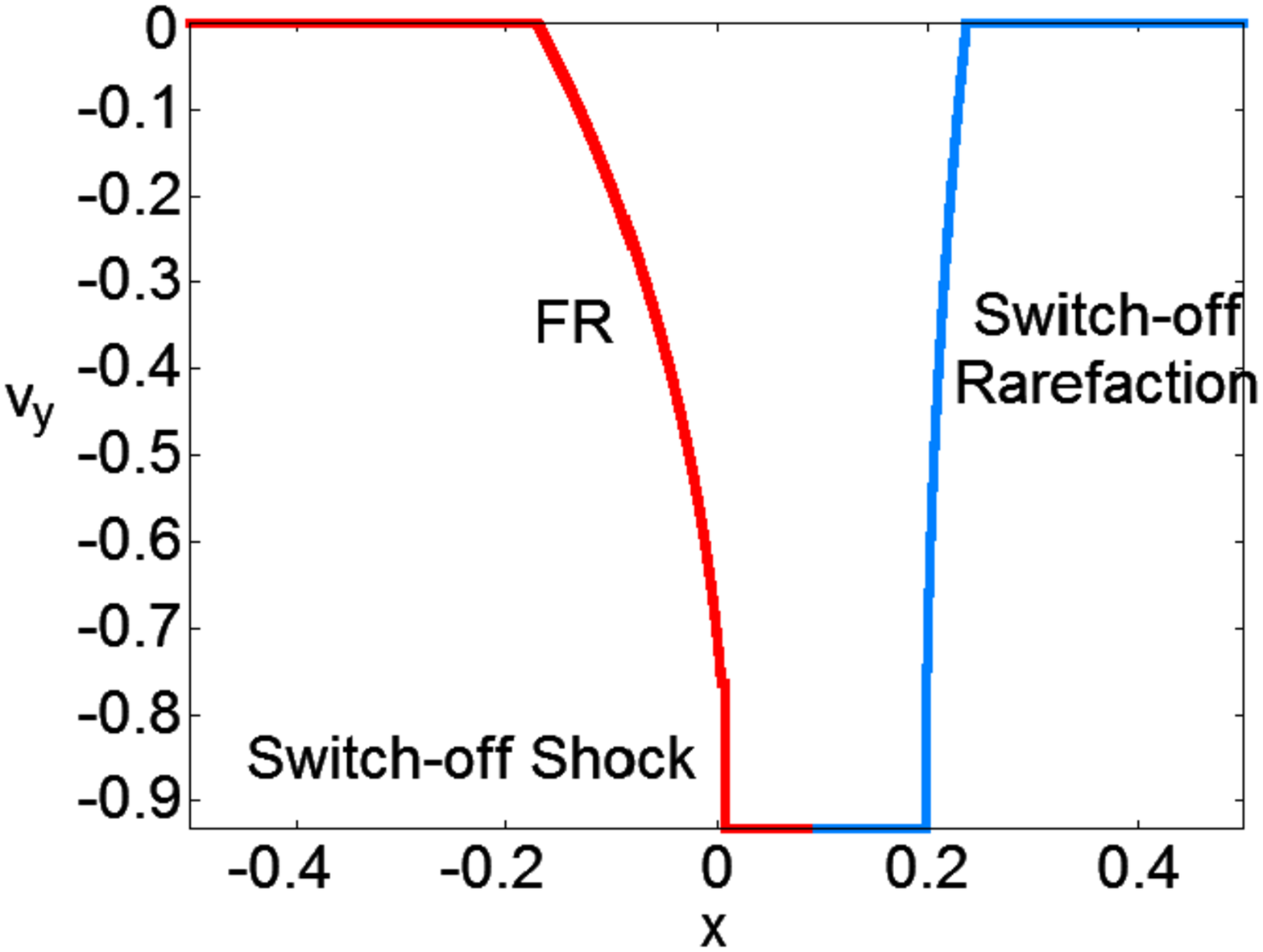}
\end{center}
\end{minipage} \\
\begin{minipage}{0.45\hsize}
\begin{center}
\includegraphics[scale=0.26]{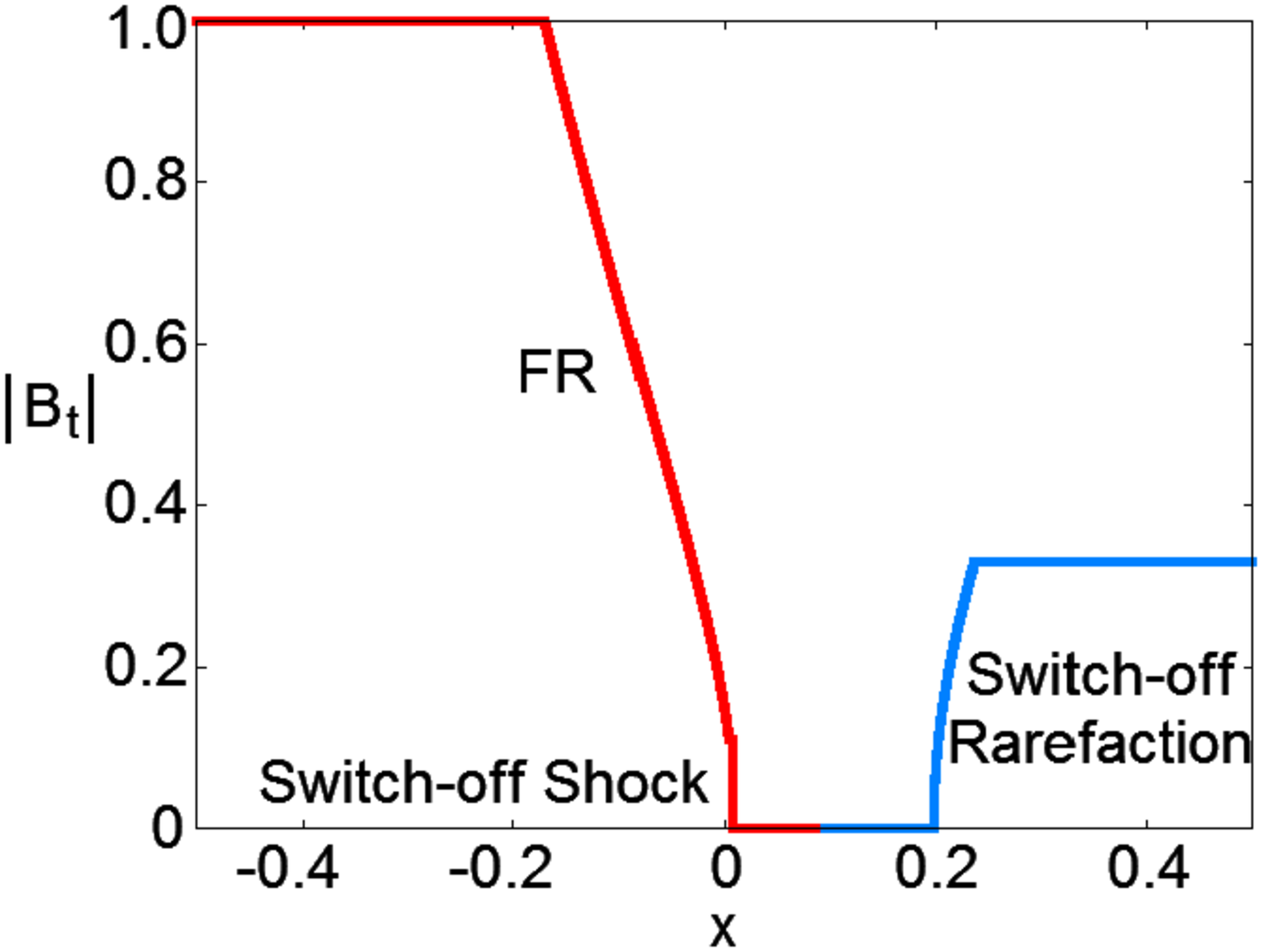}
\end{center}
\end{minipage} &
\begin{minipage}{0.45\hsize}
\begin{center}
\includegraphics[scale=0.26]{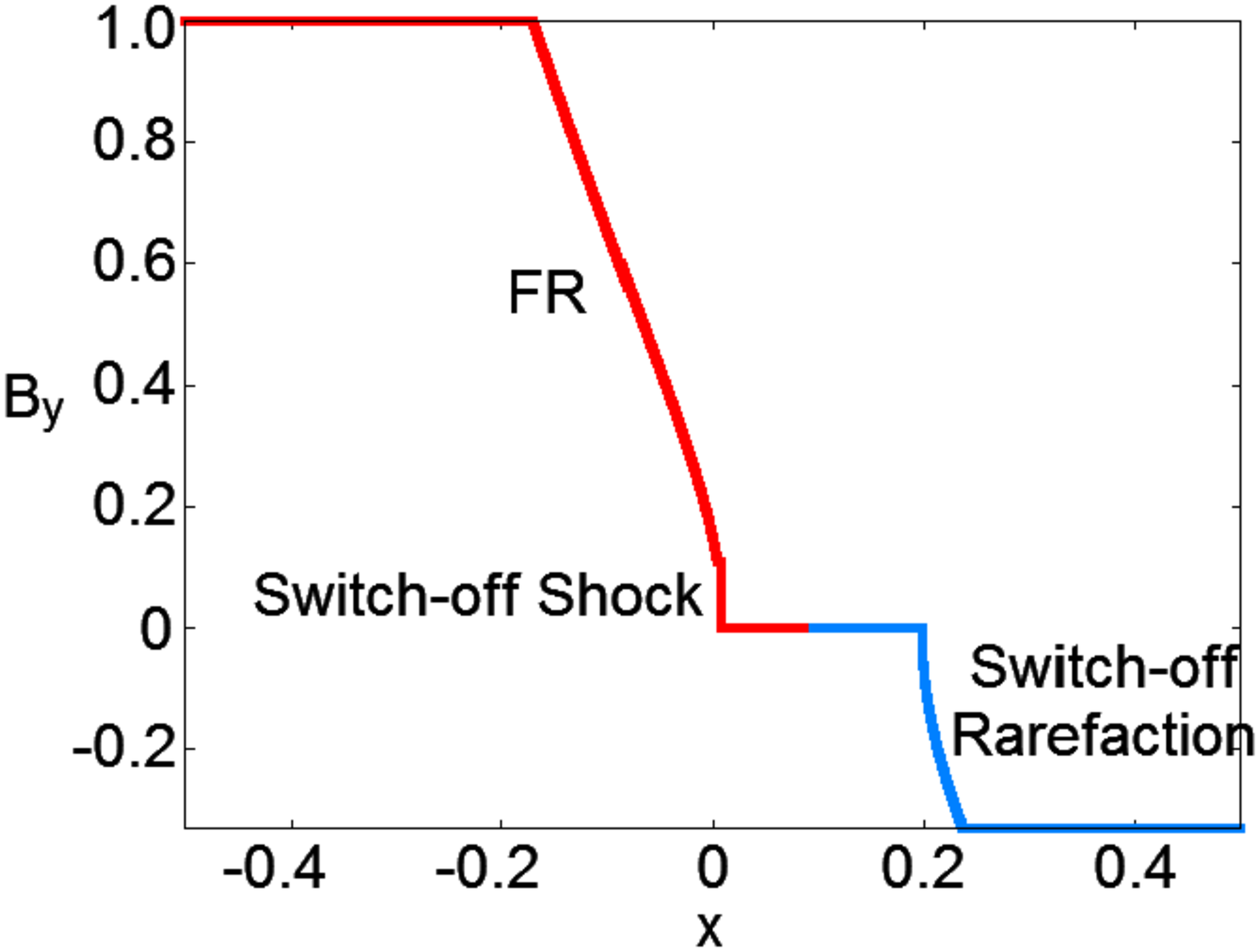}
\end{center}
\end{minipage} 
\end{tabular}
\caption{
The non-regular solution with both a switch-off shock and switch-off rarefaction wave.
The notation is the same as in Fig.~\ref{BW compound} except for the designation ES, 
which stands for the Euler shock. The initial condition is the same 
as for the Brio \& Wu problem except for the strength of the transverse magnetic field of 
the right state. See \S\ref{sec5.2} of the main body for the precise value.
}
\label{BW switch-off}
\end{figure}

The initial condition is given by the following:
\begin{eqnarray}
(\rho_L,\ p_L,\ v_{xL},\ v_{yL},\ v_{zL},\ B_{yL},\ B_{zL}) =& (1,\ 1,\ 0,\ 0,\ 0,\ 1,\ 0), \\
(\rho_R,\ p_R,\ v_{xR},\ v_{yR},\ v_{zR},\ B_{yR},\ B_{zR}) =& (0.125,\ 0.1,\ 0,\ 0,\ 0,\ -0.329875,\ 0),
\end{eqnarray}
in addition to $B_n = 0.75$ and $\gamma = 5/3$. This differs from the Brio \& Wu problem only on the magnitude of the 
transverse magnetic field ($-B_{yR} = B_\mathrm{crit} = 0.329875$) of the right state\footnote{The ratio of the specific 
heats is not set to $2$, the value adopted by \citet{BW88}, but to $5/3$ for the same reason as given for the first case.}.
This is again a coplanar configuration. The only solution our code obtained is a non-regular solution, which 
is presented in Fig.~\ref{BW switch-off} at $t=0.1$. Note that $v_z$ and $B_z$ are identically zero in this solution.
As can be seen, the solution consists of a fast rarefaction wave and switch-off shock, both of which go 
leftward, and a switch-off rarefaction wave and Euler shock, which proceed rightward, as well as a contact discontinuity.
Although it may seem that the shocks and rarefaction waves form compound waves on both sides of the contact discontinuity, 
they are actually detached. As a matter of fact, it is theoretically shown that the switch-off waves, irrespective of shock or
rarefaction, never form compound waves. As mentioned earlier, the switch-off shock is a $2,3 \rightarrow 4$ shock, 
with the \Alfven \ speed being equal to the upstream flow speed. Preceded by the switch-off rarefaction, which is actually one of the
fast rarefaction waves, the fast speed and \Alfven \ speeds are degenerate ahead of the Euler shock, which is itself an evolutionary 
shock not only in the full system but also in the reduced MHD system in this case although it is possible to be non-evolutionary 
in principle. It should be stressed that it is the advantage of our code that can treat these degenerate cases. In fact, other codes, 
e.g. developed by \citet{ATJweb}, ignore the switch-off waves entirely.
It is normally impossible to prove the non-existence of particular solutions by numerical computations. We think, however, that 
no regular solution exists indeed to the current problem. In order to demonstrate why we think so and elucidate the special place 
that the current initial condition occupies, we study the neighborhood of this problem in detail.

\begin{figure}
\begin{tabular}{cc}
\begin{minipage}{0.45\hsize}
\begin{center}
\includegraphics[scale=0.26]{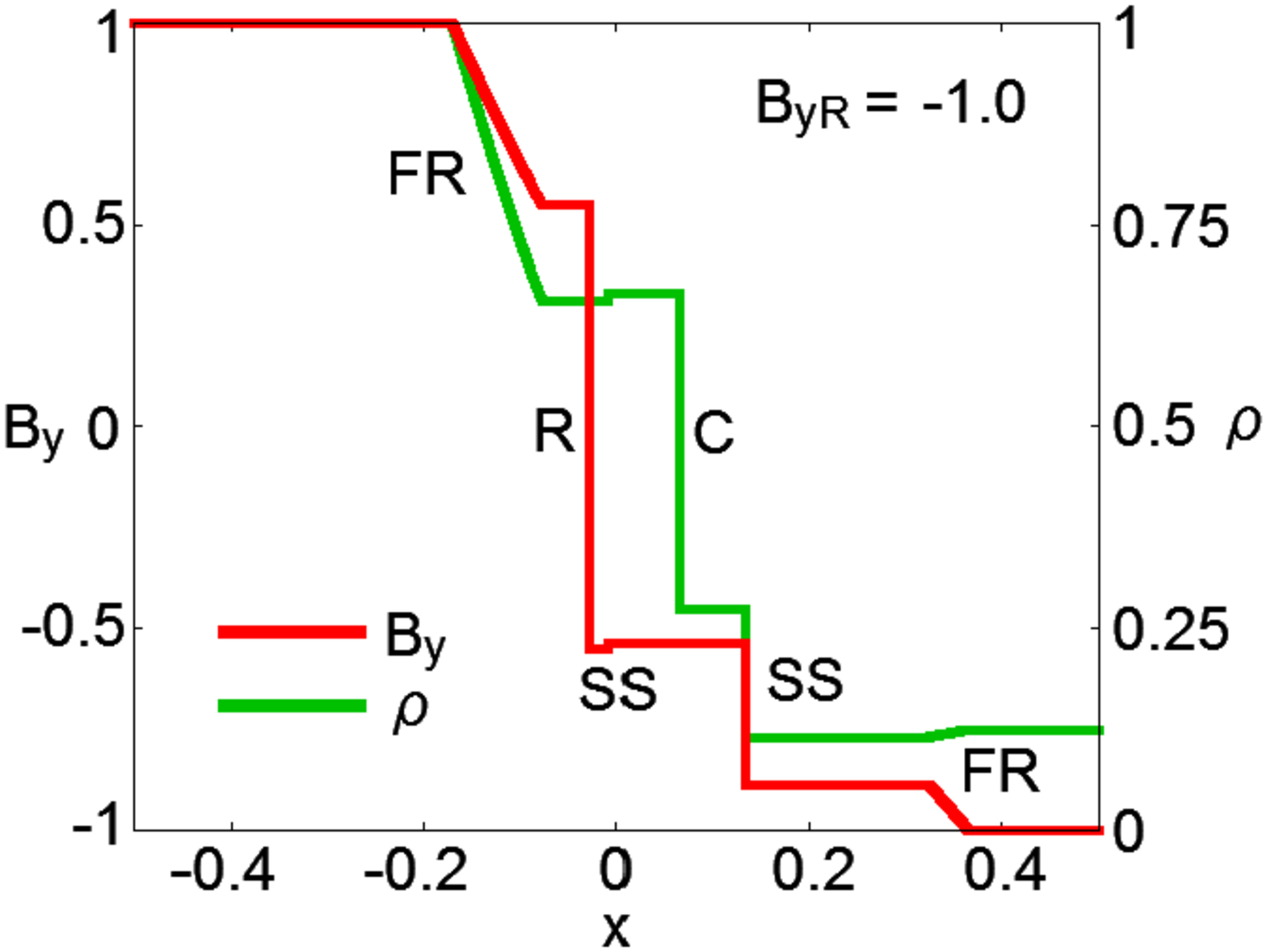}
\end{center}
\end{minipage} &
\begin{minipage}{0.45\hsize}
\begin{center}
\includegraphics[scale=0.26]{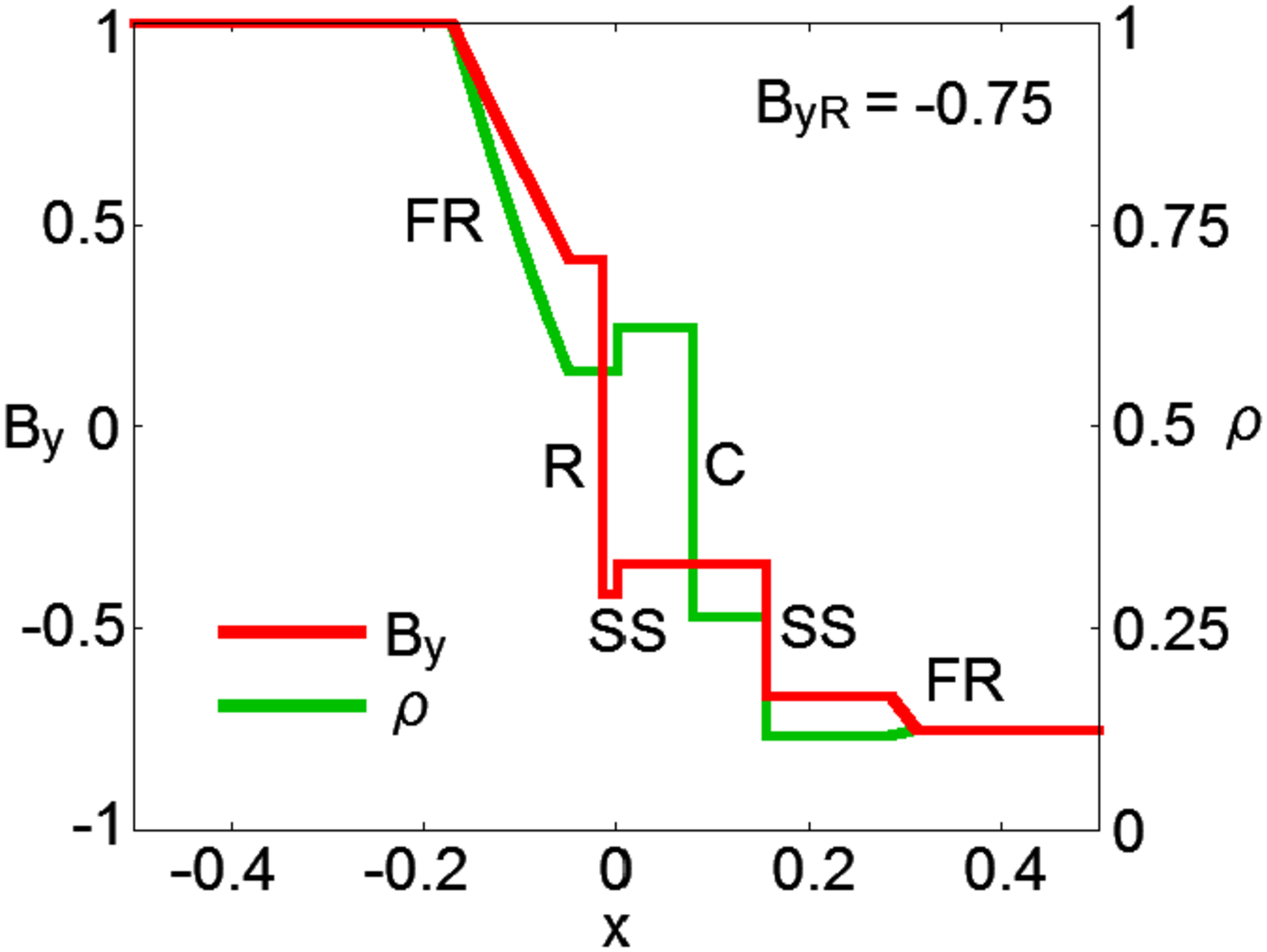}
\end{center}
\end{minipage} \\
\begin{minipage}{0.45\hsize}
\begin{center}
\includegraphics[scale=0.26]{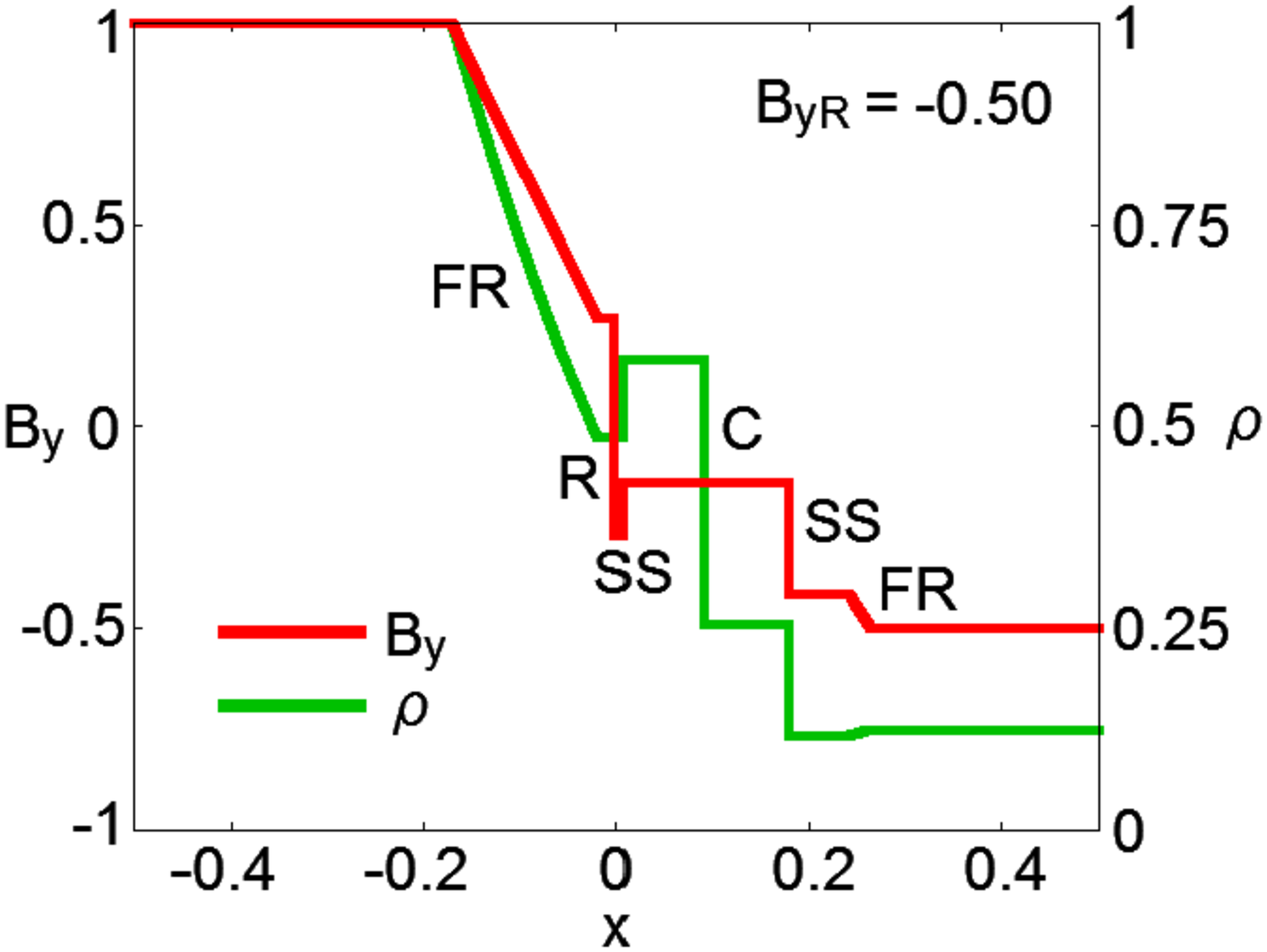}
\end{center}
\end{minipage} &
\begin{minipage}{0.45\hsize}
\begin{center}
\includegraphics[scale=0.26]{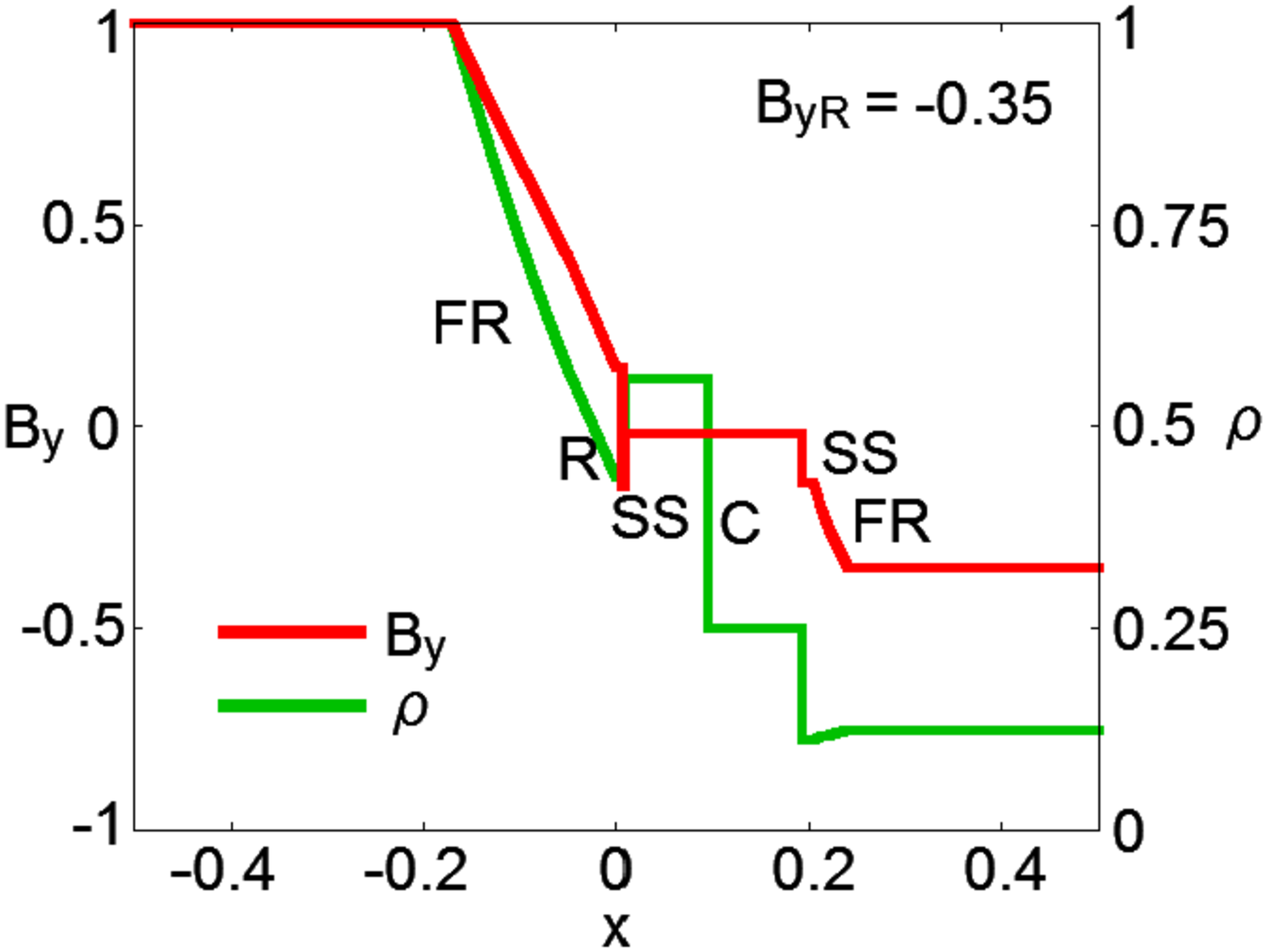}
\end{center}
\end{minipage} \\
\begin{minipage}{0.45\hsize}
\begin{center}
\includegraphics[scale=0.26]{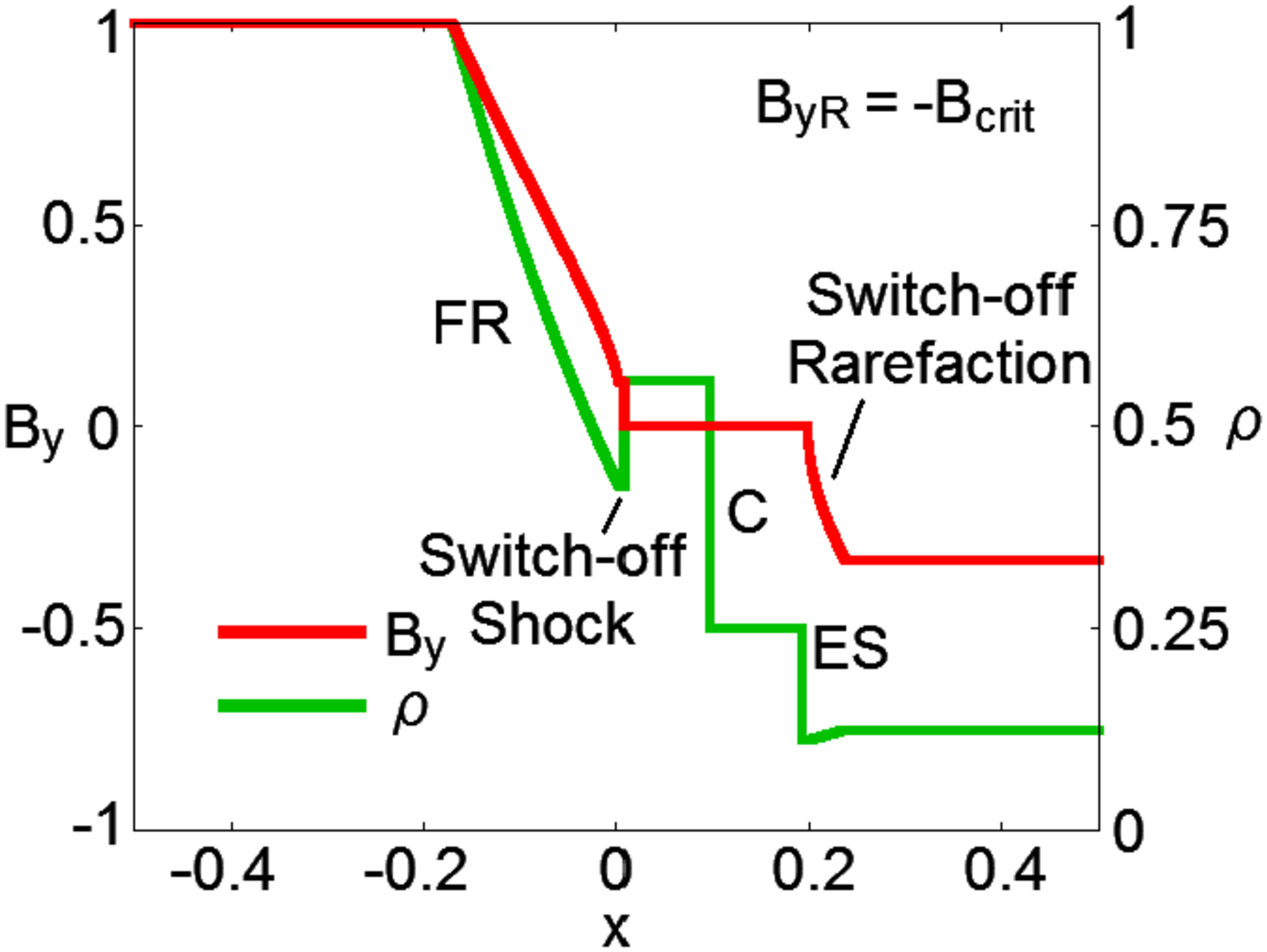}
\end{center}
\end{minipage} &
\begin{minipage}{0.45\hsize}
\begin{center}
\includegraphics[scale=0.26]{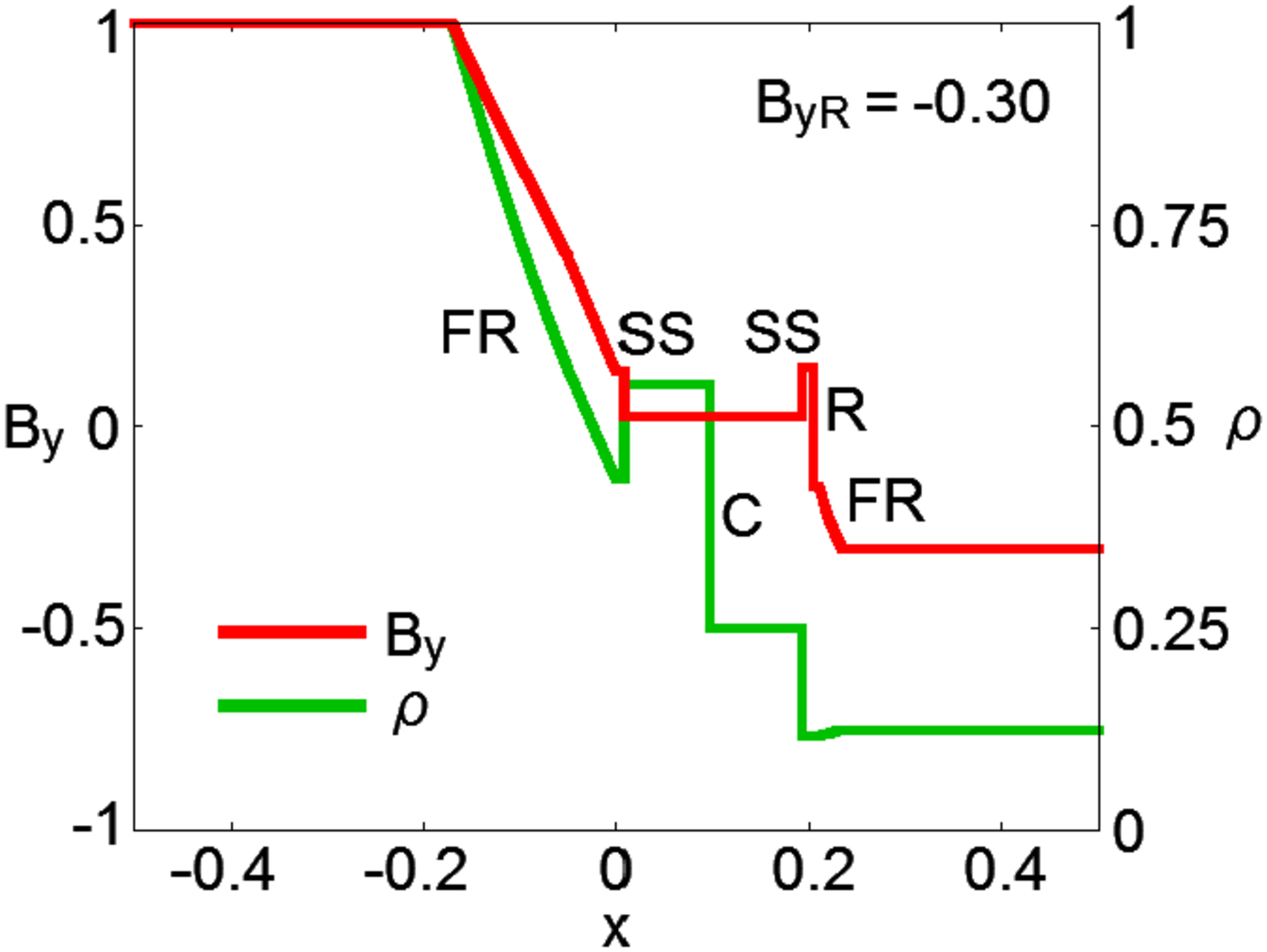}
\end{center}
\end{minipage} \\
\begin{minipage}{0.45\hsize}
\begin{center}
\includegraphics[scale=0.26]{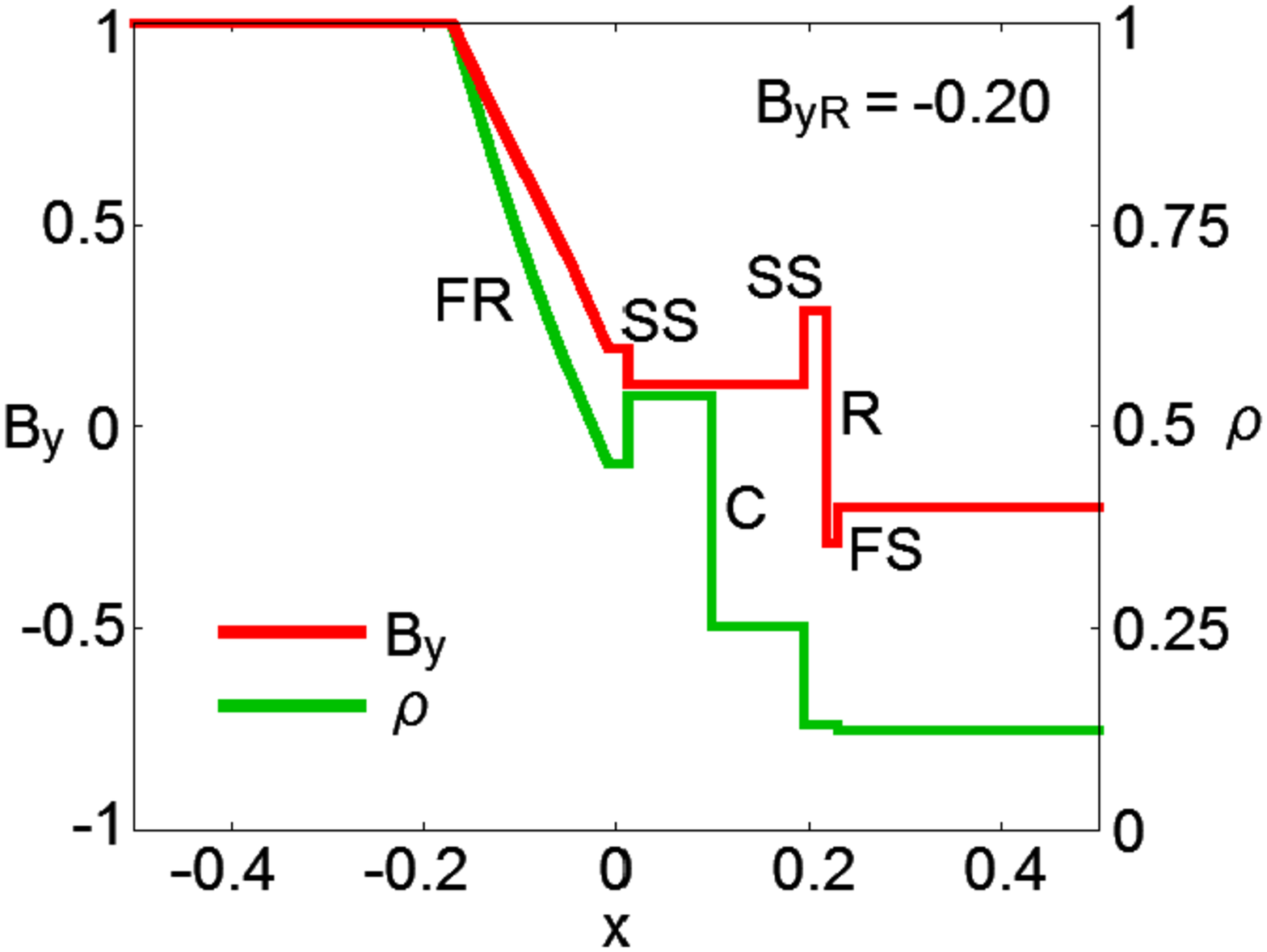}
\end{center}
\end{minipage} &
\begin{minipage}{0.45\hsize}
\begin{center}
\includegraphics[scale=0.26]{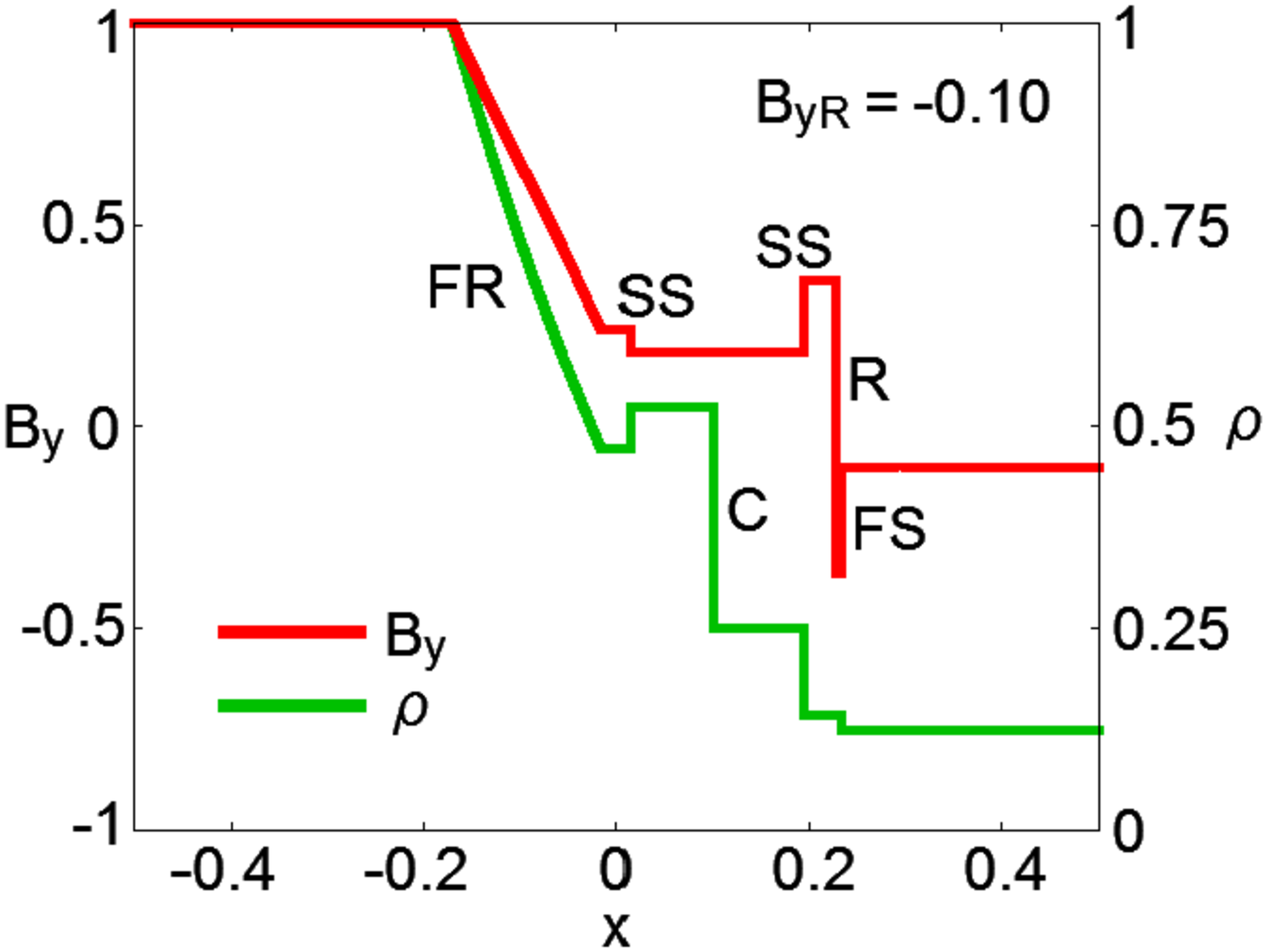}
\end{center}
\end{minipage} 
\end{tabular}
\caption{Coplanar neighboring solutions to the solution presented in Fig.~\ref{BW switch-off}. 
The red line represents the transverse magnetic field whereas the green line displays the 
density. Only the strength of the transverse magnetic field of the right state is varied. 
See \S\ref{sec5.2} of the main body for details. 
}
\label{BW neighborhood 1}
\end{figure}

Presented in Fig.~\ref{BW neighborhood 1} are the solutions to coplanar neighboring problems. More
specifically, we vary the strength of the transverse magnetic field of the right state, $B_{yR}$. The critical value $B_\mathrm{crit}$ 
is approached both from above and from below. We are concerned here only with the regular solutions. As shown in the figures, 
in the case of $|B_{yR}| > B_\mathrm{crit}$ the solutions consist of a fast rarefaction wave, rotational discontinuity and slow shock 
going leftward and a fast rarefaction wave and slow shock going to the right in addition to a contact discontinuity. As the initial
strength of the transverse magnetic field in the right state becomes smaller, 
the left-going rotational discontinuity and slow shock come closer to each other. 
Among the right-going waves, the fast rarefaction wave approaches the 
slow shock. At the critical point the left-going rotational discontinuity and slow shock merge into the switch-off shock and the 
right-going fast rarefaction wave and slow shock become the switch-off rarefaction wave and the Euler shock, respectively. 
  
As the initial strength of the transverse magnetic field becomes even smaller and $|B_{yR}| < B_\mathrm{crit}$ is satisfied, the right-going 
rotational discontinuity emerges and is responsible for the inversion of the transverse magnetic field. The Euler shock is replaced by the
rotational discontinuity and a slow shock whereas the switch-off rarefaction is modified to an ordinary fast rarefaction wave again at first
as long as $|B_{yR}|$ is not very different from $B_\mathrm{crit}$. A fast shock appears instead of the fast rarefaction wave when $|B_{yR}|$ 
becomes smaller than a certain value. On the left side of the contact discontinuity, on the other hand, the switch-off shock is changed to 
an ordinary slow shock and no rotational discontinuity exists. It is stressed that the unique non-regular solution at the critical point is
connected continuously to the sequences of the regular solutions obtained for $|B_{yR}| > B_\mathrm{crit}$ and $|B_{yR}| < B_\mathrm{crit}$,
respectively.

\begin{figure}
\begin{tabular}{cc}
\begin{minipage}{0.45\hsize}
\begin{center}
\includegraphics[scale=0.31]{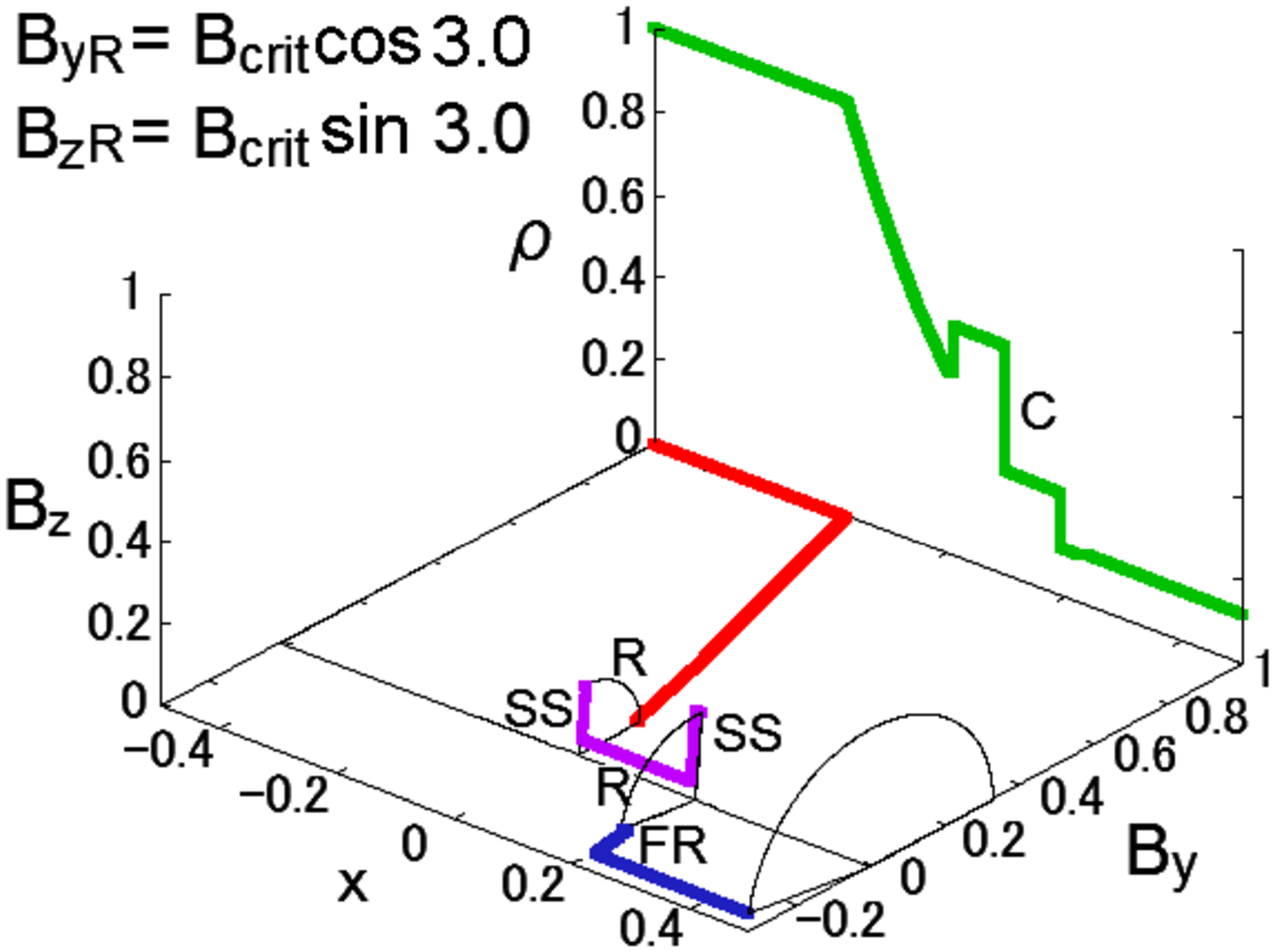}
\end{center}
\end{minipage} &
\begin{minipage}{0.45\hsize}
\begin{center}
\includegraphics[scale=0.31]{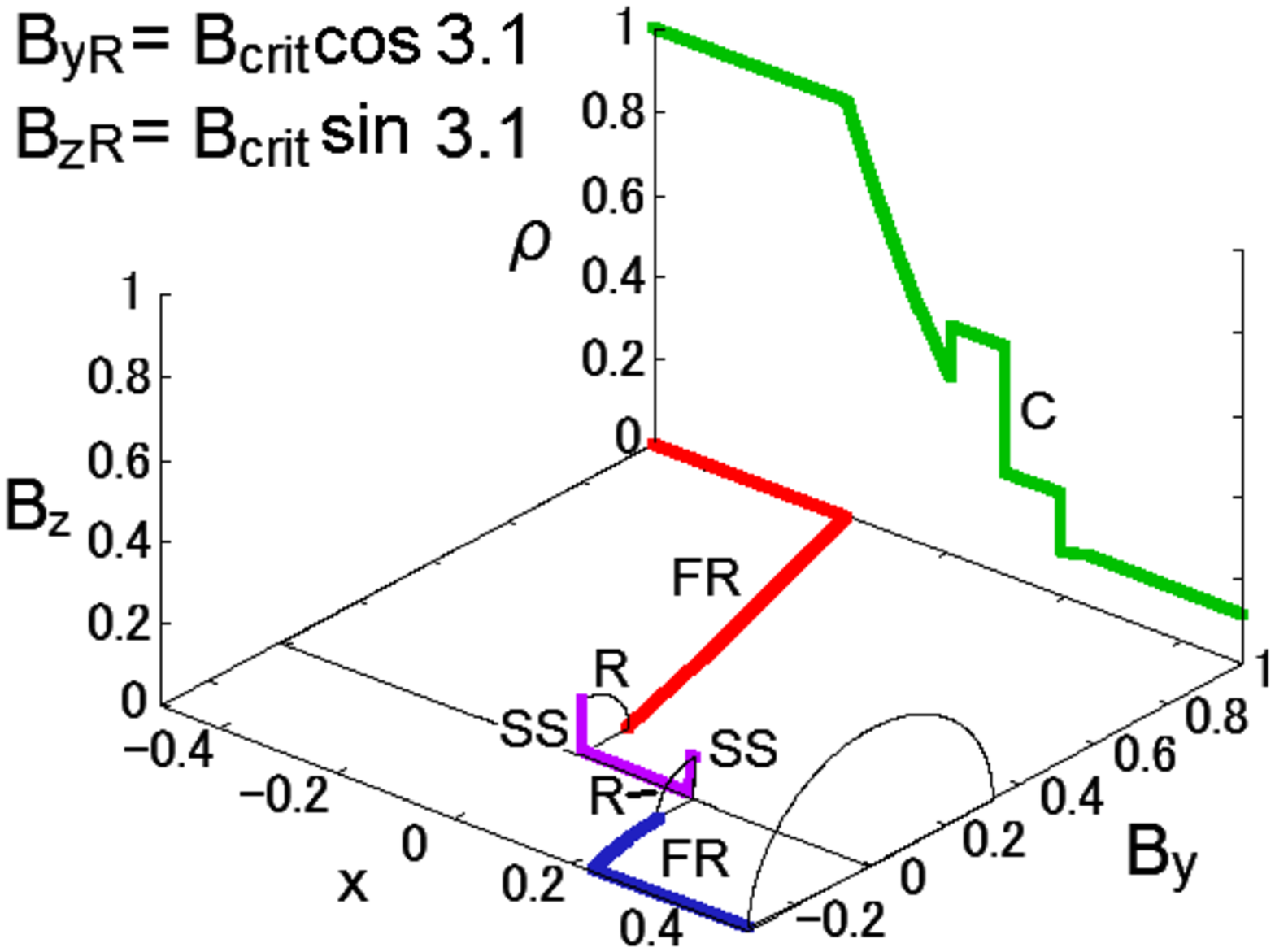}
\end{center}
\end{minipage} \\
\begin{minipage}{0.45\hsize}
\begin{center}
\includegraphics[scale=0.31]{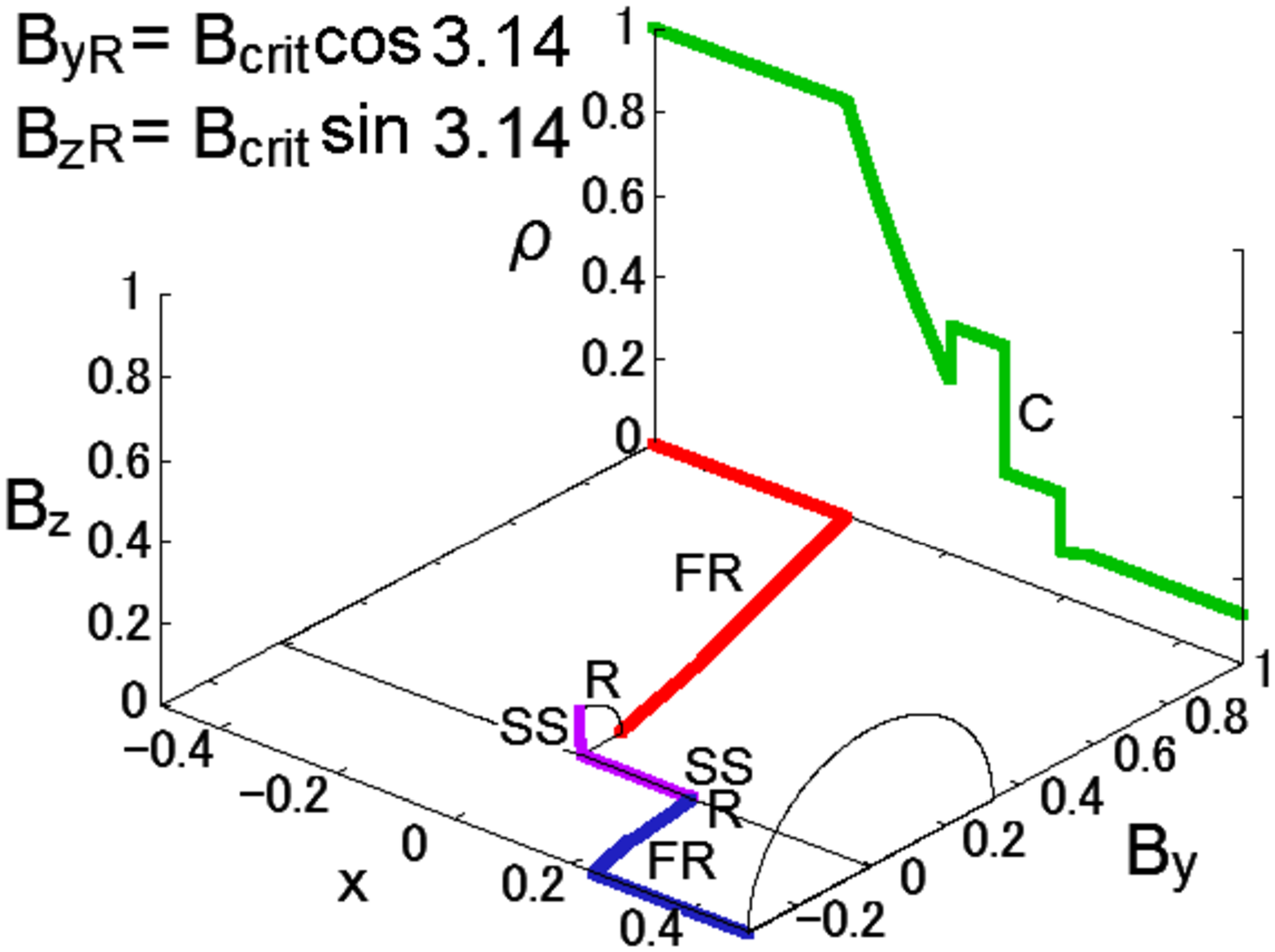}
\end{center}
\end{minipage} &
\begin{minipage}{0.45\hsize}
\begin{center}
\includegraphics[scale=0.31]{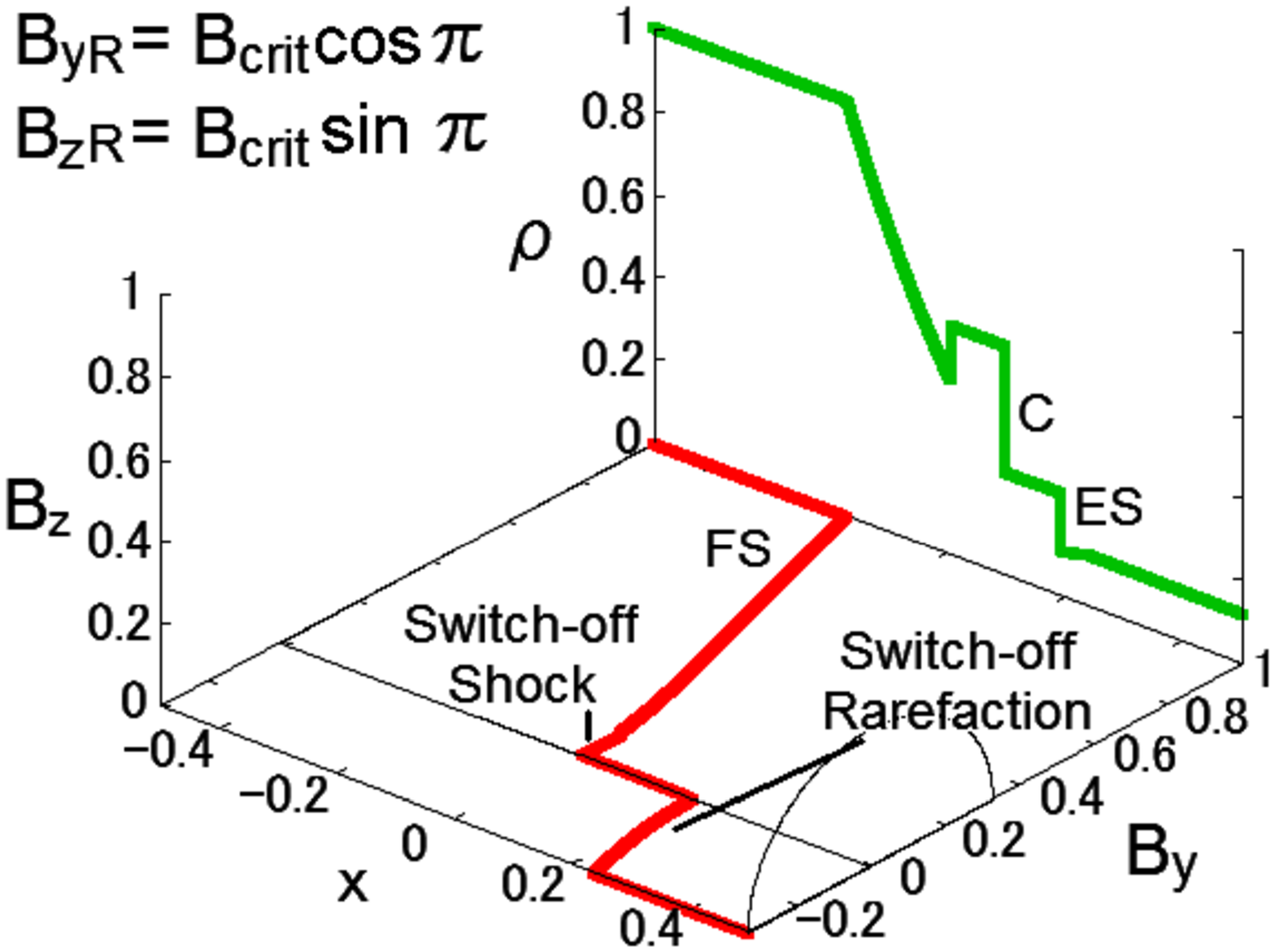}
\end{center}
\end{minipage} 
\end{tabular}
\caption{Non-coplanar neighboring solutions to the solution presented in Fig.~\ref{BW switch-off} as well as 
in the bottom right panel in this figure. The green line displays the density whereas the red, purple and blue 
lines represent the transverse magnetic fields, each on a different plane, the angle of which is shown by an arc. 
Only the direction of the initial transverse magnetic field of the right state is varied in these computations. 
See \S\ref{sec5.2} of the main body for details.}
\label{BW neighborhood 2}
\end{figure}

Next we turn to non-coplanar neighboring problems. This time we change the direction of the initial transverse magnetic field in the right state,
keeping its strength at the critical value $B_{tR} = B_\mathrm{crit} $. In this case we always found regular solutions alone, which are displayed
in Fig.~\ref{BW neighborhood 2}. As is evident, there appear left- and right-going rotational discontinuities when the transverse magnetic
fields are misaligned initially. They are preceded by a fast rarefaction wave and followed by a slow shock running in each direction. As the 
misalignment gets smaller, or the critical point is approached, the amplitudes of the transverse magnetic fields rotated by these rotational discontinuities become smaller. And 
they disappear at the critical point and we recover the solution including the switch-off shock and rarefaction wave presented above. Again
this clearly demonstrates that the non-regular solution at the critical point is a limit of the sequence of the regular solutions. Put another
way, there are only regular solutions in the sufficiently close vicinity of the non-regular solution at the critical point. This is the
main reason why we think that there is no regular solution at the critical point. Note, however, that there might be another regular solution 
that are not included in the neighborhood and eluded our search. 

It is interesting to see how a coplanar neighboring solution is approached by non-coplanar solutions. Note that they are both regular
solutions. As observed above, the left- and right-going rotational discontinuities cooperate to rotate the transverse magnetic field
by the angle imposed by the initial condition. It is found that it is always the right-going rotational discontinuity that rotates 
the transverse magnetic field by a larger angle if $0 < B_{tR} < B_\mathrm{crit}$ and vice versa. As the misalignment becomes smaller and
the coplanar configuration is approached, one of the rotational discontinuities that gives a smaller rotation gets weaker and disappears
for the coplanar configuration. For a non-vanishing but very small misalignment, the relative amplitude of the two rotational discontinuities 
as a function of $B_{tR}$ changes very rapidly near $B_{tR} = B_\mathrm{crit}$. In this sense the non-coplanar solutions shown in 
Fig.~\ref{BW neighborhood 2} for $B_{tR} = B_\mathrm{crit}$ are rather special themselves. 
 
It is finally a legitimate question to ask what has become of the coplanar non-regular solutions including $2 \rightarrow 3$ intermediate shocks 
that are observed near the Brio \& Wu problem in the previous section. This is an interesting issue in its  own right indeed.
As long as $|B_{yR}| > B_\mathrm{crit}$ is satisfied, we find uncountably many non-regular solutions that include a $2 \rightarrow 3$ 
intermediate shock. The sequence of solutions is terminated at one end with the regular solution including a rotational discontinuity. 
The other end of the sequence, on the other hand, is initially the $2 \rightarrow 3,4$ intermediate shock. As  $|B_{yR}|$ becomes smaller, 
however, the $2 \rightarrow 3,4$ intermediate shock cannot be reached at some point and the sequence is terminated with a $2 \rightarrow 3$ 
intermediate shock. At the same time, another non-regular solution with a $2 \rightarrow 4$ intermediate shock appears. This solution 
moves up the slow locus as $|B_{yR}|$ gets smaller, becomes a $1,2 \rightarrow 4$ intermediate shock eventually at 
$B_{yR} \approx -B_{S} =-0.50303$ and disappears. 
In the other case of $|B_{yR}| < B_\mathrm{crit}$, there are again uncountably many non-regular solutions, which consist of a 
right-going $2 \rightarrow 3$ intermediate shock and a slow shock, in addition to the regular solutions that have right-going rotational 
discontinuities. The latter is an end point of the sequence of the former. No solution with a compound wave is found in this case 
and the other end point of the sequence is the $2 \rightarrow 3$ intermediate shock, which approaches the regular solution as $|B_{yR}|$ 
increases and seems to merge at $B_{yR} \approx -B_{Q} = -0.25$. Thereafter only the regular solutions exist. Note finally that all these solutions 
disappear at $B_{yR} =- B_\mathrm{crit}$ and only the non-regular solution with the switch-off shock waves exists.

\section{Summary and Discussions} \label{sec.conclusions}
In order to explore non-unique solutions to the MHD Riemann problem, we have developed the numerical code to obtain weak 
solutions that satisfy the entropy condition, the minimum requirement that the physically relevant solutions have to meet. 
Unlike other preceding ones~\citep[e.g.][]{ATJweb}, our code can treat both regular and non-regular solutions and can also 
handle the initial condition, in which
no transverse magnetic field exists. We have confirmed that the shock waves included in solution satisfy the Rankine-Hugoniot
jump conditions and the rarefaction waves keep the Riemann invariants constant indeed. Details of the code will be presented 
elsewhere~\citep{TY12}. The code will be useful not only for the study of
various solutions as demonstrated in this paper but also for numerical simulations, in which the code will provide the exact solutions to be
compared with and may be implemented to evaluate numerical fluxes. In this paper we have applied the new code to two initial conditions, 
which we believe are useful for the consideration of which weak solutions are really physical, the problem with a long history of controversy. 

The first problem is one of the best known problems for
developers of MHD codes: the Brio \& Wu problem~\citep{BW88}. This particular problem has been frequently utilized for the validation of newly 
developed numerical codes partly because it has a coplanar configuration and 2-dimensional codes can be applied. It has been also 
known for a long time that numerical solutions always involve the compound wave, which is a $2 \rightarrow 3,4$ intermediate shock, 
to which a slow rarefaction wave is attached. This is a non-regular solutions, in which the compound wave does not satisfy the evolutionary
condition and its reality has been a subject of debate. There are two competing arguments so far. Wu and his 
collaborators~\citep{Wu87, BW88, Wu88a, Wu88b, Wu90, Wu&Kennel92} insist that the intermediate shocks are stable and it is the evolutionary
conditions to blame. On the other hand, Falle \& Komissarov~\citep{FK97,FK01} claim that the existence of intermediate shocks in numerical 
solutions are due to the coplanarity, which is inherent in the intermediate shocks. Unless the symmetry is broken somehow, the solution 
will retain the symmetry (coplanarity or planarity) and, as a consequence, we are dealing with the reduced MHD system. Then the point is 
that the $2 \rightarrow 3,4$ intermediate shock that comprises the compound wave in the numerical solution of the Brio \& Wu problem 
becomes regular whereas the regular solution that contains rotational discontinuities, which was shown to exist by \citet{ATJweb} and has 
been also confirmed in this paper, is now non-regular in the reduced MHD system.

It seems that the existence of the uncountably many other solutions that are non-regular both in the full and reduced systems supports
the claim by Falle \& Komissarov~\citep{FK97,FK01}, since they have never been realized in numerical simulations. It is intriguing
to point out, however, that these solutions do not satisfy the evolutionary conditions not because the number of outgoing characteristics
is wrong but because the requirement on the linear independence is not met. This implies that there exit too many neighboring solutions 
to these non-regular solutions and we cannot single out a physically relevant one. This under-determinacy is certainly different from the 
over-determinacy that we commonly find for non-evolutionary shocks and needs further investigations. It should be also mentioned that 
some authors~\citep{Hada94, Markovskii98, I&I07} insist that eventually all intermediate shocks satisfy the evolutionary conditions
if dissipations are taken into account. 
Although this may explain the realization of the compound wave
in the solution of the Brio \& Wu problem, it remains to be addressed why other non-regular solutions that are also evolutionary 
in their analyses but are not produced numerically. They may be linearly unstable. Further investigations are certainly needed.

In the second application we have picked up the initial condition, which appears to possess no regular solution. This initial condition was
found by continuously modifying the original Brio \& Wu problem in various ways. In fact, the initial magnetic field has a coplanar 
configuration and is obtained from that in the Brio \& Wu problem by reducing the strength of the transverse magnetic field in the right
state. We have obtained two sequences of regular solutions as we approach this initial condition from both directions, keeping the 
coplanarity. They are continuously connected at this point by the non-regular solution, which includes both the switch-off shock and 
rarefaction waves. The switch-off shock is non-regular and is located at the boundary between the regular slow shocks and intermediate 
shocks on the slow locus. It is true that we cannot prove the non-existence of regular solutions by these numerical calculations but
the fact that all the neighboring regular solutions, coplanar or non-coplanar, are terminated at this non-regular solution suggests
strongly that regular solution is unlikely to exist. If true, the conventional expectation that there is always a unique regular 
solution is wrong. We are certainly interested in if this non-regular solution is what is realized indeed for this initial condition. 
This will be a difficult task for numerical simulations, however, since the initial condition has to be prepared precisely.  

This initial condition is singular also in the following sense: as the coplanar configurations with $|B_{yR}| > B_\mathrm{crit}$ approach this
critical initial condition, the rotational discontinuity is always left-going whereas the right-going one emerges in the opposite case, 
$|B_{yR}| < B_\mathrm{crit}$, and at the critical condition these rotational discontinuities disappear and the switch-off shock and 
rarefaction wave occur; if the initial transverse magnetic fields are misaligned slightly, then the sequence of the regular solutions 
exhibits a rapid and drastic change of the configuration in the vicinity of the critical initial condition: the left- and right-going 
rotational discontinuities exchange their roles. Such sudden changes of configuration may have a ramification for the stability of 
the regular solution for the non-coplanar initial configurations with $B_{tR} \approx B_\mathrm{crit}$. 

We have also found that the sequence of non-regular solutions behaves in an intriguing manner.
When we approach the critical initial condition from the regime with $|B_{yR}| > B_{\mathrm{crit}}$, we always find uncountably many 
non-regular solutions including a $2 \rightarrow 3$ intermediate shock, which is terminated at one end with the regular solution that 
consists of a rotational discontinuity. The other end point is initially the $2 \rightarrow 3,4$ intermediate shock
that forms a compound wave. For smaller $|B_{yR}|$, however, the end point becomes a $2 \rightarrow 3$ intermediate shock. At the same
time another non-regular solution including a $2 \rightarrow 4$ intermediate shock emerges, moving along the slow Hugoniot loci toward
a $1,2 \rightarrow 4$ intermediate shock, at which this type of solution ceases to exist. When the critical initial condition is
approached from the opposite direction, $|B_{yR}| < B_{\mathrm{crit}}$, we again find uncountably many non-regular solutions including 
a $2 \rightarrow 3$ intermediate shock unless $|B_{yR}|$ is very close to $B_{\mathrm{crit}}$. In contrast to the opposite case, no 
$2 \rightarrow 3,4$ intermediate shock is realized. The sequence of the non-regular solutions shrinks as $|B_{yR}|$ gets larger
and at some point it reduces to the regular solution with a rotational discontinuity and no non-regular solution exists any longer.
For the critical initial condition even this regular solution disappears and only the non-regular solution with the switch-off shock 
exists. These interesting behaviors certainly warrant further investigations.

After all, it seems that the stability of various intermediate shocks needs major reanalysis. The conventional evolutionary conditions are
certainly unsatisfactory. Maybe the introduction of dissipations and its zero limits should be scrutinized. In so doing our new code will
be useful to generate all possible solutions. We believe that the survey and investigation of other initial conditions and the corresponding 
solutions, which manifest some singularities, will provide us with new insights and eventually a clue to the understanding of which solutions
are physically relevant.


\bibliographystyle{jpp}

\bibliography{bibliography}

\end{document}